\documentclass[11pt]{article}

\pdfoutput=1

\usepackage[dvipsnames]{xcolor}
\usepackage{amssymb,amsmath,amscd}
\usepackage{epsfig}
\usepackage{epstopdf}
\usepackage{latexsym}
\usepackage{graphicx}
\usepackage{booktabs}
\usepackage{bbm}
\usepackage[margin=15pt,small]{caption}
\usepackage{subcaption}
\usepackage{enumitem}
\usepackage{float}
\usepackage[toc]{appendix}
\usepackage[numbers,compress]{natbib}
\usepackage{tikz}
\usepackage[labelfont=bf]{caption}

\usepackage{braket,cellspace}
\usepackage{mathrsfs}
\usepackage{braket}
\usepackage{chngpage}
\usepackage{tabu}

\usetikzlibrary{positioning}
\definecolor{darkspringgreen}{rgb}{0.09, 0.45, 0.27}

\usepackage{datetime}
\usepackage[
      colorlinks=true,
      linkcolor=darkspringgreen,  
      urlcolor=darkspringgreen,    
      filecolor=darkspringgreen,     
      citecolor=darkspringgreen,
      linktocpage=true,
      pdfstartview=FitV,
      bookmarksopen=true    
      ]{hyperref}

\setlist{nolistsep}
\ifpdf
\fi

\let\oldbibliography\thebibliography
\renewcommand{\thebibliography}[1]{\oldbibliography{#1}
\setlength{\itemsep}{0pt}} 

\numberwithin{equation}{section} 

\begin{document}  

\begin{titlepage}

\begin{center} 

\vspace*{23mm}

{\LARGE \bf Global aspects of conformal symmetry
 \\[7 pt] and the ANEC in dS and AdS}

\bigskip
\bigskip
\bigskip
\bigskip

{\bf Felipe Rosso\\ }
\bigskip
Department of Physics and Astronomy \\
University of Southern California \\
Los Angeles, CA 90089, USA  \\
\bigskip
\tt{felipero@usc.edu}  \\
\end{center}

\bigskip

\begin{abstract}
\noindent Starting from the averaged null energy condition (ANEC) in Minkowski we show that conformal symmetry implies the ANEC for a conformal field theory (CFT) in a de Sitter and anti-de Sitter background. A similar and novel bound is also obtained for a CFT in the Lorentzian cylinder. Using monotonicity of relative entropy, we rederive these results for dS and the cylinder. As a byproduct we obtain the vacuum modular Hamiltonian and entanglement entropy associated to null deformed regions of CFTs in (A)dS and the cylinder. A third derivation of the ANEC in dS is shown to follow from bulk causality in AdS/CFT. Finally, we use the Tomita-Takesaki theory to show that Rindler positivity of Minkowski correlators generalizes to conformal theories defined in dS and the cylinder.
\end{abstract}

\vfill

\end{titlepage}


\newpage

\setcounter{tocdepth}{2}
\tableofcontents

\section{Introduction and summary}
\label{sec:intro}

The main focus of this work is the averaged null energy condition (ANEC), defined for an arbitrary quantum field theory (QFT) on a fixed space-time $g_{\mu \nu}$ as
\begin{equation}\label{eq:210}
\int_{-\infty}^{+\infty}d\lambda\,
  k^\mu k^\nu 
  T_{\mu \nu}\ge  0\ ,
\end{equation}
where $T_{\mu \nu}$ is the stress tensor operator and $k^\mu$ is the tangent vector over a complete null geodesic with affine parameter $\lambda$. The original motivation for considering this condition comes from general relativity, where it is a reasonable substitute for the null energy condition $k^\mu k^\nu T_{\mu \nu}\ge 0$, known to fail in quantum theories. The ANEC can be used to rule out space-times with certain unwanted features~\cite{Morris,Ori:1993eh,Alcubierre:1994tu}, as well as for proving classic theorems in general relativity \cite{Tipler:1978zz,Borde:1987qr,Roman:1986tp}.\footnote{See \cite{Freivogel:2018gxj,Leichenauer:2018tnq} for a related but different bound recently proposed in semi-classical gravity.} Even in the simplest case of a QFT in Minkowski, the ANEC has been applied to obtain very interesting results such as the conformal collider bounds of Ref. \cite{Hofman:2008ar}.

Although the ANEC in Minkowski has been proven for general QFTs in Refs. \cite{Faulkner:2016mzt,Hartman:2016lgu,Longo:2018obd}, the question still remains whether it is a true statement of quantum theories defined in more general backgrounds. In this work we take a few steps in this direction and prove the ANEC for arbitrary conformal field theories (CFTs) defined on fixed de Sitter and anti-de Sitter space-times. Moreover, for a CFT in the Lorentzian cylinder $\mathbb{R}\times S^{d-1}$ we obtain a similar condition given by
\begin{equation}\label{eq:209}
\int_{-\pi/2}^{\pi/2}d\bar{\lambda}\,
  \cos^d(\bar{\lambda})
  k^\mu k^\nu 
  T_{\mu \nu}\ge  0\ ,
\end{equation}
where $\bar{\lambda}$ is affine and the null geodesic is not complete but goes between antipodal points in the spatial sphere $S^{d-1}$. The stress tensor in (\ref{eq:209}) is vacuum subtracted $T_{\mu \nu}\equiv T_{\mu \nu}-\bra{0}T_{\mu \nu}\ket{0}$ in order to avoid a trivial violation due to some constant Casimir energy.\footnote{Given that (A)dS are maximally symmetric space-times, this is not necessary for the ANEC in (\ref{eq:210}). For more details see discussion around (\ref{eq:181}).} 

We start in Sec. \ref{sec:ANEC_mapping}, where we derive the three constraints in (A)dS and the cylinder in a simple way. Given that the ANEC in Minkowski has been well established for general QFTs \cite{Faulkner:2016mzt,Hartman:2016lgu,Longo:2018obd}, we start from this condition and apply certain conformal transformations from Minkowski to these space-times.\footnote{See Refs. \cite{Visser:1994jb,Urban:2009yt} for previous studies on the behavior of the ANEC under conformal transformations.} After the mapping, the resulting constraint gives the ANEC in (A)dS and the bound~(\ref{eq:209}) for the cylinder. To implement these transformations appropriately we must carefully deal with the fact that the conformal group is only globally well defined in the Lorentzian cylinder.\footnote{For a pedagogical introduction see David Simmons-Duffin's lecture notes on TASI 2019 (although to this date the notes are not complete, they are still very useful). Another useful explanation is given in the first secion of Ref. \cite{Brunetti:1992zf}.} Since this plays an important role in this work, let us briefly explain its significance.

\begin{figure}[t]
\begin{center}
\includegraphics[height=2.45 in]{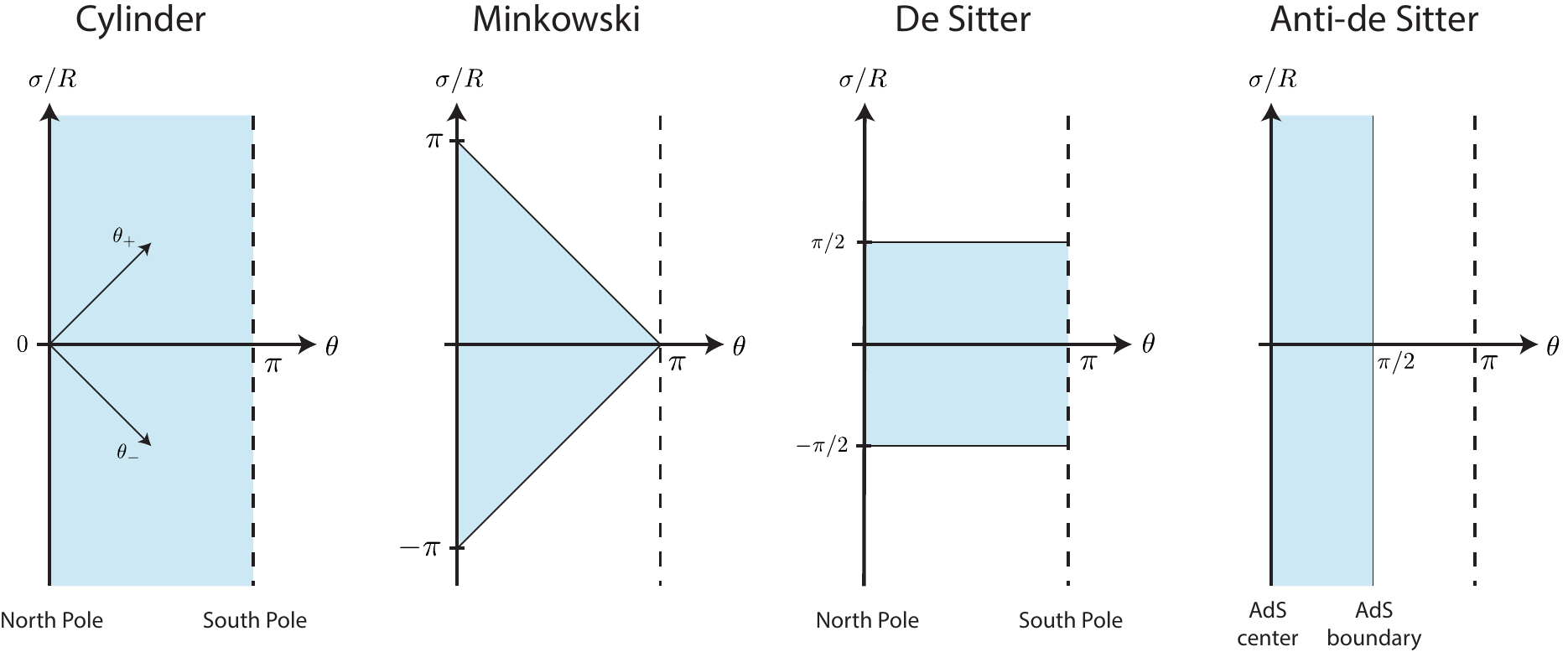}
\caption{Diagrams illustrating the effect of the conformal transformations given in table \ref{table:2} when applied to the Lorentzian cylinder. The whole infinite strip on the left diagram corresponds to the cylinder, with the North and South pole at $\theta=0$ and $\theta=\pi$ respectively.}
\label{fig:14}
\end{center}
\end{figure}

The Lorentzian cylinder $\mathbb{R}\times S^{d-1}$ can be represented by an infinite strip in the $(\sigma/R,\theta)$ plane, where $\sigma\in \mathbb{R}$ is the time coordinate and $\theta\in[0,\pi]$, with the end points corresponding to the poles of the spatial sphere $S^{d-1}$ of radius $R$, see Fig. \ref{fig:14}. The conformal transformations relating the cylinder, Minkowski and (A)dS are essentially given by different ways of cutting out regions of this infinite strip. When mapping a curve (or surface) from one space-time into another it is crucial that we keep track of this, since a given curve may not fit inside some of the sections of the strip shown in Fig.~\ref{fig:14}. The key technical feature of (A)dS that enables the derivation of the ANEC is that a complete and affinely parametrized null geodesic in Minkowski is also complete and affine in (A)dS. Since this is not true for the Lorentzian cylinder, we do not obtain the ANEC in this case but the constraint in~(\ref{eq:209}).

In Sec. \ref{sec:null_energy_bounds} we investigate whether an independent proof of these results can be obtained from monotonicity of relative entropy, as done in Ref. \cite{Faulkner:2016mzt} for the Minkowski ANEC. We do so by first computing the vacuum modular Hamiltonians of null deformed regions in these space-times, which we obtain by conformally mapping the Minkowski modular operator associated to null deformations of Rindler \cite{Casini:2017roe}.  The appropriate conformal transformations are a slight modification from the ones used in Sec. \ref{sec:ANEC_mapping}. The case of dS is particularly simple, where we show that the modular Hamiltonian associated to null deformations of the static patch is given by
\begin{equation}\label{eq:213}
K_{\rm dS}=
  2\pi R^{d-2}
  \int_{S^{d-2}}d\Omega(\vec{x}_\perp)
  \int_{\bar{A}(\vec{x}_\perp)}^{+\infty}
  d\eta\,
  \left(
  \eta-\bar{A}(\vec{x}_\perp)
  \right)
  T_{\eta \eta}(\eta,\vec{x}_\perp)\ ,
\end{equation}
where for fixed $\vec{x}_\perp$, $\eta$ is an affine parameter in dS and the stress tensor is projected along this direction. For $\bar{A}(\vec{x}_\perp)=0$ the integral is over the future horizon of the de Sitter static patch, while arbitrary~$\bar{A}(\vec{x}_\perp)$ corresponds to null deformations. Using this together with monotonicity of relative entropy gives the ANEC in dS. Although a similar procedure results in the bound in the cylinder~(\ref{eq:209}), it does not generalize to the AdS case due to some technical issues related to our previous comment on the global definition of the conformal group. We finish Sec. \ref{sec:null_energy_bounds} by computing the universal terms of the entanglement entropy associated to the null deformed modular Hamiltonians in (A)dS and the cylinder. The details of the computations are summarized in App. \ref{zapp:entanglement}, where we build on some results of Ref. \cite{Casini:2018kzx} using AdS/CFT.

We continue in Sec. \ref{sec:ref_positivity}, where we explore some aspects that would be necesary to generalize the causality proof of the Minkowski ANEC \cite{Hartman:2016lgu} to these curved space-times. In particular, we study one of its crucial ingredients, the ``wedge reflection positivity" or ``Rindler positivity", which for two scalar operators can be written as
\begin{equation}\label{eq:212}
\bra{0}
  \mathcal{O}^\dagger(\widetilde{X}^\mu)
  \mathcal{O}(X^\mu)\ket{0}>0\ ,
  \qquad \qquad
  \widetilde{X}^\mu(X^\mu)=(-T,-X,\vec{Y})\ ,
\end{equation}
where $X^\mu=(T,X,\vec{Y})$ are Cartesian coordinates in Minkowski and $X^\mu$ must satisfy $X>|T|$. This property was derived in Ref. \cite{Casini:2010bf} from the Tomita-Takesaki theory \cite{Haag:1992hx,Witten:2018lha}. Using the conformal transformations of Sec. \ref{sec:null_energy_bounds} we map the Bisognano-Wichmann Tomita operator \cite{Bisognano:1976za} to the CFTs in the Lorentzian cylinder and de Sitter, and show that a generalized version of (\ref{eq:212}) holds in these backgrounds. The resulting property for the cylinder is particularly interesting since unlike (\ref{eq:212}), the transformation $\widetilde{X}^\mu$ is non-linear.\footnote{The wedge reflection positivity for the CFT in the Lorentzian cylinder and de Sitter for operators of arbitrary even spin are given in (\ref{eq:211}) and (\ref{eq:205}) respectively.}

The third (and last) independent proof of the ANEC in de Sitter is based on AdS/CFT and given in appendix \ref{zapp:ANEC_holography}. We show that the approach of Ref. \cite{Kelly:2014mra} used to derive the Minkwoski ANEC for holographic theories described by Einstein gravity can be naturally extended to de Sitter. We should mention that while this work was in preparation Ref. \cite{Iizuka:2019ezn} used a similar method to derive the bound in the Lorentzian cylinder (\ref{eq:209}) for space-time dimensions $d=3,4,5$ and holographic CFTs dual to Einstein gravity.

We finish in Sec. \ref{sec:Discussion} with a discussion of our results and several future research directions. In particular we comment on the connection between these bounds and the quantum null energy condition~(QNEC). Using the modular Hamiltonian in (\ref{eq:213}), we point out that the QNEC in de Sitter can be written in terms of the second order variation of relative entropy.  

\vspace{5pt}

\noindent \textbf{Note added:} In the follow up work of Ref. \cite{Rosso:2020cub}, the proofs of the achronal ANEC for CFTs in de Sitter and anti-de Sitter given in this paper were extended to arbitrary QFTs (beyond conformal symmetry) using the methods of Ref. \cite{Faulkner:2016mzt}.

\section{ANEC in (A)dS from conformal symmetry}
\label{sec:ANEC_mapping}

In this section we map the null plane in Minkowski to the Lorentzian cylinder, de Sitter and anti-de Sitter. After describing the geometric aspects of the transformation we apply it to the ANEC operator in Minkowski space-time. This allows us to obtain the ANEC for CFTs in (A)dS and a similar novel bound for theories defined in the cylinder. 

\subsection{Taking the null plane on a conformal journey - Take I}

The conformal transformations relating Minkowski, the cylinder and (A)dS have been known for a long time \cite{Candelas:1978gf}. The simplest way to introduce them is to start from the metric in the Lorentzian cylinder $\mathbb{R}\times S^{d-1}$ written as
\begin{equation}\label{eq:170}
ds^2_{LC}=
  -d\sigma^2+R^2\left(
  d\theta^2+\sin^2(\theta)d\Omega^2(\vec{v}\,)
  \right)\ ,
\end{equation}
where $\sigma\in \mathbb{R}$ is the time coordinate and $\theta\in [0,\pi]$, with the end points corresponding to the North and South pole of the spatial sphere $S^{d-1}$ of radius $R$. The line element $d\Omega^2(\vec{v}\,)$ is given by 
\begin{equation}\label{eq:152}
d\Omega^2(\vec{v}\,)=
  \left(
  \frac{2L}{L^2+|\vec{v}\,|^2}
  \right)^2d\vec{v}.d\vec{v}\ ,
\end{equation}
which corresponds to a unit sphere $S^{d-2}$ in stereographic coordinates $\vec{v}\in \mathbb{R}^{d-2}$. The length scale $L$ can be any, not necessarily related to $R$.\footnote{To obtain the $S^{d-2}$ in terms of the usual angles we describe the vector $\vec{v}\in \mathbb{R}^{d-2}$ in spherical coordinates and then parametrize its radius according to $|\vec{v}\,|=L\tan(\phi/2)$ with $\phi\in[0,\pi]$.} This cylinder manifold can be represented by an infinite strip in the $(\sigma/R,\theta)$ plane, as shown in the first diagram of Fig. \ref{fig:14}, where the North and South pole are given by the vertical lines at $\theta=0$ and $\theta=\pi$ respectively. Other values of $\theta\in(0,\pi)$ in this diagram corresponds to a unit sphere $S^{d-2}$. 

Conformal transformations in the cylinder are essentially given by different ways of cutting this infinite strip. The cutting is implemented by a change of coordinates which puts the metric of the cylinder in the form $ds_{LC}^2=w^2d\bar{s}^2$, followed by a Weyl rescaling which removes the conformal factor~$w^2$. Effectively, this maps a section of the Lorentzian cylinder to the space-time $d\bar{s}^2$. Through this procedure we can obtain Minkowski and (A)dS.\footnote{Starting from the Lorentzian cylinder, Ref. \cite{Candelas:1978gf} discusses some additional conformal relations. Although in this work we restrict to Minkowski and (A)dS, a similar treatment is possible in these other cases.} The appropriate change of coordinates and conformal factors in each case are indicated in table \ref{table:2}. From this it is straightforward to see that each of the transformations cuts the infinite strip as given in Fig. \ref{fig:14}. For instance, in the Minkowski case we see that $r_\pm \in \mathbb{R}$ translates into $\theta_\pm \in [-\pi,\pi]$ together with the implicit constraint $\theta\in [0,\pi]$.

\begin{table}
\begin{adjustwidth}{-.8in}{-.8in}  
\centering
\setlength{\tabcolsep}{5pt}
\scalebox{0.90}{%
{\tabulinesep=1.0mm
\begin{tabu}{ Sl | Sl | Sl | Sl }
\specialrule{.13em}{0em}{0em}
\textbf{Map to} &
\textbf{New coordinates} &
\textbf{Conformal factor} $w^2$ &
\textbf{Transformed space-time}  \\
\specialrule{.13em}{0em}{0em}
$\mathbb{R}\times \mathbb{R}^{d-1}$ &
$\displaystyle 
  r_\pm=R\tan(\theta_\pm/2)$ &
$\displaystyle 
 \frac{(R^2+r_+^2)(R^2+r_-^2)}
  {4R^2}$ &
$\displaystyle 
 -dt^2+dr^2+r^2d\Omega^2(\vec{v}\,)$
\\
\specialrule{.05em}{0em}{0em}
dS &
$\displaystyle 
  \cosh(t_s/R)=1/\cos(\sigma/R)$ &
$\displaystyle 
 \cos^2(\sigma/R)$  &
$\displaystyle 
 -dt_s^2+R^2\cosh^2(t_s/R)
  \left[
  d\theta^2+\sin^2(\theta)d\Omega^2(\vec{v}\,)
  \right]$ \\
\specialrule{.05em}{0em}{0em}
AdS &
$\displaystyle 
  \rho=R\tan(\theta)$ & 
$\displaystyle 
 \cos^2(\theta)$  &
$\displaystyle 
  -
  \frac{\rho^2+R^2}{R^2}
  d\sigma^2
  +
  \frac{R^2}{\rho^2+R^2}d\rho^2
  +\rho^2d\Omega^2(\vec{v}\,)
  $ 
\\ 
\specialrule{.13em}{0em}{0em}
\end{tabu}}}
\end{adjustwidth} 
\caption{Details of the conformal transformations relating the Lorentzian cylinder to various space-times. We indicate the new coordinates, the conformal factor $w^2$ obtained from the change of coordinates $ds^2_{LC}=w^2d\bar{s}^2$ and the metric of the transformed space-time. The null coordinates in the cylinder are $\theta_\pm=\sigma/R\pm \theta$, while in Minkowski we define $r_\pm=r\pm t$ with the radius $r\ge 0$ and $t\in \mathbb{R}$. For the (A)dS space-times we have $t_s\in \mathbb{R}$ and $\rho\ge 0$. }\label{table:2}
\end{table}

The way in which we have written the metrics in (A)dS in table \ref{table:2} is (probably) the most familiar form but not the most convenient to describe null surfaces, which is ultimately what we are interested in. A more suitable description of these space-times is given directly in terms of the coordinates in the cylinder
\begin{equation}\label{eq:154}
\begin{aligned}
ds^2_{\rm dS}&=
  \frac{-d\sigma^2+R^2\left(
  d\theta^2+\sin^2(\theta)d\Omega^2(\vec{v}\,)
  \right)}{\cos^2(\sigma/R)}\ ,\\
ds^2_{\rm AdS}&=
  \frac{-d\sigma^2+R^2\left(
  d\theta^2+\sin^2(\theta)d\Omega^2(\vec{v}\,)
  \right)}{\cos^2(\theta)}\ .
\end{aligned}
\end{equation}
Changing to $t_s$ and $\rho$ given in table \ref{table:2}, we obtain the more familiar forms of (A)dS. Notice that due to the denominators in (\ref{eq:154}) the range of $\sigma$ is restricted to $|\sigma/R|\le \pi/2$ for dS while $\theta\in[0,\pi/2]$ in AdS. This implements the cutting of the infinite strip as sketched in Fig. \ref{fig:14}.

Let us now consider the null plane in $d$-dimensional Minkowski and analyze its transformation properties under these mappings. Taking Cartesian coordinates $X^\mu=(T,X,\vec{Y})$ in Minkowski, the null plane ${X_-=X-T=0}$ can be parametrized in terms of $(\lambda,\vec{x}_\perp)$ as
\begin{equation}\label{eq:50}
\mathcal{N}_{\rm plane}=
  \left\lbrace
  X^\mu \in \mathbb{R}\times \mathbb{R}\times \mathbb{R}^{d-2}:
  \quad
  X^\mu(\lambda,\vec{x}_\perp)=
  (\lambda,\lambda,\vec{x}_\perp)\ ,
  \quad
  (\lambda,\vec{x}_\perp)\in \mathbb{R}\times \mathbb{R}^{d-2}
  \right\rbrace .
\end{equation}
For fixed $\vec{x}_\perp$ the curve $X^\mu(\lambda)$ trivially satisfies the geodesic equation
\begin{equation}\label{eq:63}
\frac{d^2X^\mu }{d\lambda^2}+
  \Gamma^\mu_{\alpha \beta}
  \frac{dX^\alpha}{d\lambda}
  \frac{dX^\beta}{d\lambda}=0\ ,
\end{equation}
since the connection $\Gamma^\mu_{\alpha \beta}$ vanishes in these coordinates. This means that $\lambda$ is an affine parameter while we can think of $\vec{x}_\perp$ as a label going through the different geodesics.

Since the transformation from Minkowski to the cylinder in table \ref{table:2} is given in terms of radial null coordinates $r_\pm=r\pm t$, it is convenient to first change from the Cartesian spatial coordinates~$(X,\vec{Y})$ to spherical. We can do this by defining $(r,\vec{v}\,)$ according to\footnote{The inverse transformation is given by
$(X,\vec{Y})=r
\left(|\vec{v}\,|^2-4R^2,4R\vec{v}
\right)/\left(|\vec{v}\,|^2+4R^2\right)$.}
\begin{equation}\label{eq:187}
r=\big(X^2+|\vec{Y}|^2\big)^{1/2}\ ,
  \qquad \qquad
  \vec{v}=
  \frac{2R\vec{Y}}{X
  +\big(X^2+|\vec{Y}|^2\big)^{1/2}}\ .
\end{equation}
Using this together with (\ref{eq:50}) we can write the null plane in spherical coordinates, where the Minkowski metric is ${ds^2=-dt^2+dr^2+r^2d\Omega^2(\vec{v}\,)}$.\footnote{The metric in the unit sphere $d\Omega^2(\vec{v}\,)$ is given in (\ref{eq:152}) with $L=2R$.} The conformal mapping from Minkowski to the cylinder is then applied by writing $r_\pm=R\tan(\theta_\pm/2)$ with ${\theta_\pm=\theta\pm \sigma/R}$, so that the null surface in the cylinder coordinates $v^\mu=(\theta_+,\theta_-,\vec{v}\,)$ becomes
\begin{equation}\label{eq:188}
v^\mu(\lambda,\vec{x}_\perp)=
  \left(
  \theta_+(\lambda,\vec{x}_\perp),
  \theta_-(\lambda,\vec{x}_\perp),
  \frac{2R\vec{x}_\perp}
  {\lambda+\sqrt{\lambda^2+|\vec{x}_\perp|^2}}
  \right)\ ,
  \qquad 
  (\lambda,\vec{x}_\perp)\in \mathbb{R}\times \mathbb{R}^{d-2}\ ,
\end{equation}
where
\begin{equation}\label{eq:189}
\theta_\pm(\lambda,\vec{x}_\perp)=
  2\,{\rm arctan}\left(
  \frac{\sqrt{\lambda^2+|\vec{x}_\perp|^2}\pm \lambda}{R}
  \right)\ .
\end{equation}
If we evaluate the conformal factor associated to this transformation and given in table \ref{table:2} along the surface we find
\begin{equation}\label{eq:191}
w^2(\lambda,\vec{x}_\perp)=
  \frac{4R^2\lambda^2+(R^2+|\vec{x}_\perp|^2)^2}
  {4R^4}\ .
\end{equation}

To understand the surface let us analyze its behavior for fixed values of $\vec{x}_\perp$. The geodesic equation~(\ref{eq:63}) is not invariant under the conformal transformations since the connection transforms with an additional term under the Weyl rescaling, and becomes
\begin{equation}\label{eq:64}
\frac{d^2v^\mu}{d\lambda^2}+
  \bar{\Gamma}^\mu_{\alpha \beta}
  \frac{dv^\alpha}{d\lambda}
  \frac{dv^\beta}{d\lambda}=
  \left[
  -2\frac{d}{d\lambda}
  \ln\left(w(\lambda)\right)
  \right]\frac{dv^\mu }{d\lambda}\ ,
\end{equation}
where $\bar{\Gamma}^\mu_{\alpha \beta}$ is the connection in the cylinder. One can explicitly check that the curve (\ref{eq:188}) has a null tangent vector which satisfies this equation for any value of $\vec{x}_\perp$. Altogether, this means that $v^\mu(\lambda,\vec{x}_\perp)$ is (as expected) a null geodesic, even though $\lambda$ is not affine anymore due to the non-vanishing term on the right-hand side of (\ref{eq:64}). This additional term can be canceled by defining an appropriate affine parameter $\bar{\lambda}(\lambda)$ according to
\begin{equation}\label{eq:78}
\bar{\lambda}''(\lambda)=
\left[
  -2\frac{d}{d\lambda}
  \ln\left(w(\lambda)\right)
  \right]
  \bar{\lambda}'(\lambda)
  \qquad \Longrightarrow \qquad
  \bar{\lambda}(\lambda)=
  c_0
  \int \frac{d\lambda}{w^2(\lambda)}+
  c_1\ ,
\end{equation}
where $c_0$ and $c_1$ are integration constants which can depend on the transverse coordinates $\vec{x}_\perp$. Using~(\ref{eq:191}) we can evaluate this explicitly and obtain an affine parameter in the cylinder
\begin{equation}\label{eq:190}
  \lambda(\bar{\lambda},\vec{x}_\perp)=
  \frac{R^2+|\vec{x}_\perp|^2}{2R}
  \tan(\bar{\lambda})\ ,
  \qquad \qquad
  |\bar{\lambda}|\le \pi/2\ ,
\end{equation}
where we have conveniently fixed the integration constants $c_0$ and $c_1$.

Let us analyze the behavior of each of these geodesics. For any value of $\vec{x}_\perp$ all the curves begin and end at the same space-time points, given by
\begin{equation}\label{eq:194}
(\sigma/R,\theta,|\vec{v}|\,)
  \big|_{\rm initial}=
  \left(
  -\frac{\pi}{2},\frac{\pi}{2},+\infty
  \right)\ ,
  \qquad \qquad
  \left(\sigma/R,\theta,|\vec{v}|\,\right)
  \big|_{\rm final}=
  \left(
  \frac{\pi}{2},\frac{\pi}{2},0
  \right)\ .
\end{equation}
Remember that the $S^{d-2}$ in the cylinder metric (\ref{eq:170}) is parametrized in stereographic coordinates~$\vec{v}$, so that $|\vec{v}|$ equal to zero and infinity correspond to antipodal points in the $S^{d-2}$. This means that both the initial and final points lie on the equator $\theta=\pi/2$ of the spatial sphere $S^{d-1}$, but on opposite sides. As the affine parameter takes values in $\bar{\lambda}\in(-\pi/2,\pi/2)$, the curves travel between these points without intersecting and covering the whole sphere. 

Some special values of $\vec{x}_\perp$ have particularly simple trajectories. For instance, the geodesics with~$|\vec{x}_\perp|=R$ always stay on the equator $\theta=\pi/2$, and are parametrized according to
\begin{equation}\label{eq:192}
{\rm For\,\,}|\vec{x}_\perp|=R
  \qquad \Longrightarrow \qquad
  v^\mu(\bar{\lambda},\vec{x}_\perp)=
  \left(
  \frac{\pi}{2}+\bar{\lambda},
  \frac{\pi}{2}-\bar{\lambda},
  \frac{2\vec{x}_\perp}
  {\tan(\bar{\lambda})+\sec(\bar{\lambda})}
  \right)\ .
\end{equation}
Other simple curves are given by $|\vec{x}_\perp|$ equal to zero or infinity, which corresponds to trajectories that go through the North and South pole of $S^{d-1}$ respectively. Their motion in the $\vec{v}$ coordinate is always constant expect at the pole where it discontinuously changes from zero to infinity. 

\begin{figure}[t]
\begin{center}
\includegraphics[height=2.7 in]{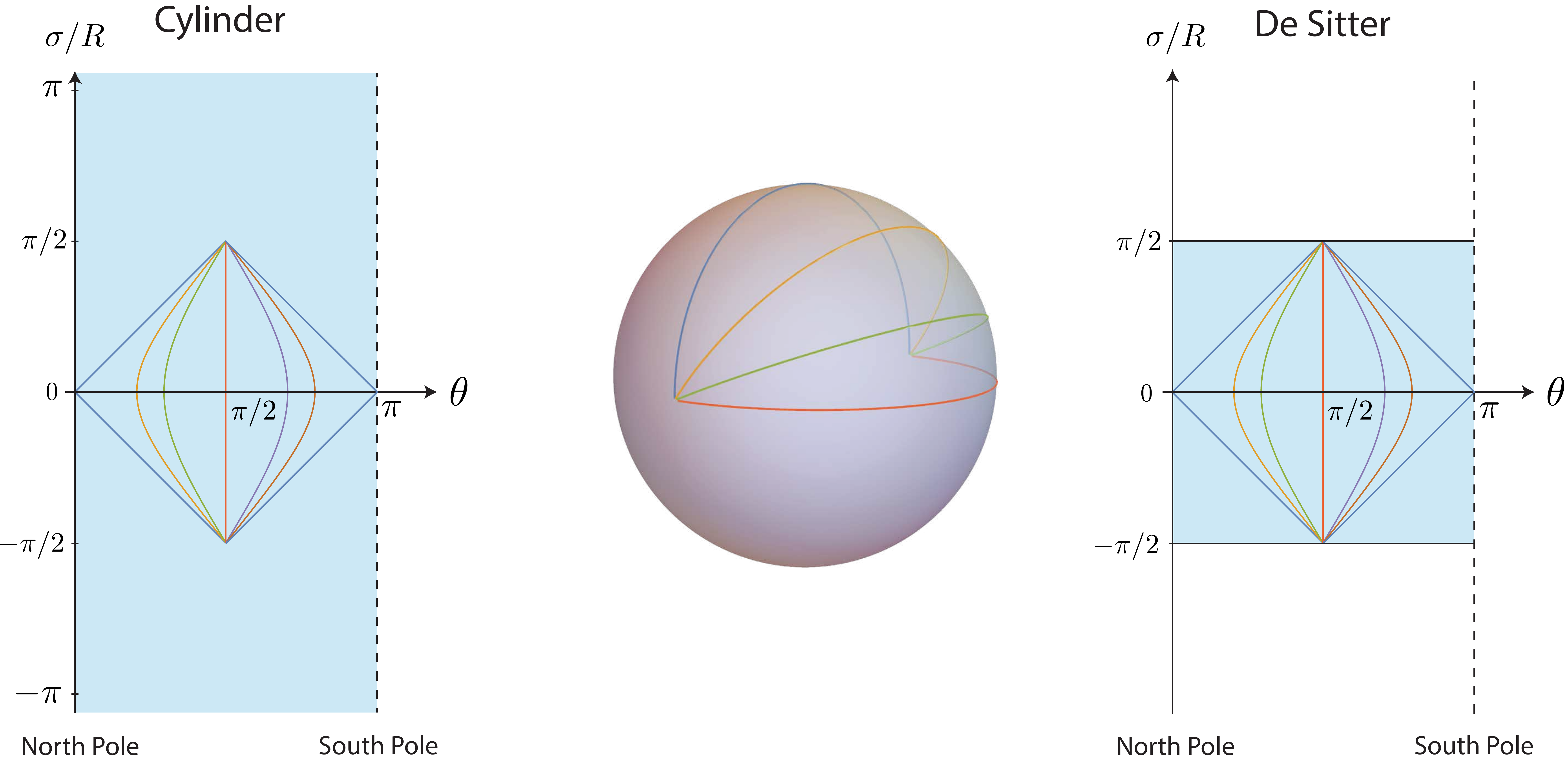}
\caption{Geodesics in the Lorentzian cylinder and de Sitter for several values of $\vec{x}_\perp$. The diagram in the left does not contain all the information since it is missing the motion in the coordinates $\vec{v}$ on the~$S^{d-2}$. In the center diagram, we plot some trajectories for the case $d=3$ where the spatial section of the cylinder is given by $S^2$. Equal colors in each diagram correspond to the same geodesics. To the right we have the geodesics in the $(\sigma/R,\theta)$ plane together with the region covered by de Sitter. Since the topology of dS is the same as the cylinder, the trajectories in dS are also given by the center diagram.}
\label{fig:15}
\end{center}
\end{figure}

For all other values of $|\vec{x}_\perp|$ the curves travel along other possible paths in the sphere without intersecting. In the center diagram of Fig. \ref{fig:15} we show some trajectories for the case $d=3$, where the spatial section of the cylinder is an $S^2$.\footnote{To plot the curves on the $S^2$ it is useful to write the stereographic coordinate $v\in \mathbb{R}$ as $v=2R\tan(\phi/2)$ with~$|\phi|\le \pi$ and then consider Cartesian coordinates~$(x,y,z)$ in terms of the spherical angles $(\theta,\phi)$. Using (\ref{eq:188}) this gives $(x,y,z)$ in terms of $(\lambda,\vec{x}_\perp)$ so that the curves always lie on the surface of the $S^2$.} For higher dimensions we can represent the geodesics in the~$(\sigma/R,\theta)$ plane as shown in the left diagram of that figure. Although all these curves are null, they are not necessarily at an angle of $\pi/4$ since they have a non-trivial motion in the coordinate $\vec{v}$. Only for $|\vec{x}_\perp|$ equal zero and infinity the coordinate $\vec{v}$ remains constant and the curves have an angle of~$\pi/4$ in the $(\sigma/R,\theta)$ plane. 

The mapping of this surface to (A)dS is straightforward since it only involves the Weyl rescaling in (\ref{eq:154}). Using this, the conformal factors connecting Minkowski to (A)dS evaluated along the null surface can be computed from (\ref{eq:189}) and (\ref{eq:191})
\begin{equation}\label{eq:197}
w^2_{\rm dS}(\vec{x}_\perp)=
  \left(
  \frac{R^2+|\vec{x}_\perp|^2}{2R^2}
  \right)^2\ ,
  \qquad \qquad
  w^2_{\rm AdS}(\vec{x}_\perp)=\left(
  \frac{R^2-|\vec{x}_\perp|^2}{2R^2}
  \right)^2\ .
\end{equation}
Note that in both cases the results are independent of $\lambda$. This apparently innocent observation will have very deep consequences. In particular, it means that the affine parameter $\lambda$ in the null plane is also affine in (A)dS, since the right-hand side of the geodesic equation (\ref{eq:64}) automatically vanishes.

For de Sitter we plot the geodesics in the $(\sigma/R,\theta)$ plane in the right diagram of Fig. \ref{fig:15}. All curves fit exactly inside in the space-time, traveling from the boundary at past infinity to future infinity. Since the topology of de Sitter is the same as the cylinder $\mathbb{R}\times S^{d-1}$, with a time dependent radius~$S^{d-1}$, the trajectories are the same as for the cylinder shown in the center diagram of Fig.~\ref{fig:15}. The difference is that the curves in de Sitter cannot be extended beyond their initial and final points, since they encounter the dS boundaries at $|\sigma/R|=\pi/2$.

\begin{figure}[t]
\begin{center}
\qquad
\includegraphics[height=3.0	 in]
{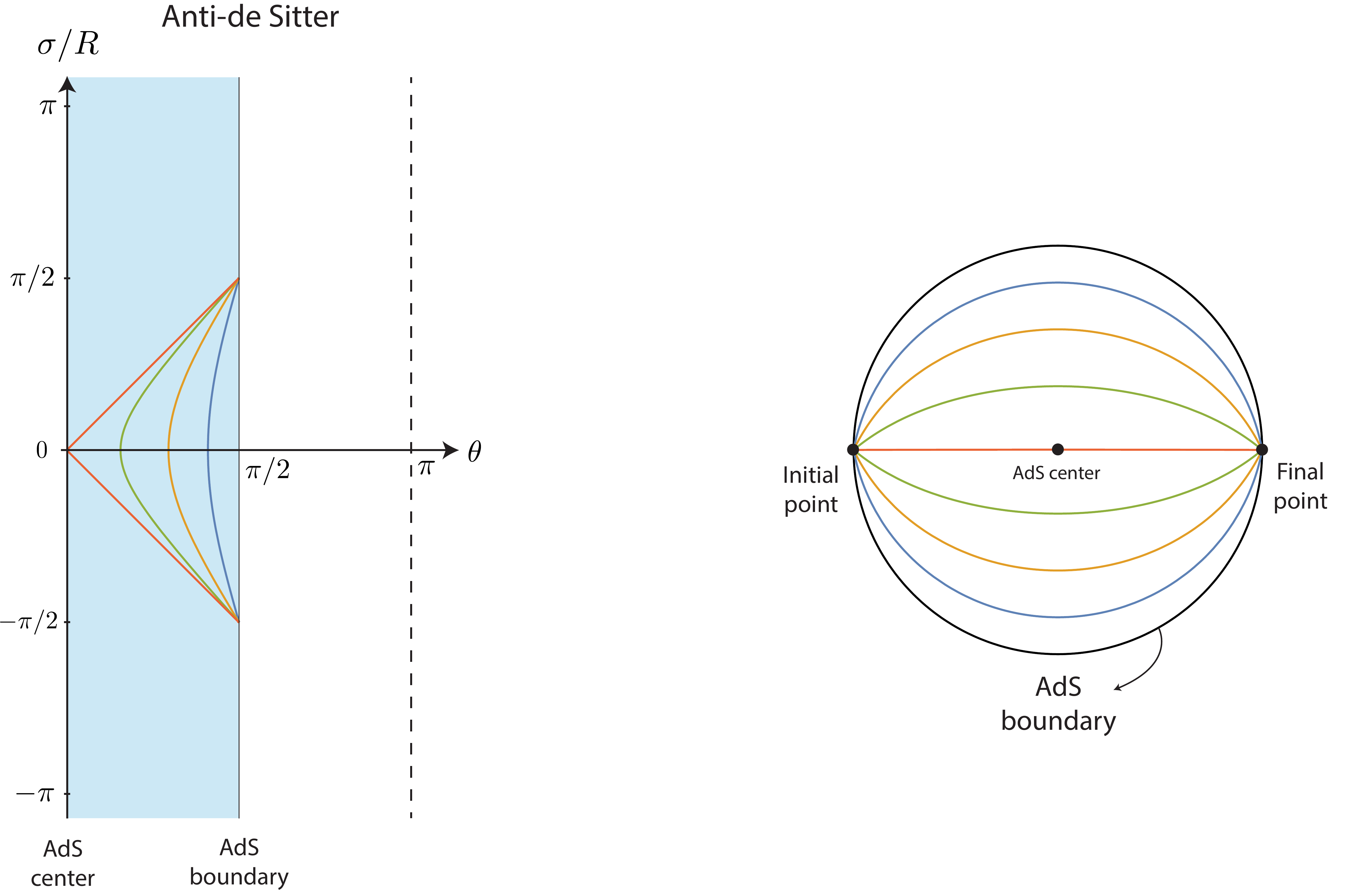}
\caption{In the left diagram we plot the AdS geodesics in the $(\sigma/R,\theta)$ plane. Comparing with Fig.~\ref{fig:15} we see that only half of the curves with $|\vec{x}_\perp|<R$ fit inside the space-time. In the right diagram we plot the trajectories in a cross section of the solid cylinder for the case of $d=3$. Equal colors in each diagram correspond to the same geodesics.}
\label{fig:16}
\end{center}
\end{figure}

The AdS case is quite different, since there are geodesics that lie outside the space-time, as we see in the left diagram of Fig. \ref{fig:16}. Only curves with $|\vec{x}_\perp|<R$ lie inside AdS. The critical geodesic that has a vertical path in the $(\sigma/R,\theta)$ plane is given by $|\vec{x}_\perp|=R$ in (\ref{eq:192}), and travels exactly along the AdS boundary. This is in accordance with the vanishing of the conformal factor in (\ref{eq:197}), which is signaling something important since the conformal transformation is not invertible around that point. 

For $d=3$ we plot the trajectories of the AdS geodesics in a cross section of the solid cylinder, so that we get the right diagram in Fig. \ref{fig:16}.\footnote{To obtain this plot we write the Cartesian coordinates $(x,y)$ as $(x,y)=\theta(\cos(\phi),\sin(\phi))$ where $\phi$ is obtained from~${v=2R\tan(\phi/2)}$. Using the description of the geodesics in (\ref{eq:188}) we get $(x,y)$ as a function of $(\lambda,\vec{x}_\perp)$.} Different values of $\vec{x}_\perp$ follow distinct paths in AdS. This is in contrast to the cylinder and dS where all the geodesics are equivalent up to a rotation of the sphere $S^{d-1}$. The maximum depth in AdS reached by each geodesic is given at $\lambda=0$, and can be written in terms of the AdS radial coordinates $\rho=R\tan(\theta)$ in table \ref{table:2} as
$$\rho_{\rm min}=
  \left(
  \frac{2R^2}{R^2-|\vec{x}_\perp|^2}
  \right)
  |\vec{x}_\perp|\ , \qquad
  \qquad |\vec{x}_\perp|\in[0,R]\ .
  $$
The maximum depth corresponds to $|\vec{x}_\perp|=0$ where the geodesic reaches the center of AdS, while for~$|\vec{x}_\perp|=R$ the geodesics travel along the AdS boundary and $\rho_{\rm min}$ diverges.

\subsection{Mapping the Minkowski ANEC}

Let us now apply the mapping of the Minkowski null plane to obtain some interesting results regarding the energy measured along null geodesics. Consider the ANEC in Minkowski, proven for general QFTs in Refs. \cite{Faulkner:2016mzt,Hartman:2016lgu,Longo:2018obd} and given by
\begin{equation}\label{eq:193}
\mathcal{E}(\vec{x}_\perp)\equiv
  \int_{-\infty}^{+\infty}d\lambda\,
  T_{\lambda \lambda}(\lambda,\vec{x}_\perp)\ge 0\ .
\end{equation}
The integral is over a null geodesic in the null plane (\ref{eq:50}), parametrized by $\lambda$ and labeled by $\vec{x}_\perp$. The stress tensor $T_{\mu \nu}$ is projected along this null path according to
\begin{equation}\label{eq:200}
T_{\lambda \lambda}=
  \frac{d X^\mu }{d\lambda}
  \frac{d X^\nu }{d\lambda}
  T_{\mu \nu}\ ,
\end{equation}
where $X^\mu=X^\mu(\lambda,\vec{x}_\perp)$ in (\ref{eq:50}). 

To map the integral operator in (\ref{eq:193}) we require the transformation of the stress tensor. Given the Hilbert space $\mathcal{H}$ associated to the field theory in Minkowski, the unitary operator $U:\mathcal{H}\rightarrow \bar{\mathcal{H}}$ implements the mapping to $\bar{\mathcal{H}}$, the Hilbert space of the transformed CFT. Since $T_{\mu \nu}$ is a quasi-primary operator with spin $\ell=2$ and scaling dimension $\Delta=d$ it transforms under the adjoint action of $U$ as
\begin{equation}\label{eq:184}
U T_{\mu \nu}U^\dagger=
  |w(v^\mu)|^{2-d}
  \frac{\partial v ^\alpha}{\partial X^\mu}
  \frac{\partial v ^\beta}{\partial X^\nu}
  \left(
  \bar{T}_{\alpha \beta}-\bar{S}_{\alpha \beta}
  \right)\ .
\end{equation}
The anomalous term $\bar{S}_{\alpha \beta}$ is proportional to the identity operator and non-vanishing for even $d$. For~${d=2}$ it can be written in terms of the Schwartzian derivative. Assuming that $T_{\mu \nu}$ has vanishing expectation value in the Minkowski vacuum $\ket{0}$,\footnote{For our purpose this assumption is not strictly necessary. Although Poincare symmetry of the vacuum only implies~$\bra{0}T_{\mu \nu}\ket{0}\propto \eta_{\mu \nu}$, when projecting the stress tensor along the null direction $T_{\lambda \lambda}$ this constant factor drops out.} we can determine the anomalous contribution $\bar{S}_{\alpha \beta}$ as
\begin{equation}\label{eq:27}
0=\bra{0}T_{\mu \nu}\ket{0}
  \qquad \Longrightarrow \qquad
  \bar{S}_{\alpha \beta}=
  \bra{\bar{0}}\bar{T}_{\alpha \beta}
  \ket{\bar{0}}\ ,
\end{equation}
where we have used that $\bar{S}_{\alpha \beta}$ is proportional to the identity operator. The effect of the anomalous term is to ensure that the mapped stress tensor $\bar{T}_{\alpha\beta}$ vanishes when evaluated in the new vacuum state $\ket{\bar{0}}$. For the most part we leave this vacuum substraction implicit and simply write~${\bar{T}_{\alpha \beta}\equiv \bar{T}_{\alpha \beta}-\bra{\bar{0}}\bar{T}_{\alpha \beta}\ket{\bar{0}}}$. 

Using this we can write the transformation of the operator $T_{\lambda \lambda}$ appearing in (\ref{eq:193}) as 
\begin{equation}\label{eq:5}
UT_{\lambda \lambda}(\lambda,\vec{x}_\perp)
U^\dagger=
|w(\lambda,\vec{x}_\perp)|^{2-d}
\left[
\frac{d v^\alpha }{d\lambda}
\frac{dv^\beta }{d\lambda}
\bar{T}_{\alpha \beta}(\lambda,\vec{x}_\perp)
\right]=
|w(\lambda,\vec{x}_\perp)|^{2-d}
\bar{T}_{\lambda \lambda}(\lambda,\vec{x}_\perp)\ ,
\end{equation}
where the components of $\bar{T}_{\lambda \lambda}$ are now computed from the null surface $v^\mu(\lambda,\vec{x}_\perp)$ in (\ref{eq:188}). In this way, the mapping of the Minkowski ANEC in (\ref{eq:193}) is in general given by
\begin{equation}\label{eq:198}
U\mathcal{E}(\vec{x}_\perp)U^\dagger=
  \int_{-\infty}^{+\infty}d\lambda\,
  |w(\lambda,\vec{x}_\perp)|^{2-d}
  \bar{T}_{\lambda\lambda}(\lambda,\vec{x}_\perp)
  \ge 0
  \ .
\end{equation}
This gives a non-trivial constraint for the CFTs defined on the cylinder and (A)dS implied by conformal symmetry and the ANEC in Minkowski. Since we are using the same coordinates~${v^\mu=(\theta_+,\theta_-,\vec{v}\,)}$ to describe all of these space-times, the geodesics are always given by (\ref{eq:188}).

\subsubsection{Weighted average in Lorentzian cylinder}

For the case of the Lorentzian cylinder the conformal factor is given by (\ref{eq:191}). Since it has a non-trivial dependence in $\lambda$, we change the integration variable to $\bar{\lambda}(\lambda)$, the affine parameter in (\ref{eq:190}), which gives
$$
U\mathcal{E}(\vec{x}_\perp)U^\dagger=
  \frac{1}{R}
  \left(
  \frac{2R^2}{R^2+|\vec{x}_\perp|^2}
  \right)^{d}
  \int_{-\pi/2}^{\pi/2}d\bar{\lambda}\,
  \cos^{d}(\bar{\lambda})
  \bar{T}_{\bar{\lambda}\bar{\lambda}}
  (\bar{\lambda},\vec{x}_\perp)
  \ ,
$$
where we remember to consider the hidden factors of $d\lambda$ in the definition of $T_{\lambda \lambda}$ when changing the integration variable. The positivity of the Minkowski ANEC implies a novel bound for the null energy of a CFT in the cylinder\footnote{While this work was in preparation Ref. \cite{Iizuka:2019ezn} appeared where this inequality was derived for $d=3,4,5$ and strongly coupled holographic CFTs described by Einstein gravity. This derivation show that the bound is valid in a more general setup.}
\begin{equation}\label{eq:195}
\int_{-\pi/2}^{\pi/2}d\bar{\lambda}\,
  \cos^{d}(\bar{\lambda})
  \bar{T}_{\bar{\lambda}\bar{\lambda}}
  (\bar{\lambda},\vec{x}_\perp)
  \ge 0\ .
\end{equation}
Before analyzing its features, let us rewrite it in a more convenient way. 

Even though this inequality seems simple enough, the coordinate description of the geodesics in (\ref{eq:188}) is complicated. However their trajectories in Fig. \ref{fig:15} are very simple. A more convenient description of the same geodesics can be obtained by taking advantage of the rotation symmetry of the sphere. In particular we can rotate the coordinates in $S^{d-1}$ such that the initial and final points~(\ref{eq:194}) are instead given by the North and South pole. This has the advantage that every geodesic has a constant value of $\vec{v}$ along its trajectory, instead of the complicated dependence in (\ref{eq:188}). The geodesics in the rotated frame are described in terms of the space-time coordinates~$v^\mu=(\theta_+,\theta_-,\vec{v}\,)$ as
\begin{equation}\label{eq:196}
v^\mu(\bar{\lambda},\vec{x}_\perp)=
  (2\bar{\lambda}+\pi/2,\pi/2,\vec{x}_\perp)\ ,
  \qquad \qquad
  (\bar{\lambda},\vec{x}_\perp)\in 
  [-\pi/2,\pi/2]\times \mathbb{R}^{d-2}\ .
\end{equation}
These curves start and end at the same time as (\ref{eq:194}) but at different spatial points of the sphere, given by the North and South pole. The tangent vector is clearly null and one can check that it satisfies the geodesic equation with affine parameter $\bar{\lambda}$. In Sec. \ref{sec:null_energy_bounds} we rederive the bound (\ref{eq:195}) from relative entropy directly in terms of a geodesic equivalent to (\ref{eq:196}). Let us now comment on the most interesting features of (\ref{eq:195}).

The bound (\ref{eq:195}) is not equivalent to the ANEC in the Lorentzian cylinder. To start, the condition is along a finite length geodesic which is not complete. Although we can obtain a bound for a complete geodesic going around the sphere $S^{d-1}$ an infinite number of times by applying (\ref{eq:195}) to each section, it is not equivalent to the ANEC due to the non trivial weight function $\cos^d(\bar{\lambda})$.\footnote{It is important that the inequality (\ref{eq:195}) is written in terms of the \textit{affine} parameter of the geodesic, since we could always define a new parameter which absorbs the weight function $\cos^d(\bar{\lambda})$ in the integral.} This weight function is required so that the operator (\ref{eq:195}) is well defined. In the integration range~${|\bar{\lambda}|\le \pi/2}$, the function $\cos^d(\bar{\lambda})$ is non-negative, smooth and vanishes at the boundaries. The rapid decay of the function at $|\bar{\lambda}|=\pi/2$ is crucial, given that it is precisely at the boundary of a sharply integrated operator, where large amounts of negative energy can acumulate.\footnote{See section 4.2.4 of Ref. \cite{Fewster:2004nj} for an explicit example of this feature in two dimensional CFTs.}

Let us also recall that the stress tensor appearing in (\ref{eq:195}) is normalized so that it vanishes in the vacuum state of the cylinder. This arises due to the anomalous transformation of the stress tensor under the conformal map (see the discussion around (\ref{eq:27})). The operator in the inequality is then given by $\bar{T}_{\mu \nu}\equiv \bar{T}_{\mu \nu}-\bra{\bar{0}}\bar{T}_{\mu \nu}\ket{\bar{0}}$, where $\ket{\bar{0}}$ is the vacuum of the CFT in the cylinder. This vacuum contribution has been explicitly computed in Ref. \cite{Herzog:2013ed} for arbitrary CFTs, where it is shown to vanish when $d$ is odd while for even $d$ it is given by
\begin{equation}\label{eq:22}
\bra{\bar{0}}\bar{T}_{\bar{\lambda} \bar{\lambda}}\ket{\bar{0}}=
  \frac{4(-1)^{d/2}A_d}
  {(d-1)R^{d-2}{\rm Vol}(S^d)}\ ,
\end{equation}
with $A_d$ the trace anomaly coefficient, see Ref. \cite{Myers:2010tj} for conventions. The vacuum substraction ensures that the inequality (\ref{eq:195}) is not trivially violated by come constant negative Casimir energy. 

Finally let us comment in the large $d$ limit, which is particularly interesting since the function~$\cos^d(\bar{\lambda})$ localizes at $\bar{\lambda}=0$. Although $\bar{\lambda}=0$ in (\ref{eq:196}) corresponds to the equator of $S^{d-1}$ we can always rotate the coordinates system so that the integral localized around an arbitrary point. This means we can write the bound directly in terms of the space-time coordinates $v^\mu$ in the large $d$ limit as the following local constraint
\begin{equation}\label{eq:169}
\bar{T}_{--}
  (\theta_+,\theta_-,\vec{v}\,)
  \ge 
  \lim_{d\rightarrow +\infty}
  \frac{(-1)^{d/2}A_d}
  {(d-1)R^{d-2}{\rm Vol}(S^d)}\ ,
\end{equation}
where we have projected the stress tensor in the null coordinate $\theta_-$. 

Evaluating the limit on the right-hand side is not as simple as it might seem since the coefficient~$A_d$ vanishes for $d$ odd and has a non-trivial dependence when $d$ is even. Explicit expressions for $A_d$ can be written for free or holographic theories \cite{Cappelli:2000fe,Myers:2010tj}. Regardless of the particular value of $A_d$, there are only two possible outcomes for the limit in (\ref{eq:169}): it is either undetermined or it converges to zero. While an undetermined result means that there is something funny going on with large $d$ limit in~(\ref{eq:195}), if it goes to zero it implies that the stress tensor is locally a positive operator in the cylinder. This is an interesting result which we hope to further investigate in future work.

\subsubsection{ANEC in (A)dS}

Let us now consider the mapping to (A)dS, where the conformal factors evaluated on the null surface are given in (\ref{eq:197}). Since these are independent of $\lambda$ the mapping of the Minkowski ANEC (\ref{eq:198}) is given by
$$
U\mathcal{E}(\vec{x}_\perp)U^\dagger=
  |w_{\rm (A)dS}(\vec{x}_\perp)|^{2-d}
  \int_{-\infty}^{+\infty}d\lambda\,
  \bar{T}_{\lambda\lambda}(\lambda,\vec{x}_\perp)\ ,
$$
which implies
\begin{equation}\label{eq:199}
\mathcal{E}_{\rm (A)dS}(\vec{x}_\perp)\equiv
  \int_{-\infty}^{+\infty}d\lambda\,
  \bar{T}_{\lambda \lambda}(\lambda,\vec{x}_\perp)\ge 0\ .
\end{equation}
Let us explain what are the features that allows us to identify this as the ANEC in both de Sitter and anti-de Sitter.

The first crucial fact is that $w_{\rm (A)dS}(\vec{x}_\perp)$ is independent of $\lambda$, so that the right hand side of the geodesic equation (\ref{eq:64}) vanishes and implies that $\lambda$ is an affine parameter in (A)dS.\footnote{It important that the integral in the ANEC is written in terms of an affine parameter. While the condition in (\ref{eq:193}) is clearly invariant under affine transformations $\lambda \rightarrow a\lambda+b$, it changes its form under a more general transformation, \textit{e.g.} $\lambda \rightarrow L\sinh(\lambda/L)$.} Moreover, this allows to remove it from the $\lambda$ integral in (\ref{eq:198}) so that there is no weight function along the trajectory, as we had for the case of the Lorentzian cylinder (\ref{eq:195}). Another important feature is that the geodesics in both dS and AdS are complete, \textit{i.e.} they cannot be extended beyond $\lambda \in \mathbb{R}$. This is certainly the case as the curves start and end at the (A)dS boundaries. Altogether, this allows us to identify (\ref{eq:199}) as the ANEC in (A)dS, valid for any conformal theory.

Similarly to the case of the cylinder, for dS we can use the spatial symmetry to describe the null geodesics in (\ref{eq:188}) in a more convenient way. Since de Sitter space-time is topologically given by~$\mathbb{R}\times S^{d-1}$, we can use the same reasoning around (\ref{eq:196}) to describe the geodesics in de Sitter as
$$v^\mu(\lambda,\vec{x}_\perp)=
  \left(
  2\arctan(\lambda)+\pi/2,\pi/2,\vec{x}_\perp
  \right)\ ,
  \qquad 
  (\lambda,\vec{x}_\perp)\in \mathbb{R}\times
  \mathbb{R}^{d-2}\ .$$
In Sec. \ref{sec:null_energy_bounds} we rederive the ANEC in de Sitter from relative entropy directly in terms of a null geodesic equivalent to this one.

For AdS we do not have a symmetry argument to simplify the description of the geodesics in~(\ref{eq:188}). As we see in the right diagram of Fig. \ref{fig:16} the geodesics for different values of $\vec{x}_\perp$ are distinct and travel through the space-time in different ways.

Before moving on let us recall that the stress tensor appearing in (\ref{eq:199}) contains a substraction with respect to the (A)dS vacuum, \textit{i.e.} $\bar{T}_{\mu \nu}\equiv \bar{T}_{\mu \nu}-\bra{0_{\rm (A)dS}}\bar{T}_{\mu \nu}\ket{0_{\rm (A)dS}}$. However, there is an important distinction in this case given by the fact that (anti-)de Sitter is a maximally symmetric space-time. This implies that the vacuum expectation value of the stress tensor is proportional to the (A)dS metric,\footnote{We can explicitly check this from equation (21) in Ref. \cite{Herzog:2013ed} using that the Riemann tensor of (A)dS is determined from its metric.} which results in
\begin{equation}\label{eq:181}
\bra{0_{\rm (A)dS}}
  \bar{T}_{\mu \nu}\ket{0_{\rm (A)dS}}\propto
  g_{\mu \nu}^{\rm (A)dS}
  \qquad \Longrightarrow \qquad
  \bra{0_{\rm (A)dS}}
  \bar{T}_{\lambda \lambda}
  \ket{0_{\rm (A)dS}}\propto
  g_{\mu \nu}^{\rm (A)dS}
  \frac{dv^\mu }{d\lambda}
  \frac{dv^\nu }{d\lambda}=0\ .
\end{equation}
Therefore, the Casimir energy of (A)dS makes no contribution to the ANEC in (\ref{eq:199}).

\section{Null energy bounds from relative entropy}
\label{sec:null_energy_bounds}

In the previous section we showed that the ANEC in (A)dS and a similar bound for the Lorentzian cylinder follow from the Minkowski ANEC and conformal symmetry. The aim of this section is to investigate whether these results can also be obtained from relative entropy, as done in Ref. \cite{Faulkner:2016mzt} for the Minkowski ANEC. Let us start by briefly review the approach used in that paper.

\begin{figure}[t]
\begin{center}
\qquad \qquad \qquad
\includegraphics[height=2.0 in]{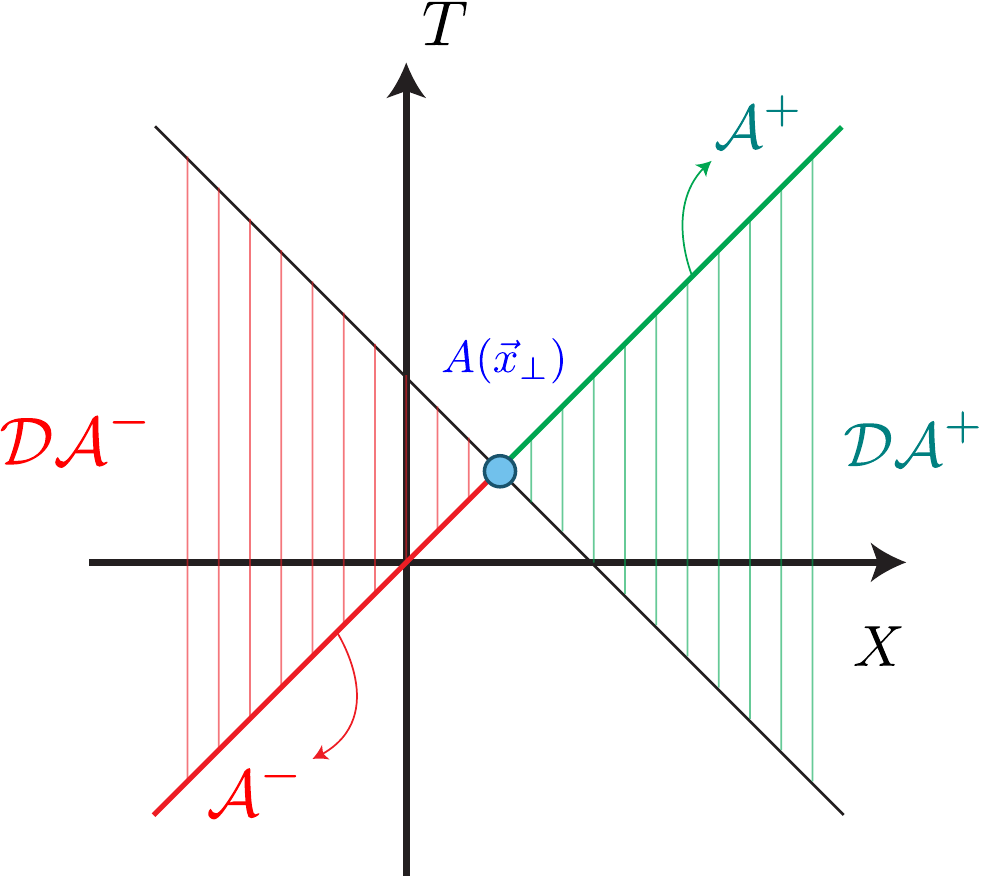}
\qquad \qquad \qquad \quad 
\caption{Plot of $\mathcal{DA}^+$ (green) and $\mathcal{DA}^-$ (red) in the $(T,X)$ plane in Minkowski space-time. The function $A(\vec{x}_\perp)$ (blue dot) changes along the null plane $X-T=0$ as we move through the transverse directions $\vec{x}_\perp$ outside of the page. A very nice 3D picture of this region can be found in Fig. 1 of Ref.~\cite{Casini:2018kzx}.}
\label{fig:3}
\end{center}
\end{figure}

Consider a smooth curve in the null plane (\ref{eq:50}) defined by $\lambda=A(\vec{x}_\perp)$ which splits the surface in two regions ${\mathcal{N}_{\rm plane}=\mathcal{A}^+\cup \mathcal{A}^-}$, where $\mathcal{A}^\pm$ are given by $\lambda\ge \pm A(\vec{x}_\perp)$. Given a QFT in $d$-dimensional space-time $X^\mu$ we take the space-time region $\mathcal{DA}^+$ for which $\mathcal{A}^+$ is its future horizon, and analogously for $\mathcal{DA}^-$. A diagram of the setup is given in Fig. \ref{fig:3}. For these space-time regions let us consider the reduced density operator $\rho_{\mathcal{A}^{\pm}}$ associated to the vacuum state $\ket{0}$. We can define $\rho_{\mathcal{A}^\pm}$ as the operator which satisfies the following property
\begin{equation}\label{eq:67}
\bra{0}\mathcal{O}_{\mathcal{A}^\pm}\ket{0}=
  {\rm Tr}\big(
  \rho_{\mathcal{A}^\pm}\,
  \mathcal{O}_{\mathcal{A}^\pm}
  \big)\ ,
\end{equation}
for $\mathcal{O}_{\mathcal{A}^\pm}$ any operator (not necessarily local) supported exclusively in $\mathcal{DA}^\pm$. Given a reduced density operator its logarithm defines the modular Hamiltonian $K_{\mathcal{A}^\pm}=-\ln(\rho_{\mathcal{A}^\pm})+{\rm const}$, where the constant is fixed by normalization.

For this setup the modular Hamiltonian of the vacuum state was computed in Ref. \cite{Casini:2017roe} (see also Refs. \cite{Faulkner:2016mzt,Wall:2011hj,Koeller:2017njr}) and shown to have the following simple local expression
\begin{equation}\label{eq:66}
K_{\mathcal{A}^\pm}=\pm 2\pi
  \int_{\mathcal{A}^\pm}dS
  \left(\lambda-A(\vec{x}_\perp)\right)
  T_{\lambda \lambda}(\lambda,\vec{x}_\perp)\ ,
\end{equation}
where $dS=d\vec{x}_\perp d\lambda$ is the induced surface element on the null plane and $T_{\lambda \lambda}$ is defined in (\ref{eq:200}). When~$A(\vec{x}_\perp)=0$ the regions in Fig. \ref{fig:3} corresponds to the Rindler wedge and its complement, so that~(\ref{eq:66}) follows from the Bisognano-Wichmann theorem \cite{Bisognano:1976za}. In this case the modular Hamiltonian can be written as a local integral over any Cauchy surface in $\mathcal{DA}^\pm$, not necesarily along the null horizons. This is not true when $A(\vec{x}_\perp)$ is a non-trivial function, since the operator has a local expression only along the null surface $\mathcal{A}^\pm$ \cite{Casini:2017roe}.

It is useful to also consider the full modular Hamiltonian $\hat{K}_{\mathcal{A}^+}$, defined for a generic space-time region $V$ as
\begin{equation}\label{eq:161}
\hat{K}_V=K_V-K_{V'}\ ,
\end{equation}
where $V'$ is the causal complement of $V$. Using the expressions in (\ref{eq:66}) we find
\begin{equation}\label{eq:177}
\hat{K}_{\mathcal{A}^+}=
  2\pi \int_{\mathcal{N}_{\rm plane}}
  dS\,\left(
  \lambda-A(\vec{x}_\perp)
  \right)
  T_{\lambda \lambda}(\lambda,\vec{x}_\perp)\ ,
\end{equation}
where the integral is now over the full null plane. This operator has the advantage that it is globally defined in the Hilbert space, without any ambiguities that can arise in (\ref{eq:66}) from the boundary of integration. In the context of the Tomita-Takesaki theory that we review in Sec. \ref{sec:ref_positivity}, $\hat{K}_V$ determines the modular operator.

To prove the Minkowski ANEC, Ref. \cite{Faulkner:2016mzt} combined the full modular Hamiltonian in (\ref{eq:177}) together with relative entropy, that is defined as 
\begin{equation}\label{eq:180}
S(\rho||\sigma)=
  {\rm Tr}\left(
  \rho\ln(\rho)
  \right)-
  {\rm Tr}\left(
  \rho\ln(\sigma)
  \right)\ge 0\ ,
\end{equation}
where $\rho$ and $\sigma$ are any two density operators. The monotonicity property of relative entropy implies that given any two space-time regions such that $A \supseteq B$, the reduced operators satisfy the inequality~$S\left(\rho_A||\sigma_A\right)\ge S\left(\rho_B||\sigma_B\right)$. Taking $\sigma$ as a pure state and starting from this inequality and an analogous one for the complementary regions, it is straightforward to prove following constraint \cite{Blanco:2013lea}
\begin{equation}\label{eq:1}
\hat{K}_{A}-\hat{K}_B\ge 0\ ,
  \qquad {\rm for} \qquad
  A \supseteq B\ ,
\end{equation} 
where $\hat{K}_{A/B}$ is the full modular Hamiltonian of $\sigma$.\footnote{The inequality implied by relative entropy is more general than (\ref{eq:1}) and given by
$${\rm Tr}\big[
  \rho\big(\hat{K}_A-\hat{K}_B\big)\big]
  \ge 2S_f(A,B)\ ,$$
where $S_f(A,B)$ is the free entropy of the state $\rho$. This is a non-negative and UV finite quantity constructed from the entanglement entropy  $2S_f(A,B)=\left(S_A-S_{A'}\right)-\left(S_B-S_{B'}\right)$, see Ref. \cite{Blanco:2013lea}. If $\rho$ is a pure state, the free entropy vanishes and we recover (\ref{eq:1}).} Using (\ref{eq:177}) we can explicitly write the inequality for null deformations of Rindler, which gives
$$\hat{K}_A-\hat{K}_B=
  2\pi\int_{-\infty}^{+\infty}d\lambda
  \int_{\mathbb{R}^{d-2}}
  d\vec{x}_\perp\left(
  B(\vec{x}_\perp)-A(\vec{x}_\perp)
  \right)
  T_{\lambda \lambda}(\lambda,\vec{x}_\perp)\ge 0\ , $$
where we must have $B(\vec{x}_\perp)\ge A(\vec{x}_\perp)$. Taking $A(\vec{x}_\perp)=0$ and $B(\vec{x}_\perp)=\delta(\vec{x}_\perp-\vec{x}_\perp^{\,0})$ for any $\vec{x}_\perp^{\,0}$, gives the ANEC in Minkowski (\ref{eq:193}) as derived in Ref. \cite{Faulkner:2016mzt}.

Our strategy for extending this proof is simple. Using conformal transformations we map the modular Hamiltonian in (\ref{eq:177}) to (A)dS and the Lorentzian cylinder. From this we can explicitly write the inequality (\ref{eq:1}) coming from relative entropy and obtain a bound for the energy along null geodesics. We shall see that this procedure is non-trivial and while it works for de Sitter and the Lorentzian cylinder, it fails to give the ANEC in the anti-de Sitter case. Along the way we obtain several new modular Hamiltonians and compute their associated entanglement entropy.

\subsection{Taking the null plane on a conformal journey - Take II}
\label{sec:geometry}

Since our aim is to map the modular Hamiltonian (\ref{eq:66}), given by an integral over a region of the null plane, we start by discussing the geometric transformation of the null plane. Although we have already analyzed this in the previous section, the resulting surface (\ref{eq:188}) has a complicated coordinate description which is not the most convenient. We now consider a slightly different conformal transformation that is more useful for writing the modular Hamiltonians. 

Instead of mapping the null plane directly to the cylinder, we first consider a conformal transformation mapping the Minkowski space-time $X^\mu=(T,X,\vec{Y}\,)$ into itself $x^\mu=(t,x,\vec{y}\,)$. This transformation is given by
\begin{equation}\label{eq:3}
x^\mu(X^\mu)=
  \frac{X^\mu+(X\cdot X)C^\mu}
  {1+2(X\cdot C)+(X\cdot X)(C\cdot C)}
  -D^\mu\ ,
\end{equation}
where $(X\cdot X)=\eta_{\mu \nu}X^\mu X^\nu$. It gives a space-time translation in the $D^\mu=(R,R,\vec{0}\,)$ direction together with a special conformal transformation with parameter $C^\mu=(0,1/(2R),\vec{0}\,)$. The Minkowski metric in the new coordinates becomes $ds^2=w^2(x^\mu)\eta_{\mu \nu}dx^\mu dx^\nu$, where the conformal factor is given by
$$
w^2(X^\mu)=
  \big(1+2(X\cdot C)+(X\cdot X)(C\cdot C)\big)^2\ .
$$
Evaluating this along the null plane (\ref{eq:50}) we find
\begin{equation}\label{eq:79}
w^2(\lambda,\vec{x}_\perp)=
  \left(
  \frac{\lambda+ p(\vec{x}_\perp)}{R}
  \right)^2
  \qquad \,\, {\rm where} \qquad \,\,
  p(\vec{x}_\perp)=
  \frac{|\vec{x}_\perp|^2+4R^2}{4R}\ .
\end{equation} 
The mapped suface can be found by evaluating (\ref{eq:3}) in the parametrization of the null plane in (\ref{eq:50})
\begin{equation}\label{eq:23}
x^\mu(\lambda,\vec{x}_\perp)=
R\left(
  \frac{p(\vec{x}_\perp)}
  {\lambda+p(\vec{x}_\perp)}
  \right)
  \big(-1,\vec{n}(\vec{x}_\perp)\big)\ ,
  \qquad \quad
  \vec{n}(\vec{x}_\perp)=\left(
  \frac{|\vec{x}_\perp|^2-4R^2}
  {|\vec{x}_\perp|^2+4R^2},
  \frac{4R\vec{x}_\perp}
  {|\vec{x}_\perp|^2+4R^2}
  \right)\ ,
\end{equation}
where $\vec{n}\in \mathbb{R}^{d-1}$ is a unit vector $|\vec{n}(\vec{x}_\perp)|=1$. This surface corresponds to a future and past null cone starting from the origin $x^\mu=0$. Although $\lambda$ is not affine anymore, we can define an affine parameter~$\alpha$ according to $\lambda(\alpha)=p(\vec{x}_\perp)(R/\alpha-1)$,\footnote{This expression for $\alpha$ can be obtained from the integral in (\ref{eq:78}), using $w^2(\lambda,\vec{x}_\perp)$ in (\ref{eq:79}) and conveniently fixing the integration constants $c_1$ and $c_0$.} so that the surface is given by
\begin{equation}\label{eq:14}
x^\mu(\alpha,\vec{x}_\perp)=\alpha(-1,\vec{n}(\vec{x}_\perp))\ ,
  \qquad \qquad
  (\alpha,\vec{x}_\perp)\in \mathbb{R}\times \mathbb{R}^{d-2}\ .
\end{equation}
Positive $\alpha$ corresponds to the past null cone of the origin $x^\mu=0$, while negative $\alpha$ gives the future cone. The transverse coordinates $\vec{x}_\perp$ parametrize a unit sphere $S^{d-2}$ in stereographic coordinates, as can be seen by computing the induced metric on the surface and finding $\alpha^2d\Omega^2(\vec{x}_\perp)$ with $d\Omega(\vec{x}_\perp)$ in~(\ref{eq:152}) (where $L=2R$).

There is a subtlety in this transformation that we must be careful with. As we can see from the description in terms of $\lambda$ in (\ref{eq:23}), there is a discontinuity in the mapping when $\lambda=-p(\vec{x}_\perp)$, that is precisely where the conformal factor (\ref{eq:79}) vanishes. Similarly to the previous mapping to AdS in (\ref{eq:197}), this is signaling a failure of the transformation, which is somewhat expected given that special conformal transformations are not globally defined in Minkowski but on its conformal compactification, the Lorentzian cylinder. To properly interpret the surface (\ref{eq:14}) we must go to the cylinder.

\begin{figure}[t]
\begin{center}
\includegraphics[height=2.3 in]{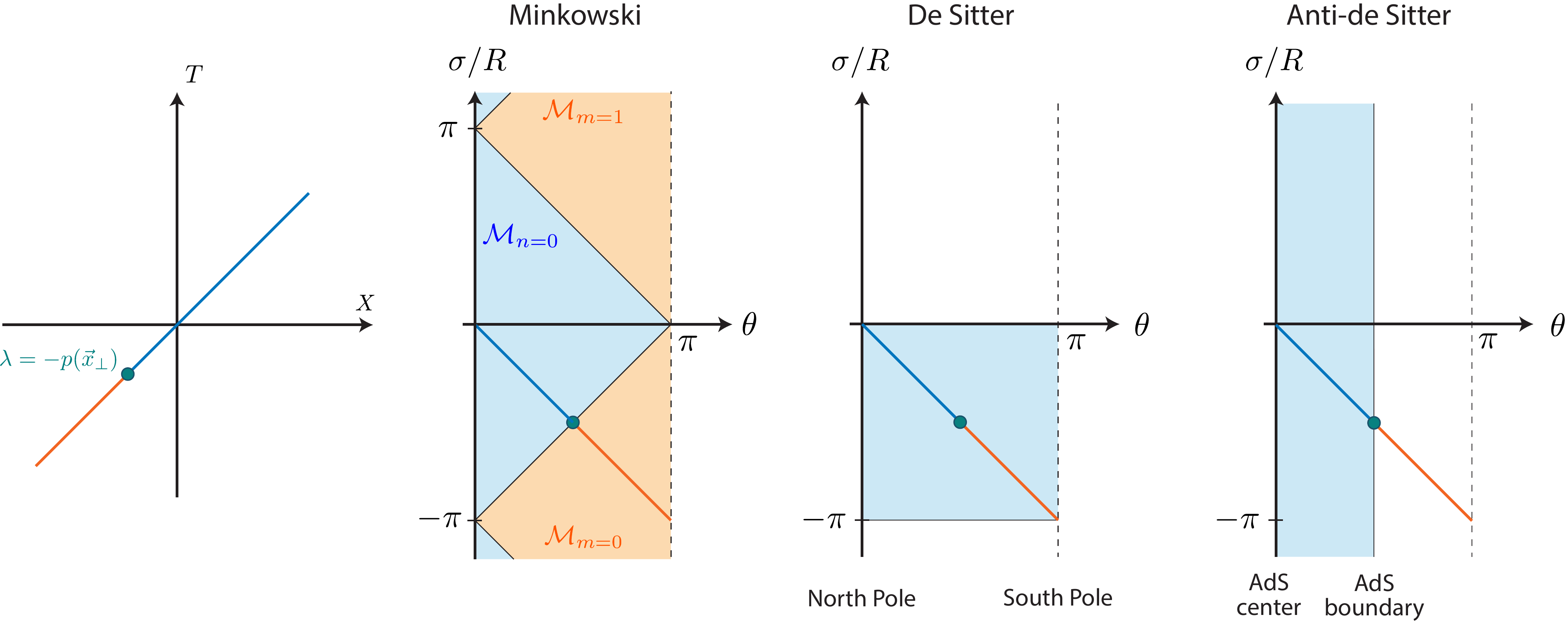}
\caption{Mapping of the null plane in Minkowski to the null surfaces in the $(\sigma/R,\theta)$ plane. The different diagrams indicate the sections of the cylinder covered by each of the space-times. We see that discontinuity at $\lambda=-p(\vec{x}_\perp)$ in the Minkowski case is nothing more than the surface going from the Minkowski patch $\mathcal{M}_{n=0}$ to $\mathcal{M}_{m=0}$ in the second diagram. For de Sitter we consider a slight variation of the Weyl rescaling so that the surface fits in the space-time, see discussion above (\ref{eq:202}).}\label{fig:9}
\end{center}
\end{figure}

Since a single copy of Minkowski is not enough to cover the whole cylinder, we consider an infinite number of Minkowski space-times $\mathcal{M}_{n}$ and $\mathcal{M}_m$ labeled by the integers $(n,m)$, so that the whole cylinder manifold $\mathcal{M}_{LC}$ is obtained from
$$\mathcal{M}_{LC}=
  \Big(\bigcup_{n\in \mathbb{Z}}
  \mathcal{M}_{n}\Big)
  \cup
  \Big(\bigcup_{m\in \mathbb{Z}}
  \mathcal{M}_{m}\Big)\ .$$
To each of the Minkowski copies we apply a slightly different conformal transformation
\begin{equation}\label{eq:10}
r_\pm^{(n)}(\theta_\pm)=
  R\tan(\theta_\pm/2)\ ,
\qquad \qquad
r_\pm^{(m)}(\theta_\pm)=
  -R\tan(\theta_\mp/2)\ ,
\end{equation}
where the domain of the coordinates $\theta_\pm$ in each case is given by
\begin{equation}\label{eq:114}
\begin{aligned}
  D_n&=
  \big\lbrace
  \theta_\pm \in \mathbb{R}:\quad
  (\theta_\pm\mp 2n\pi)\in[-\pi,\pi]\ ,
  \quad
  \theta\in[0,\pi]
  \big\rbrace\ ,
  \\[8pt]
  D_m&=
  \big\lbrace
  \theta_\pm \in \mathbb{R}:\quad
  (\theta_\pm\mp (2m-1)\pi)\in[0,2\pi]\ ,
  \quad
  \theta\in[0,\pi]
  \big\rbrace\ .
\end{aligned}
\end{equation}
In every case, the transformations acts in the same way as in table \ref{table:2} but mapping $\mathcal{M}_{(n,m)}$ to different sections of the Lorentzian cylinder. These are given in the $(\sigma/R,\theta)$ plane by the shaded blue and orange regions in the second diagram of Fig. \ref{fig:9}. The main difference between the $n$ (blue) and $m$ (orange) patches is that the $n$ series maps the Minkowski origin to the North pole, while for $m$ the origin is mapped to the South. 

Let us now use these relations to map the null plane across the special conformal transformation and into the cylinder. From (\ref{eq:23}) we can write the null radial coordinates $r_\pm$ on the surface as 
\begin{equation}\label{eq:56}
r_\pm(\lambda,\vec{x}_\perp)=
  \frac{2Rp(\vec{x}_\perp)}
  {|\lambda+p(\vec{x}_\perp)|}
  \Theta\big[
  \mp \left(\lambda+p(\vec{x}_\perp)\right)
  \big]\ .
\end{equation}
Applying the transformation associated to the patch $n=0$ in (\ref{eq:10}) to the region of the null plane~${\lambda>-p(\vec{x}_\perp)}$ and $m=0$ to $\lambda<-p(\vec{x}_\perp)$ we find
\begin{equation}\label{eq:57}
\begin{aligned}
\lambda & >-p(\vec{x}_\perp)
  \qquad \Longrightarrow \qquad
  r_\pm^{(n=0)}:
  \left(
  \theta_+=0,
  \tan(\theta_-/2)=\frac{2p(\vec{x}_\perp)}
  {\lambda+p(\vec{x}_\perp)}
  \right)\ ,
  \qquad
  \theta_-\in [0,\pi]\ ,\\[4pt]
\lambda & <-p(\vec{x}_\perp)
  \qquad \Longrightarrow \qquad
  r_\pm^{(m=0)}:
  \left(
  \theta_+=0,
  \tan(\theta_-/2)=
  \frac{2p(\vec{x}_\perp)}
  {\lambda+p(\vec{x}_\perp)}
  \right)\ ,
  \qquad
  \theta_-\in[\pi,2\pi]\ ,
\end{aligned}
\end{equation}
where the range of $\theta_-$ in each case is obtained from (\ref{eq:114}). Notice that the surface across the two patches is continuous as $\theta_-\rightarrow \pi$. Moreover, the singularity that is present in the Minkowski space $x^\mu$ at $\lambda\rightarrow -p(\vec{x}_\perp)$ is smoothed out in the cylinder by the tangent function. This completely determines the mapping of the null plane in the Minkowski coordinates $X^\mu$ to the Lorentzian cylinder, which we sketch in Fig. \ref{fig:9}. 

We can now reinterpret the discontinuity in the Minkowski null cone $x^\mu$ in (\ref{eq:23}) from the perspective of the Lorentzian cylinder. As we see in Fig. \ref{fig:9}, this discontinuity is nothing more than the null surface going from the Minkowski copy $\mathcal{M}_{n=0}$ to $\mathcal{M}_{m=0}$. The future null cone in appears to come from infinity, that is precisely what happens from the perspective of $\mathcal{M}_{m=0}$ in Fig. \ref{fig:9}. This means that the future and past null cones in (\ref{eq:23}) are not in the same Minkowski patch, since the mapping of the full null plane does not fit in the Minkowski space-time $x^\mu$. Shortly, this will play an important role when computing the modular Hamiltonian associated to the null cone.

The conformal factor relating the Minkowski space-time $X^\mu$ with the cylinder is obtained by taking the product of (\ref{eq:79}) and the expression in table \ref{table:2} evaluated at (\ref{eq:56}), which gives
\begin{equation}\label{eq:201}
w(\lambda,\vec{x}_\perp)^2=
  \left[\frac{(\lambda+p(\vec{x}_\perp))^2}{R^2}
  \right]
  \times
  \left[\frac{4p^2(\vec{x}_\perp)+(\lambda+p(\vec{x}_\perp))^2}
  {4(\lambda+p(\vec{x}_\perp))^2}
  \right]=
  \frac{4p^2(\vec{x}_\perp)+(\lambda+p(\vec{x}_\perp))^2}
  {4R^2}\ .
\end{equation}
Using this to solve the integral in (\ref{eq:78}), we find an affine parameter $\beta=\beta(\lambda)$ for the surface in the cylinder
\begin{equation}\label{eq:116}
\tan(\beta)=\frac{2p(\vec{x}_\perp)}{\lambda+p(\vec{x}_\perp)}\ ,
\end{equation}
where we have conveniently fixed the integration constants to $c_0$ and $c_1$. Comparing with (\ref{eq:57}) we identify $\beta=\theta_-/2$, so that the null surface in the cylinder coordinates $v^\mu=(\theta_+,\theta_-,\vec{v}\,)$ has the following simple description
\begin{equation}\label{eq:62}
v^\mu(\beta,\vec{x}_\perp)=
  \left(
  0,2\beta,\vec{x}_\perp
  \right)\ ,
  \qquad \qquad
  (\beta,\vec{x}_\perp) \in [0,\pi]\times \mathbb{R}^{d-2}\ .
\end{equation}
The surface goes from the South pole of the $S^{d-1}$ all the way to the North pole. Up to a time translation and rotation of the $S^{d-1}$, it is equivalent to the surface obtained through the mapping of the previous section in (\ref{eq:188}) (see Fig. \ref{fig:15}) but with a much simpler description.

Let us now apply the transformation to (A)dS given by the Weyl rescaling in (\ref{eq:154}). Since the surface in the cylinder (\ref{eq:62}) has a range in $\sigma/R$ given by $\sigma/R\in[-\pi,0]$ we consider a slightly different Weyl rescaling for the de Sitter case, given by changing the conformal factor in~(\ref{eq:154}) to~${\cos^2(\sigma/R)\rightarrow \sin^2(\sigma/R)}$. This allows us to take the range of the time coordinate in dS~${\sigma/R\in[-\pi,0]}$ so that the surface (\ref{eq:62}) fits in the space-time, as we see in Fig. \ref{fig:9}. In the same figure we see that the null surface does not fit in a single copy of AdS. Evaluating the conformal factor relating the Minkowski space-time $X^\mu$ to (A)dS
using (\ref{eq:201}) and (\ref{eq:62}) we find
\begin{equation}\label{eq:202}
w_{\rm dS}(\vec{x}_\perp)=
  p(\vec{x}_\perp)/R\ ,
\qquad \qquad
w_{\rm AdS}(\lambda,\vec{x}_\perp)^2=
\left(
  \frac{\lambda+p(\vec{x}_\perp)}{2R}
  \right)^2\ .
\end{equation}

For de Sitter the conformal factor is independent of $\lambda$ and similar to the one obtained from the conformal transformation in Sec. (\ref{eq:197}). This means that $\lambda$ is an affine parameter in dS. We still find it convenient to apply an affine transformation by defining $\eta$ according to $\lambda(\eta)=p(\vec{x}_\perp)(2\eta-1)$ so that using (\ref{eq:116}) the surface in dS has a simple description. Writing $\beta$ in (\ref{eq:62}) in terms of $\eta$ we find
\begin{equation}\label{eq:203}
v^\mu(\eta,\vec{x}_\perp)=
  \left(
  0,2\,{\rm arccot}(\eta),\vec{x}_\perp
  \right)\ ,
  \qquad \qquad
  (\eta,\vec{x}_\perp) \in 
  \mathbb{R}\times \mathbb{R}^{d-2}\ ,
\end{equation}
where since $\beta={\rm arccot}(\eta)$, the image of ${\rm arccot}(\eta)$ is taken in $[0,\pi]$.

For anti-de Sitter the conformal factor depends on $\lambda$, which means $\lambda$ is not affine after the transformation. This is quite different to the mapping considered in the previous section, where it was independent of $\lambda$ (\ref{eq:197}). An affine parameter in AdS $\zeta$ can be easily found by solving the integral in (\ref{eq:78}), which gives $\lambda(\zeta)=p(\vec{x}_\perp)(2/\zeta-1)$. Writing $\beta$ in (\ref{eq:62}) in terms of $\zeta$, the surface in AdS is given by
$$
v^\mu(\zeta,\vec{x}_\perp)=
  \left(
  0,2\,{\rm arctan}(\zeta),\vec{x}_\perp
  \right)\ ,
  \qquad \qquad
  (\eta,\vec{x}_\perp) \in 
  \mathbb{R}_{\ge 0}\times \mathbb{R}^{d-2}\ .
$$
As $\zeta\rightarrow +\infty$ the surface reaches $\theta=\pi/2$ corresponding to the AdS boundary and the conformal factor (\ref{eq:202}) vanishes. The full surface does not fit in a single copy of AdS.

\begin{table}[t]\setlength{\tabcolsep}{8.5pt}
\centering
\begin{tabular}{ Sl | Sl | Sl | Sl }
\specialrule{.13em}{0em}{0em}
\textbf{Mapping of}  &
\textbf{Affine parameter} &
\textbf{Induced} &
\textbf{Fits inside} \\ 
\textbf{null plane to}  &
\textbf{along geodesic} &
\textbf{metric} &
\textbf{space-time?} \\
\specialrule{.05em}{0em}{0em}
Minkowski null cone  &
$\lambda(\alpha)=p(\vec{x}_\perp)(R/\alpha-1)$ &
$\alpha^2 d\Omega^2(\vec{x}_\perp)$ &
\textcolor{red}{No}
\\
Lorentzian cylinder  &
$\lambda(\beta)=p(\vec{x}_\perp)(2\cot(\beta)-1)$ &
$R^2\sin^2(\beta) d\Omega^2(\vec{x}_\perp)$ &
\textcolor{ForestGreen}{Yes}
\\
De Sitter  &
$\lambda(\eta)=p(\vec{x}_\perp)(2\eta-1)$ &
$
R^2d\Omega^2(\vec{x}_\perp) $ &
\textcolor{ForestGreen}{Yes} 
\\ 
Anti-de Sitter  &
$\lambda(\zeta)=
  p(\vec{x}_\perp)(2/\zeta-1)$ &
$R^2\zeta^2 d\Omega^2(\vec{x}_\perp)$ &
\textcolor{red}{No}
\\ 
\specialrule{.13em}{0em}{0em}
\end{tabular}
\caption{Summary of the mapping of the Minkowski null plane under the conformal transformations discussed in this section. We indicate the relation between the affine parameter in the null plane $\lambda$ and the one in the mapped surface, the induced metric and whether the surface fits in the mapped space-time. The metric on the unit sphere $S^{d-2}$ in stereographic coordinates is given by~${d\Omega^2(\vec{x}_\perp)=d\vec{x}_\perp.d\vec{x}_\perp/p^2(\vec{x}_\perp)}$, where $p(\vec{x}_\perp)=(|\vec{x}_\perp|^2+4R^2)/4R$.}\label{table:1}
\end{table}

\subsection{Modular Hamiltonians of null deformed regions in curve backgrounds}

Now that we have a simple description of the mapping of the null plane we can apply these conformal transformations on the modular Hamiltonian $K_{\mathcal{A}^\pm}$ in (\ref{eq:66}) and explicitly write the constraint (\ref{eq:1}) coming from relative entropy. We summarize the most important aspects of the mapping of the null plane in table \ref{table:1}.

A general conformal transformation given by a change of coordinates $z^\mu(X^\mu)$ induces a geometric transformation of the null surface $\mathcal{A}^\pm\rightarrow \bar{\mathcal{A}}^\pm$, while the Hilbert space is mapped by a unitary operator~$U:\mathcal{H}\rightarrow \bar{\mathcal{H}}$. Consider an arbitrary primary operator $\mathcal{O}_a(X^\mu)$ of spin $\ell\in \mathbb{N}_0$, where the label $a$ contains all the Lorentz indices, \textit{i.e.} $a=(\mu_1,\dots,\mu_\ell)$. An arbitrary matrix $R_a^{\,\,\,b}$ is obtained from $R_\mu^{\,\,\,\nu}$ as
\begin{equation}\label{eq:82}
R_a^{\,\,\,b}\equiv R_{\mu_1}^{\,\,\,\nu_1}\dots
  R_{\mu_\ell}^{\,\,\,\nu_\ell}\ .
\end{equation}
Since $\mathcal{O}_a(X^\mu)$ is primary, it transforms according to 
\begin{equation}\label{eq:88}
U\mathcal{O}_a(X^\mu)U^\dagger=
  |w(z^\mu)|^{\ell-\Delta}
  \frac{\partial z^b}{\partial X^a}
  \bar{\mathcal{O}}_b(z^\mu)\ ,
\end{equation}
where $\bar{\mathcal{O}}_a(z^\mu)$ acts on the Hilbert space $\bar{\mathcal{H}}$.

To obtain the transformation property of the reduced density operator $\rho_{\mathcal{A}^\pm}$ we consider its defining property (\ref{eq:67}). Writing this relation for a primary operator $\mathcal{O}_a(X^\mu)$ and using its simple transformation law (\ref{eq:88}) we find
\begin{equation}\label{eq:68}
  \bra{\bar{0}}
  \bar{\mathcal{O}}_b(z^\mu)
  \ket{\bar{0}}=
  {\rm Tr}\big(
  U
  \rho_{\mathcal{A}^\pm}
  U^\dagger 
  \bar{\mathcal{O}}_b(z^\mu)
  \big)\ ,
  \qquad \quad 
  z^\mu \in \mathcal{D\bar{A}}^\pm\ ,
\end{equation}
where $\ket{\bar{0}}=U\ket{0}$ is the vacuum state in the mapped CFT. We have canceled the conformal factors appearing on both sides as well as the Jacobian matrices, which are invertible since conformal transformations can be inverted. The location of the mapped operator is given by $z^\mu \in \mathcal{D\bar{A}}^\pm$.

This relation allows us to identify the reduced density operator associated to the causal domain of the mapped null surface $\bar{\mathcal{A}}^\pm$ as $\bar{\rho}_{\mathcal{\bar{A}}^\pm}=U\rho_{\mathcal{A}^\pm}U^\dagger$. Although (\ref{eq:68}) only involves primary operators of integer spin, we can differentiate it to obtain its descendants, while an analogous transformation property to (\ref{eq:88}) gives the equivalent relation for primary operators of half-integer spin. Altogether, this means that the modular Hamiltonian transforms in the expected way given by the adjoint action of $U$ as $\bar{K}_{\bar{\mathcal{A}}^\pm}=UK_{\mathcal{A}^\pm}U^\dagger$.

\begin{figure}[t]
\begin{center}
\includegraphics[height=2.5 in]{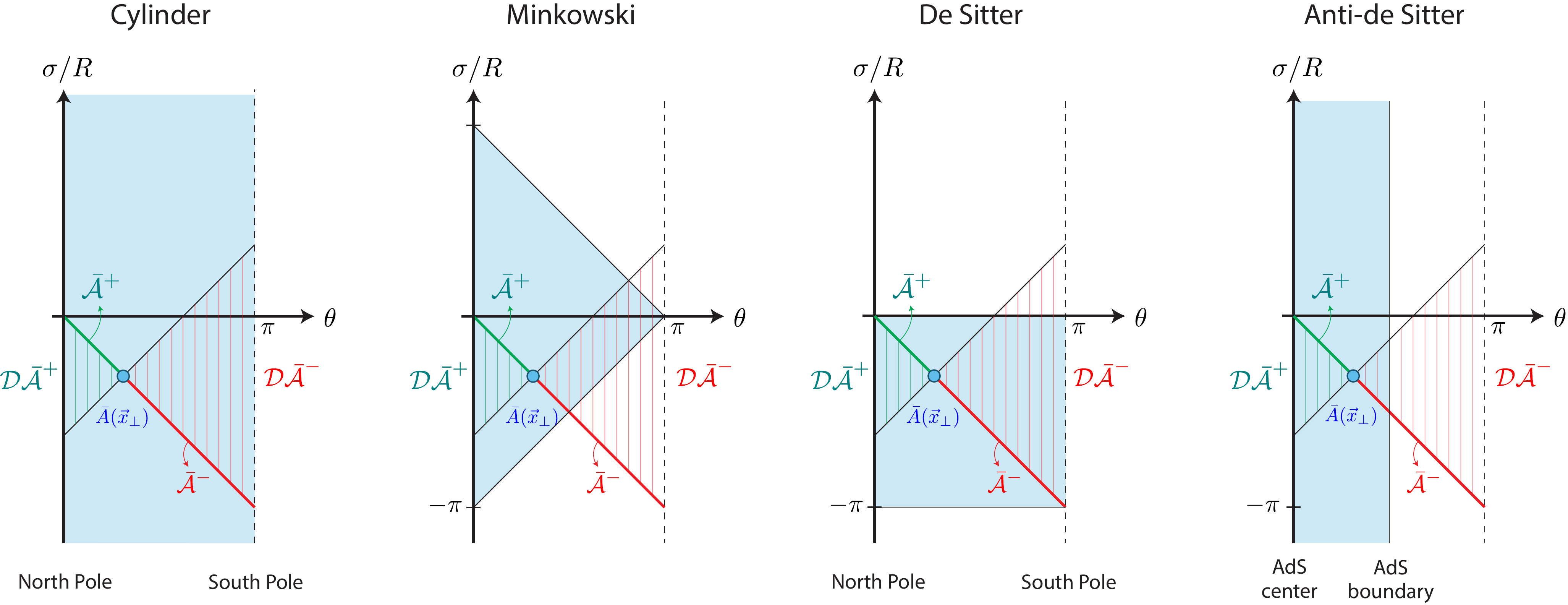}
\caption{Plot of the mapped surfaces $\bar{\mathcal{A}}^\pm$ and their causal domains $\mathcal{D\bar{A}}^\pm$ in the $(\sigma/R,\theta)$, where we indicate the region of the cylinder covered by each of the space-times.}
\label{fig:6}
\end{center}
\end{figure}

Since the modular Hamiltonian of the null plane (\ref{eq:66}) is written as an integral of the stress tensor, we can directly use the transformation of $T_{\lambda \lambda}$ in (\ref{eq:5}). The modular Hamiltonian associated to $\bar{\mathcal{A}}^\pm$ is then given by
\begin{equation}\label{eq:19}
\bar{K}_{\bar{\mathcal{A}}^\pm}=
  \pm 2\pi \int_{\bar{\mathcal{A}}^\pm}
  d\bar{S}\,
  (\lambda-A(\vec{x}_\perp))\bar{T}_{\lambda\lambda}(\lambda,\vec{x}_\perp)\ .
\end{equation}
We have absorbed the factor $|w(z)|^{2-d}$ into the surface element $d\bar{S}=d\vec{x}_\perp d\lambda\sqrt{\bar{h}}$, where $\bar{h}$ is the determinant of the induced metric of the mapped surface in the new space-time. Although $\bar{\mathcal{A}}^\pm$ is a~$(d-1)$ dimensional surface, its surface element scales as $(d-2)$ because it is null. Applying a simple change of integration variables we can write the integral in terms of a generic affine parameter~$\bar{\lambda}$ as
\begin{equation}\label{eq:71}
\bar{K}_{\bar{\mathcal{A}}^\pm}=
  \pm 2\pi \int_{\bar{\mathcal{A}}^\pm}
  d\bar{S}\,
  \left[
  \frac{\lambda(\bar{\lambda})-A(\vec{x}_\perp)}
  {\lambda'(\bar{\lambda})}
  \right]
  \bar{T}_{\bar{\lambda}\bar{\lambda}}
  (\bar{\lambda},\vec{x}_\perp)\ ,
\end{equation}
where we took into account the $\lambda$ derivatives in the definition of $T_{\lambda \lambda}$. In an analogous way, the full modular Hamiltonian in (\ref{eq:177}) transforms according to
\begin{equation}\label{eq:160}
\hat{K}_{\bar{\mathcal{A}}^+}=
  2\pi \int_{\bar{\mathcal{N}}}
  \,d\bar{S}\,\left[
  \frac{\lambda(\bar{\lambda})-A(\vec{x}_\perp)}
  {\lambda'(\bar{\lambda})}
  \right]
  \bar{T}_{\bar{\lambda}\bar{\lambda}}
  (\bar{\lambda},\vec{x}_\perp)\ ,
\end{equation}
where $\bar{\mathcal{N}}=\bar{\mathcal{A}}^+\cup \bar{\mathcal{A}}^-$. Using these relations and the results of the previous section summarized in table \ref{table:1} we can easily write these operators explicitly. In Fig. \ref{fig:6} we plot the null horizons $\bar{\mathcal{A}}^\pm$ and their causal regions for the different space-times in the $(\sigma/R,\theta)$ plane.

\subsubsection{Minkowski null cone}

Let us start by taking $\bar{\mathcal{A}}^+$ as a region of the past null cone in Minkowski (\ref{eq:14}), given by
\begin{equation}\label{eq:120}
\bar{\mathcal{A}}^+=\left\lbrace 
  x^\mu \in \mathbb{R}\times\mathbb{R}^{d-1}:
  \quad x^\mu(\alpha,\vec{x}_\perp)=
  \alpha(-1,\vec{n}(\vec{x}_\perp))\ ,
  \quad
  (\alpha,\vec{x}_\perp) \in [0,\bar{A}(\vec{x}_\perp)]\times \mathbb{R}^{d-2}
  \right\rbrace\ ,
\end{equation}
where $\vec{n}(\vec{x}_\perp)$ is given in (\ref{eq:23}). The affine parameter $\alpha$ is obtained from the relation $\lambda(\alpha)$ in table~\ref{table:1}, while the entangling surface $\lambda=\bar{A}(\vec{x}_\perp)\ge 0$ is determined from $A(\vec{x}_\perp)=p(\vec{x}_\perp)(R/\bar{A}(\vec{x}_\perp)-1)$. Using~(\ref{eq:71}) and the results in table \ref{table:1} the modular Hamiltonian associated to the region $\bar{\mathcal{A}}^+$ is given by
\begin{equation}\label{eq:72}
\bar{K}_{\bar{\mathcal{A}}^+}=
  2\pi 
  \int_{S^{d-2}}d\Omega(\vec{x}_\perp)
  \int_{0}^{\bar{A}(\vec{x}_\perp)}
  d\alpha\,\alpha^{d-2}
  \left[
  \frac{\alpha(\bar{A}(\vec{x}_\perp)-\alpha)}
  {\bar{A}(\vec{x}_\perp)}
  \right]
  \bar{T}_{\alpha \alpha}
  (\alpha,\vec{x}_\perp)\ .
\end{equation}

Fixing $\bar{A}(\vec{x}_\perp)=R$, the space-time region $\mathcal{D\bar{A}}^+$ corresponds to the causal domain of a ball of radius~$R$ centered at $t=-R$, whose modular Hamiltonian has been long known \cite{Hislop:1981uh,Casini:2011kv}. For an arbitrary function $\bar{A}(\vec{x}_\perp)$ it gives the modular Hamiltonian associated to null deformations of the ball.\footnote{For a nice 3D picture of the setup see Fig. 2 of Ref. \cite{Casini:2018kzx}.} This operator was previously considered in Ref. \cite{Casini:2017roe} but the result in that paper is incorrect, as can be seen by noting that it does not reproduce the correct result when $\bar{A}(\vec{x}_\perp)=R$.\footnote{A correct expression for the modular Hamiltonian that is equivalent to (\ref{eq:72}) was previously given in Ref. \cite{Neuenfeld:2018dim} without proof.} The integral in (\ref{eq:72}) can also be written directly in terms of the space-time coordinates using that~$r_-=2\alpha$ and~$\vec{x}_\perp=\vec{v}$.

For the complementary space-time region $\mathcal{D\bar{A}}^-$ we cannot write the modular Hamiltonian since the null surface $\bar{\mathcal{A}}^-$ does not fit inside Minkowski, see Fig. \ref{fig:6}. This means we cannot write the full modular Hamiltonian $\hat{K}_{\bar{\mathcal{A}}^+}$ and derive a null energy bound from the monotonicity of relative entropy.

An exception to this is given by the case of the ball where $\bar{A}(\vec{x}_\perp)=R$ implies $A(\vec{x}_\perp)=0$. As previously discussed, for this particular case the modular Hamiltonian becomes the Bisognano-Wichmann result, meaning that it can be written as a local integral over any Cauchy surface in the region~$\mathcal{D\bar{A}}^-$. We can use this freedom to chose a surface which fits in Minkowski, starting from~$\bar{A}(\vec{x}_\perp)=R$ (blue dot in second diagram of Fig. \ref{fig:6}) and finishing at space-like infinity ${(\sigma/R,\theta)=(0,\pi)}$. Using this we can write the modular Hamiltonian corresponding to the complementary region of a ball in Minkowski, as done for example in Ref. \cite{Blanco:2013lea}. This analysis clarifies the validity of such expression.

\subsubsection{Lorentzian cylinder}

We now consider the transformation to the Lorentzian cylinder, where the null surface $\bar{\mathcal{A}}^+$ is written in the coordinates $u^\mu=(\sigma/R,\theta,\vec{v}\,)$ as
\begin{equation}\label{eq:73}
\bar{\mathcal{A}}^+=
  \left\lbrace
  u^\mu\in \mathbb{R}\times [0,\pi]\times
  \mathbb{R}^{d-2}:
  \quad
   u^\mu=
  \left(
  -\beta,\beta,\vec{x}_\perp
  \right)\ ,
  \quad
  (\beta,\vec{x}_\perp)
   \in [0,\bar{A}(\vec{x}_\perp)]
   \times \mathbb{R}^{d-2}
  \right\rbrace\ .
\end{equation}
The entangling surface $\beta=\bar{A}(\vec{x}_\perp)\in[0,\pi]$ is given by the function $\bar{A}(\vec{x}_\perp)$, which can be written from the relation $\lambda(\beta)$ in table \ref{table:1} as $A(\vec{x}_\perp)=p(\vec{x}_\perp)(2\cot(\bar{A}(\vec{x}_\perp))-1)$. The modular Hamiltonian is obtained from (\ref{eq:71}) and table \ref{table:1}, so that we find
\begin{equation}\label{eq:75}
\bar{K}_{\bar{\mathcal{A}}^+}=
  2\pi \int_{S^{d-2}}d\Omega(\vec{x}_\perp)
  \int_0^{\bar{A}(\vec{x}_\perp)}d\beta\,
  \left(R\sin(\beta)\right)^{d-2}
  \left[
  \frac{\sin(\beta)\sin(\bar{A}(\vec{x}_\perp)-\beta)}{\sin(\bar{A}(\vec{x}_\perp))}\right]
  \bar{T}_{\beta\beta}(\beta,\vec{x}_\perp)\ .
\end{equation}
For $\bar{A}(\vec{x}_\perp)=\theta_0$ the region $\mathcal{D\bar{A}}^+$ corresponds to the causal domain of a cap region centered at the North Pole on the spatial sphere $S^{d-1}$ and agrees with the result obtained in \cite{Casini:2011kv}. The operator can be written in terms of the space-time coordinates using that $\theta_-=2\beta$ and $\vec{x}_\perp=\vec{v}$.

Since the whole null surface fits in the cylinder, we can write the operator associated to the complementary region or equivalently, we can directly express the full modular Hamiltonian using~(\ref{eq:160}) as
\begin{equation}\label{eq:162}
\hat{K}_{\mathcal{\bar{A}}^+}=
  2\pi\int_{S^{d-2}}d\Omega(\vec{x}_\perp)
  \int_0^\pi d\beta\,
  \left(R\sin(\beta)\right)^{d-2}
  \left[
  \frac{
  \sin(\beta)
  \sin(\bar{A}(\vec{x}_\perp)-\beta)}
  {\sin(\bar{A}(\vec{x}_\perp))}
  \right]
  \bar{T}_{\beta\beta}(\beta,\vec{x}_\perp)\ .
\end{equation}
From this we can explicitly write the constraint (\ref{eq:1}) coming from relative entropy and obtain a bound on the null energy. Since the two regions are determined by the functions $\bar{A}(\vec{x}_\perp)$ and $\bar{B}(\vec{x}_\perp)$ the constraint becomes
\begin{equation}\label{eq:168}
\int_0^{\pi}d\beta\,\sin^{d}(\beta)
  \int_{\mathbb{R}^{d-2}}
  d\vec{x}_\perp
  \left[
  \frac{\cot(\bar{B}(\vec{x}_\perp))-
  \cot(\bar{A}(\vec{x}_\perp))}
  {p(\vec{x}_\perp)^{d-2}}
  \right]
  \bar{T}_{\beta \beta}(\beta,\vec{x}_\perp)\ge 0\ ,
\end{equation}
where $\bar{A}(\vec{x}_\perp)\ge \bar{B}(\vec{x}_\perp)$ so that the condition for the regions in (\ref{eq:1}) is satisfied. We have also written the surface element $d\Omega(\vec{x}_\perp)$ explicitly in terms of $p(\vec{x}_\perp)$. It is now convenient to fix the functions~$\bar{A}(\vec{x}_\perp)$ and $\bar{B}(\vec{x}_\perp)$ to
\begin{equation}\label{eq:175}
\bar{A}(\vec{x}_\perp)=\pi/2\ ,
  \qquad {\rm and} \qquad
  \cot(\bar{B}(\vec{x}_\perp))=
  p(\vec{x}_\perp)^{d-2}
  \delta(\vec{x}_\perp-\vec{x}_\perp^{\,0})\ ,
\end{equation}
where $\vec{x}_\perp^{\,0}$ is any fixed vector in $\mathbb{R}^{d-2}$. Although the condition for $\bar{B}(\vec{x}_\perp)$ involving the Dirac delta might seem unusual due to the cotangent function, $\bar{B}(\vec{x}_\perp)$ is determined by the original function~$B(\vec{x}_\perp)$ from $B(\vec{x}_\perp)=p(\vec{x}_\perp)(2\cot(\bar{B}(\vec{x}_\perp))-1)$. The behavior of $\bar{B}(\vec{x}_\perp)$ implied by (\ref{eq:175}) is qualitatively given by
$$\bar{B}(\vec{x}_\perp)\sim
  \begin{cases}
  \quad \pi/2 \quad\ , 
  \quad{\rm for\,\,}
  \vec{x}_\perp \neq \vec{x}_\perp^{\,0}\\
  \quad \,\,\,0 \quad \,\,\,\, \ , 
  \quad{\rm for\,\,}
  \vec{x}_\perp = \vec{x}_\perp^{\,0}\ .
  \end{cases}$$ 
Using this we can solve the integral in $\vec{x}_\perp$ in (\ref{eq:168}) and find
\begin{equation}\label{eq:31}
\int_0^\pi d\beta\,\sin^d(\beta)
  \,
  \bar{T}_{\beta\beta}(\beta,\vec{x}_\perp)
  \ge 0\ ,
\end{equation}
where the affine parameter $\beta$ describes the geodesic in (\ref{eq:62}). Up to a translation of the geodesic, this is equivalent to the constraint derived in the previous section (\ref{eq:195}).

\subsubsection{De Sitter}

For de Sitter, the null surface $\bar{\mathcal{A}}^+$ is given in the $u^\mu=(\sigma/R,\theta,\vec{v}\,)$ coordinates by
\begin{equation}\label{eq:15}
\mathcal{\bar{A}}^+=
  \left\lbrace
  u^\mu\in [-\pi,0]\times [0,\pi]\times \mathbb{R}^{d-2}:
  \quad
   u^\mu=
  \left(
  -\beta(\eta),\beta(\eta),\vec{x}_\perp
  \right)\ ,
  \quad
  (\eta,\vec{x}_\perp)\in 
  [\bar{A}(\vec{x}_\perp),+\infty)\times \mathbb{R}^{d-2}
  \right\rbrace\ ,
\end{equation}
where $\eta(\beta)=\cot(\beta)$. The entangling surface $\eta=\bar{A}(\vec{x}_\perp)\in \mathbb{R}$ is obtained from the relation $\lambda(\eta)$ in table \ref{table:1} as $A(\vec{x}_\perp)=p(\vec{x}_\perp)(2\bar{A}(\vec{x}_\perp)-1)$. Using (\ref{eq:71}) and the results in table \ref{table:1} we can write the associated modular Hamiltonian as
\begin{equation}\label{eq:24}
\bar{K}_{\bar{\mathcal{A}}^+}=
  2\pi R^{d-2} \int_{S^{d-2}}d\Omega(\vec{x}_\perp)
  \int_{\bar{A}(\vec{x}_\perp)}^{+\infty}
  d\eta\,
  \left(
  \eta-\bar{A}(\vec{x}_\perp)
  \right)
  \bar{T}_{\eta \eta}
  (\eta,\vec{x}_\perp)\ ,
\end{equation}
which has a similar structure to that of the Minkowski null plane (\ref{eq:66}). When $\bar{A}(\vec{x}_\perp)=0$ we have~$\beta=\pi/2$ so that the space-time regions $\mathcal{D\bar{A}}^\pm$ correspond to the left and right static patches of de Sitter, see Fig. \ref{fig:6}. For general $\bar{A}(\vec{x}_\perp)$ it is given by null deformations of these regions. 

Since the whole null surface fits inside de Sitter, we can write the modular Hamiltonian of the complementary region and therefore the full modular Hamiltonian, which from (\ref{eq:160}) is given by
\begin{equation}\label{eq:164}
\hat{K}_{\bar{\mathcal{A}}^+}=
  2\pi R^{d-2}\int_{S^{d-2}}
  d\Omega(\vec{x}_\perp)
  \int_{-\infty}^{+\infty}d\eta\,
  \left(
  \eta-\bar{A}(\vec{x}_\perp)
  \right)
  \bar{T}_{\eta \eta}(\eta,\vec{x}_\perp)\ .
\end{equation}
From this we can explicitly write the constraint (\ref{eq:1}) coming from monotonicity of relative entropy. Taking the regions as determined by the two functions $\bar{A}(\vec{x}_\perp)$ and $\bar{B}(\vec{x}_\perp)$, the general inequality in~(\ref{eq:1}) implies
\begin{equation}\label{eq:26}
  \int_{-\infty}^{+\infty}d\eta
  \int_{{\mathbb{R}}^{d-2}}
  d\vec{x}_\perp\,
  \left[
  \frac{\bar{B}(\vec{x}_\perp)-
  \bar{A}(\vec{x}_\perp)}
  {p(\vec{x}_\perp)^{d-2}}
  \right]
  \bar{T}_{\eta \eta}(\eta,\vec{x}_\perp)\ge 0\ ,
\end{equation}
where $\bar{B}(\vec{x}_\perp)\ge \bar{A}(\vec{x}_\perp)$ so that the condition for the regions in (\ref{eq:1}) is satisfied. We have also written the integral over $S^{d-2}$ explicitly in terms of $\vec{x}_\perp$. Fixing the regions such that
$$\bar{A}(\vec{x}_\perp)=0\ ,
  \qquad {\rm and} \qquad
  \bar{B}(\vec{x}_\perp)=
  p(\vec{x}_\perp)^{d-2}
  \delta(\vec{x}_\perp-\vec{x}_\perp^{\,0})\ ,$$
we can trivially solve the integral and obtain the ANEC for a CFT in de Sitter
$$\mathcal{E}_{\rm dS}(\vec{x}_\perp)=
  \int_{-\infty}^{+\infty}d\eta\,
  \bar{T}_{\eta \eta}(\eta,\vec{x}_\perp)\ge 0\ ,$$
where the geodesic is given by (\ref{eq:203}).

\subsubsection{Anti-de Sitter}

Finally let us consider the conformal transformation to AdS, where the null surface $\bar{\mathcal{A}}^+$ is written in the coordinates $u^\mu=(\sigma/R,\theta,\vec{v}\,)$ as
\begin{equation}\label{eq:158}
\mathcal{\bar{A}}^+=
  \left\lbrace
  u^\mu\in \mathbb{R}\times [0,\pi/2]\times   
  \mathbb{R}^{d-2}:
  \quad
   u^\mu=
  \left(
  -\beta(\zeta),\beta(\zeta),\vec{x}_\perp
  \right)\ ,
  \quad
  (\zeta,\vec{x}_\perp)\in[0,\bar{A}(\vec{x}_\perp)]\times \mathbb{R}^{d-2}
  \right\rbrace\ ,
\end{equation}
where $\zeta(\beta)=\tan(\beta)$ and $\bar{A}(\vec{x}_\perp)\in \mathbb{R}_{\ge 0}$ is obtained from the relation $\lambda(\zeta)$ in table \ref{table:1} as~$A(\vec{x}_\perp)=p(\vec{x}_\perp)\left(2/\bar{A}(\vec{x}_\perp)-1\right)$. The modular Hamiltonian associated to $\bar{\mathcal{A}}^+$ is obtained from~(\ref{eq:71}) and the results in table \ref{table:1}
\begin{equation}\label{eq:18}
\bar{K}_{\bar{\mathcal{A}}^+}=
  2\pi R^{d-2}
  \int_{S^{d-2}}
  d\Omega^2(\vec{x}_\perp)
  \int_0^{\bar{A}(\vec{x}_\perp)}
  d\zeta\,\zeta^{d-2}
  \left[
  \frac{\zeta(\bar{A}(\vec{x}_\perp)-\zeta)}
  {\bar{A}(\vec{x}_\perp)}
  \right] 
  \bar{T}_{\zeta \zeta}(\zeta,\vec{x}_\perp)\ .
\end{equation}
Notice that it has the same structure as the modular Hamiltonian on the deformed null cone (\ref{eq:72}). If the function $\bar{A}(\vec{x}_\perp)$ is constant, the space-time region $\mathcal{D\bar{A}}^+$ corresponds to the causal domain of a ball in AdS. We can see this noting that the usual AdS radial coordinate $\rho$ in table \ref{table:2} is given by~$\rho=R\tan(\theta)$. Since the full null surface $\mathcal{\bar{N}}=\bar{\mathcal{A}}^+\cup \bar{\mathcal{A}}^-$ does not fit inside the whole AdS space-time we cannot write the full modular Hamiltonian and the constraint (\ref{eq:1}) coming from relative entropy. This means that while the ANEC in dS can be derived from relative entropy, this is not true for AdS, as a consequence of the fact that the Minkowski null plane does not fit inside AdS.

\subsection{Entanglement entropy}

Since we have derived some new modular Hamiltonians for CFTs in the Lorentzian cylinder and (A)dS, we would like to compute their associated entanglement entropy. In Ref. \cite{Casini:2018kzx} the entropy of the regions in the null plane and cone in Minkowski were computed using two independent approaches; the first one based on some symmetry considerations and the second on the HRRT holographic prescription~\cite{Ryu:2006bv,Hubeny:2007xt}. We follow the holographic approach since it is the simplest, although in future work it would be interesting to study the generalization of the other procedure.

The details of the calculations are summarized in App. \ref{zapp:entanglement}. The final result for the entanglement entropy can be written in every case as
\begin{equation}\label{eq:171}
S=\frac{\mu_{d-2}}{\epsilon^{d-2}}+
  \dots+
  a_d^*\times
\begin{cases}
\displaystyle
  \frac{4(-1)^{\frac{d-2}{2}}}
  {{\rm Vol}(S^{d-2})}
  \int_{S^{d-2}}
  d\Omega(\vec{v}\,)
  \ln\left[
  \frac{2R}{\epsilon}b_0(\vec{v}\,)
  \right]\ ,
  \quad d{\rm \,\, even}
  \vspace{6pt}\\
  \qquad \qquad \qquad
  \displaystyle
  2\pi(-1)^{\frac{d-1}{2}}
  \qquad \qquad \quad \,\,\,\ ,
  \quad \, d{\rm \,\, odd}\ ,
\end{cases}
\end{equation}
where ${\rm Vol}(S^{d-1})=2\pi^{d/2}/\Gamma(d/2)$, $\epsilon$ is a short distance cut-off and $a_d^*$ is given by \cite{Nishioka:2018khk}
\begin{equation}\label{eq:155}
a_d^*=
  \begin{cases} 
  \qquad \qquad \,\,\,\,\,
  A_d
  \qquad \quad \,\,\,\, 
  \ ,
  & {\rm for\,\,d\,\,even} \vspace{6pt}\\
  \,\,(-1)^{\frac{d-1}{2}}\ln[
  Z(S^d)]/2\pi\ ,
  &  {\rm for\,\,d\,\,odd}\ . \\
 \end{cases}
\end{equation}
The coefficient of the Euler density in the stress tensor trace anomaly is given by $A_d$ (see Ref. \cite{Myers:2010tj} for conventions) while $Z(S^d)$ is the regularized vacuum partition function of the CFT placed on a unit $d$-dimensional sphere (see Ref. \cite{Pufu:2016zxm} for some examples in free theories).

The entanglement entropy (\ref{eq:171}) has a divergent expansion in $\epsilon$ with a leading area term, whose coefficient $\mu_{d-2}$ is non-universal (depends on the regularization procedure). The only universal term is indicated in (\ref{eq:171}) and depends on the value of $d$. For odd space-times it is the same in every setup, while for even $d$ the function $b_0(\vec{v}\,)$ is given in each case by
$$b_0(\vec{v}\,)=
  \begin{cases}
   \,\,\,\,
  \displaystyle
  \sin(\bar{A}(\vec{v}\,))\in(0,1]
  \,\,\,\,\ ,
  \qquad {\rm Lorentzian\,\,cylinder}\\
  \displaystyle
  \,\,\,
  1/\bar{A}(\vec{v}\,)\in(0,+\infty)
  \,\,\,\ ,
  \qquad\, {\rm de\,\, Sitter}\\
  \displaystyle
  \quad 
  \bar{A}(\vec{v}\,)\in(0,+\infty)
  \quad\ ,\,
  \qquad {\rm anti-de\,\, Sitter}\ .
  \end{cases}  
  $$
We have indicated the range of $b_0(\vec{v}\,)$ given by the fact that the functions $\bar{A}(\vec{v}\,)$ are different in each setup, see the definition of the null surfaces above. Based on the arguments given in Ref. \cite{Casini:2018kzx}, we expect this calculation for the entanglement entropy to hold to every order in the holographic CFT.

Notice that for the case of de Sitter we have restricted $\bar{A}(\vec{v}\,)>0$ despite of the fact that the mapping of the null plane fits in the space-time for $\bar{A}(\vec{v}\,)\in \mathbb{R}$ (see (\ref{eq:15}) and Fig. \ref{fig:6}). The issue with~$\bar{A}(\vec{v})\le 0$ is that the associated space-time region $\mathcal{D\bar{A}}^+$ lies outside of de Sitter. The entanglement entropy is a non-local quantity that captures this so that the holographic calculation breaks down in this regime, see App. \ref{zapp:entanglement} for details.

\section{Wedge reflection positivity in curved backgrounds}
\label{sec:ref_positivity}

In the previous sections we derived interesting bounds for the null energy along a complete geodesic for CFTs in (A)dS and the Lorentzian cylinder. We now want to investigate whether these results can be obtained from the causality arguments used in Ref. \cite{Hartman:2016lgu} to derive the Minkowski ANEC. One of the crucial ingredients in this proof from causality is the so called ``Rindler positivity" or ``wedge reflection positivity" (we use these terms interchangeably). This is a general property proved in Ref.~\cite{Casini:2010bf} that implies the positivity of certain correlation functions in Minkowski. The aim of this section is to show that wedge reflection positivity generalizes to CFTs in dS and the Lorentzian cylinder, but not to AdS. 

Let us start by reviewing some general aspects of the Tomita-Takesaki theory \cite{Haag:1992hx,Witten:2018lha} that is the central formalism used in this section. Given a QFT and a space-time region $W$ in Minkowski we can identify a Von Neumann algebra $\mathcal{W}$, given by all the bounded operators supported in $W$ that close under hermitian conjugation and the weak operator topology.\footnote{Given a sequence of operators $\mathcal{O}_n$ the weak operator topology defines the limit $\mathcal{O}_n\rightarrow \mathcal{O}$ according to~$|\bra{\alpha}\left(\mathcal{O}_n-\mathcal{O}\right)\ket{\beta}|<\epsilon_n$, where $\ket{\alpha}$ and $\ket{\beta}$ are any two vectors in the Hilbert space.} From this algebra we can construct its commutant $\mathcal{W}'$, that is also a Von Neumann algebra formed by all the operators that commute with every element in $\mathcal{W}$. 

The Tomita-Takesaki theory starts by assuming that we can find a cyclic and separating vector~$\ket{\psi}$ with respect to the Von Neumann algebra $\mathcal{W}$.\footnote{The vector $\ket{\psi}$ is cyclic if the set $\left\lbrace \mathcal{O}\ket{\psi},\,\forall\,\mathcal{O}\in \mathcal{W} \right\rbrace$ is dense in $\mathcal{H}$, while it is separating if ${\mathcal{O}\ket{\psi}=0}$ for $\mathcal{O}\in \mathcal{W}$ implies $\mathcal{O}\equiv 0$.} For a particular choice of $\ket{\psi}$ and $\mathcal{W}$ we define the Tomita operator $S$ according to
$$S\,\mathcal{O}\ket{\psi}=
  \mathcal{O}^\dagger \ket{\psi}\ ,
  \qquad \forall \,\,\mathcal{O}\in \mathcal{W}\ .$$
Since $\ket{\psi}$ is cyclic this defines the action of $S$ on every vector of the Hilbert space. The Tomita operator can be written in terms of its polar decomposition as $S=J\Delta^{1/2}$ with $J$ anti-unitary and~$\Delta^{1/2}$ hermitian and positive semi-definite. Moreover, since $S$ has an inverse $S^{-1}=S$, the choice of $J$ is unique and $\Delta^{1/2}$ is positive definite. The operator $J$ is called the modular conjugation and $\Delta$ the modular operator. Without too much effort, they can be shown to satisfy the following properties (\textit{e.g.} see Ref. \cite{Witten:2018lha})
\begin{equation}\label{eq:76}
J=J^\dagger=J^{-1}\ ,
  \qquad \quad
  J\Delta^{1/2} J=\Delta^{-1/2}\ ,
  \qquad \quad
  J\ket{\psi}=\Delta\ket{\psi}=\ket{\psi}\ ,
\end{equation}
where the definition of the hermitian conjugate for an anti-unitary operator is $\bra{\alpha}J\ket{\beta}=\bra{\beta}J^\dagger\ket{\alpha}$. The key properties satisfied by $J$ and $\Delta$ which amounts to the Tomita-Takesaki theorem are given by
\begin{equation}\label{eq:35}
J\,\mathcal{W}J=\mathcal{W}'\ ,
\end{equation}
$\Delta^{is}\mathcal{W}\Delta^{-is}=\mathcal{W}$ and $\Delta^{is}\mathcal{W}'\Delta^{-is}=\mathcal{W}'$ where $s\in \mathbb{R}$. The modular conjugation $J$ maps the algebra into its commutant, while $\Delta^{is}$ transforms each algebra into itself. 

Given $\mathcal{O}\in \mathcal{W}$ we define the ``reflected" operator $\widetilde{\mathcal{O}}$ as $\widetilde{\mathcal{O}}\equiv J\mathcal{O}J \,\in \mathcal{W}'$. From this formalism follows a very general inequality which bounds the expectation value of $\widetilde{\mathcal{O}}\mathcal{O}$ in the state $\ket{\psi}$
$$\bra{\psi}
  \widetilde{\mathcal{O}}
  \mathcal{O}
  \ket{\psi}=
  \bra{\psi}
  \mathcal{O}
  \Delta^{1/2}S\mathcal{O}
  \ket{\psi}^*=
  \bra{\psi}
  \mathcal{O}
  \Delta^{1/2}\mathcal{O}^\dagger
  \ket{\psi}=
  \bra{\alpha}
  \Delta^{1/2}\ket{\alpha}\ ,
  $$
where we have define $\ket{\alpha}=\mathcal{O}^\dagger\ket{\psi}$. Using that $\Delta^{1/2}$ is positive definite we arrive at the central inequality
\begin{equation}\label{eq:34}
\bra{\psi}\widetilde{\mathcal{O}}\mathcal{O}\ket{\psi}>0\ ,
\qquad 
\mathcal{O}\in \mathcal{W}\ .
\end{equation}
For a generic setup the reflected operator $\widetilde{\mathcal{O}}$ is related to $\mathcal{O}$ in a very complicated way. The only certainty we have regarding $\widetilde{\mathcal{O}}$ is that it is in the commutant algebra of $\mathcal{W}$, which follows from the Tomita-Takesaki theorem (\ref{eq:35}).  This means that extracting useful information from (\ref{eq:34}) might be very challenging. 

There is however a particular setup in which the action of $J$ becomes simple enough. Taking the Minkowski space-time coordinates $X^\mu=(T,X,\vec{Y})$, consider the right Rindler wedge
\begin{equation}\label{eq:80}
W=\left\lbrace
  X^\mu \in \mathbb{R}\times \mathbb{R}\times \mathbb{R}^{d-2}:
  \quad
  X_\pm=X\pm T>0
  \right\rbrace\ .
\end{equation}
For the Von Neumann algebra associated to this wedge and the Minkowski vacuum state $\ket{0}$,\footnote{The Minkowski vacuum state is cyclic and separating as a consequence of the Reeh-Schlieder theorem.} Bisognano and Wichmann \cite{Bisognano:1976za} proved that the modular operator $\Delta$ is given by $\Delta=e^{-\hat{K}_W}$, where~$\hat{K}_W$ is the full modular Hamiltonian defined in (\ref{eq:161}), which can be written as
\begin{equation}\label{eq:172}
\hat{K}_W=
  2\pi \int_{\mathcal{N}_{\rm plane}}
  dS\,\lambda\,T_{\lambda \lambda}(\lambda,\vec{x}_\perp)\ ,
\end{equation}
where the integral is over the full null plane in (\ref{eq:50}) with $dS=d\vec{x}_\perp d\lambda$.

Moreover, they showed that the modular conjugation $J$ is obtained from the consecutive discrete transformations $J={\rm CRT}$, where the operators ${\rm T}$ and ${\rm R}$ reflect the coordinates $T$ and $X$ respectively while ${\rm C}$ implements charge conjugation. Starting from a QFT that is invariant under the Poincare group without assuming invariance under any discrete symmetry, it can be shown that the vacuum is invariant under the combination ${\rm CRT}$, \textit{i.e.} ${\rm CRT}\ket{0}=\ket{0}$. The proof is analogous to the ${\rm CPT}$ theorem for $d=4$, see the discussion in Refs. \cite{Witten:2018lha,Manoharan1969}. This gives a very simple description of the modular conjugation $J$, whose action on an arbitrary operator $\mathcal{O}_a(X^\mu)$ of integer spin $\ell$ is given by~\cite{Casini:2010bf}
\begin{equation}\label{eq:89}
J\mathcal{O}_a(X^\mu)J=
  \frac{\partial \widetilde{X}^b}
  {\partial X^a}
  \mathcal{O}_b^\dagger(\widetilde{X}^\mu)\ ,
  \qquad {\rm with} \qquad
  \widetilde{X}^\mu(X^\mu)=(-T,-X,\vec{Y})\ ,
\end{equation}
where we are using the notation $a=(\nu_1,\dots,\nu_\ell)$ and the jacobian matrix is written in the convention~(\ref{eq:82}).\footnote{For half-integer spin the action of ${\rm CRT}$ is more complicated and there are some subtelties regarding the inequality~(\ref{eq:34}), see Ref. \cite{Casini:2010bf}.} If the operator $\mathcal{O}_a(X^\mu)$ is inserted in the right wedge $W$, the action of $J$ translates it to the complementary region $W'$, the left wedge
$$W'=\left\lbrace
  X^\mu \in \mathbb{R}\times \mathbb{R}\times \mathbb{R}^{d-2}:
  \quad
  X_\pm=X\pm T<0
  \right\rbrace\ .$$
For this reason, we call the geometric action $\widetilde{X}^\mu(X^\mu)$ a reflection. 

Using this we can explicitly write the general inequality (\ref{eq:34}) and obtain Rindler positivity as derived in Ref. \cite{Casini:2010bf}
\begin{equation}\label{eq:90}
  (-1)^P
  \bra{0}
  \mathcal{O}^\dagger_{\mu_1\dots\mu_\ell}
  (\widetilde{X}^\mu)
  \mathcal{O}_{\nu_1\dots\nu_\ell}(X^\mu) 
  \ket{0}>0\ ,
  \qquad \qquad
  X^\mu \in W\ ,
\end{equation}
where $P$ is the number of $T$ indices plus $X$ indices. Although we have only written the expression for a single operator this property holds for an arbitrary number of operators, where notice that the order of the reflected operator is not inverted, \textit{i.e.} $\widetilde{\mathcal{O}_1\mathcal{O}_2}=\widetilde{\mathcal{O}}_1\widetilde{\mathcal{O}}_2$. Moreover, since the expectation values of operators in Lorentzian signature are not functions but distributions, this is a constraint on a distribution.

\subsection{Conformal transformation of Tomita operator}

The strategy for generalizing (\ref{eq:90}) is simple. Using the conformal transformations discussed in Sec.~\ref{sec:null_energy_bounds} we can map the Tomita operator, explicitly write the general inequality (\ref{eq:34}) and obtain wedge reflection positivity in these curved backgrounds.

Consider a generic conformal transformation implemented in the space-time by a change of coordinates $z^\mu(X^\mu)$ that maps the right Rindler wedge in Minkowski $W$ to some other region $\bar{W}$ in the new space-time. The transformation of the Hilbert space is implemented by a unitary operator~$U:\mathcal{H}\rightarrow \bar{\mathcal{H}}$, so that the algebra $\mathcal{W}$ is mapped by the adjoint action of $U$ according to $U\mathcal{W}U^\dagger=\bar{\mathcal{W}}$. Using that~$\mathcal{W}$ is a Von Neumann algebra it is straightforward to show that this is also true for $\bar{\mathcal{W}}$. Although every local operator in $W$ is mapped to a local operator in $\bar{W}$ under the action of $U$, only primary operators have a simple transformation law.

The vacuum state $\ket{0}\in \mathcal{H}$ is mapped to $U\ket{0}=\ket{\bar{0}}\in \bar{\mathcal{H}}$ which can be shown to be cyclic and separating with respect to $\bar{\mathcal{W}}$, using that this is true for $\ket{0}$ and $\mathcal{W}$. This means we can construct the Tomita operator $\bar{S}$ associated to $\ket{\bar{0}}$ and the algebra $\bar{\mathcal{W}}$ in the usual way
$$\bar{S}\bar{\mathcal{O}}\ket{\bar{0}}=
  \bar{\mathcal{O}}^\dagger \ket{\bar{0}}\ ,
  \qquad \qquad
  \bar{\mathcal{O}}\in \bar{\mathcal{W}}\ .$$
The mapped Tomita operator $\bar{S}$ is related to $S$ in the Rindler wedge through the adjoint action of~$U$, so that the mapped modular operator and conjugation are given by
$$\bar{\Delta}=
  \exp\big(
  -U\hat{K}_WU^\dagger
  \big)  \ ,
  \qquad \qquad
  \bar{J}=U({\rm CRT})U^\dagger\ ,$$
where $\hat{K}_W$ is the boost generator in (\ref{eq:172}). The mapping of the modular operator $\Delta$ is completely determined by the transformation of the full modular Hamiltonian $\hat{K}_W$. Since we already analyzed the mapping of this operator in Sec. \ref{sec:null_energy_bounds} we focus on the modular conjugation.\footnote{In Sec. \ref{sec:null_energy_bounds} we analyze the transformation of the full modular Hamiltonian for more general regions given by arbitrary null deformations of the Rindler wedge. Here we restrict to the case in which we have no deformations.}

The action of the modular conjugation $\bar{J}$ can be found by applying $U$ to the ${\rm CRT}$ action in (\ref{eq:89}). If we restrict to bosonic primary operators $\mathcal{O}_a(X^\mu)$ and use that they transform according to (\ref{eq:88}), we find
\begin{equation}\label{eq:46}
\bar{J}
  \mathcal{\bar{O}}_a(z^\mu)
  \bar{J}=
  \left|
  \frac{w(z^\mu)}{w(\tilde{z}^\mu)}
  \right|^{\Delta-\ell}
  \frac{\partial\tilde{z}^b}
  {\partial z^a}
  \bar{\mathcal{O}}_b^\dagger(\tilde{z}^\mu)\ ,
  \qquad {\rm with} \qquad
  \tilde{z}^\mu=z^\mu(\widetilde{X}^\mu)\ ,
\end{equation}
where we used that the jacobian matrix is invertible since this is true for the conformal mapping. The action of $\bar{J}$ is similar to that of ${\rm CRT}$, since the local operator inserted at $z^\mu$ is geometrically reflected to $\tilde{z}^\mu$. However, notice that (\ref{eq:46}) only holds for primary operators while the action of ${\rm CRT}$ in (\ref{eq:89}) is for arbitrary operators.

From this we can write the general positivity inequality (\ref{eq:34}) coming from the Tomita-Takesaki theory and find
\begin{equation}\label{eq:107}
\frac{\partial \tilde{z}^c}{\partial z^b}
  \bra{\bar{0}}
  \bar{\mathcal{O}}_c^\dagger(\tilde{z}^\mu)
  \bar{\mathcal{O}}_a(z^\mu)
  \ket{\bar{0}}>0\ ,
  \qquad
  \bar{\mathcal{O}}_a:{\rm primary}\ ,
  \quad
  z^\mu \in \bar{W}\ .
\end{equation}
This gives a positivity constraint on the correlators of the mapped CFT that is analogous to Rindler positivity in (\ref{eq:90}). In the following, we explicitly write this for CFTs in the Lorentzian cylinder and de Sitter and show that it can be expressed as in (\ref{eq:90}).

Before moving on, let us note that $\bar{J}$ gives an interesting discrete symmetry of the vacuum~${\bar{J}\ket{\bar{0}}=\ket{\bar{0}}}$ which might not be evident from first principles. In particular, it relates two point functions of primary operators according to
\begin{equation}\label{eq:108}
\bra{\bar{0}}
  \mathcal{\bar{O}}_a(z_1)
  \mathcal{\bar{O}}_b(z_2)
  \ket{\bar{0}}=
  \left|
  \frac{w(z_1)w(z_2)}
  {w(\tilde{z}_1)w(\tilde{z}_2)}
  \right|^{\Delta-\ell}
  \left(
  \frac{\partial \tilde{z}_1^c}
  {\partial z_1^a}
  \frac{\partial \tilde{z}_2^d}
  {\partial z_2^b}
  \right)
  \bra{\bar{0}}
  \bar{\mathcal{O}}_c(\tilde{z}_1)
  \bar{\mathcal{O}}_d(\tilde{z}_2)
  \ket{\bar{0}}\ .
\end{equation}
This gives a simple non-trivial way of checking our calculations.

\subsection{Lorentzian cylinder}

Let us start by considering the conformal transformation relating Minkowski to the Lorentzian cylinder. Using a more rigorous approach, the mapping of the Tomita operator under this transformation was analyzed in Ref. \cite{Hislop:1981uh} for a massless scalar and more generally in Ref. \cite{Brunetti:1992zf} for an arbitrary CFT. 

As a first step, consider the special conformal transformation in (\ref{eq:3}) with the slight modification~$D^\mu=(R,R,\vec{0}\,)\rightarrow (0,R,\vec{0}\,)$. The right Rindler wedge $W$ in (\ref{eq:80}) is mapped to the causal domain of a ball of radius $R$ centered at the origin of the $x^\mu=(t,x,\vec{y}\,)$ coordinates \cite{Haag:1992hx}
$$\mathcal{DB}
  =\left\lbrace 
  x^\mu \in \mathbb{R}\times \mathbb{R}\times 
  \mathbb{R}^{d-2}:
  \quad
  R-\sqrt{x^2+|\vec{y}\,|^2}>|t|
  \right\rbrace\ .$$
The mapping of the ${\rm CRT}$ operator is characterized by the geometric reflection $\tilde{x}^\mu=x^\mu(-T,-X,\vec{Y})$, that from the change of coordinates in (\ref{eq:3}), can be easily found to be given by
\begin{equation}\label{eq:36}
\tilde{x}^\mu(x^\mu)=
  \frac{R^2}{(x\cdot x)}
  x_\rho \delta^{\rho \mu}\ ,
\end{equation}
where $(x\cdot x)=\eta_{\mu \nu}x^\mu x^\nu$ and $x_\mu=(-t,x,\vec{y})$. As first noted in Ref. \cite{Hislop:1981uh} this corresponds to the composition of an inversion $x^\mu \rightarrow R^2x^\mu/(x\cdot x)$ with a time reflection $t\rightarrow -t$, meaning that the ${\rm CRT}$ operator is mapped to
$$U({\rm CRT})U^\dagger=
  ({\rm CIT})\ ,$$
where ${\rm I}$ is the inversion operator. In appendix \ref{zapp:Mod_conj} we show that the discrete transformation ${\rm CIT}$ is part of the Euclidean conformal group in the same way as ${\rm CRT}$ belongs to the Euclidean Poincare group. The action of ${\rm CIT}$ on a primary operator of integer spin $\ell$ can be obtained from (\ref{eq:46}) using that\footnote{The conformal factor $w(x^\mu)$ obtained from applying the conformal transformation in (\ref{eq:3}) with ${D^\mu=(0,R,\vec{0})}$ is given by
$$w^2(x^\mu)=\left[
  \frac{4R^2}{-t^2+(R-x)^2+|\vec{y}|^2}\right]^2\ .$$}
\begin{equation}\label{eq:91}
\frac{w(x^\mu)}
  {w(\tilde{x}^\mu)}=
  \frac{R^2}{|(x\cdot x)|}\ ,
  \qquad \qquad
  \frac{\partial \tilde{x}^\mu }
  {\partial x^\nu}=
  \frac{R^2}{(x\cdot x)}
  \left[
  \eta_{\rho \nu}-\frac{2x_\rho x_\nu}{(x\cdot x)}
  \right]\delta^{\rho\mu}\ .
\end{equation}

\begin{figure}[t]
\begin{center}
\quad
\includegraphics[height=2.8 in]{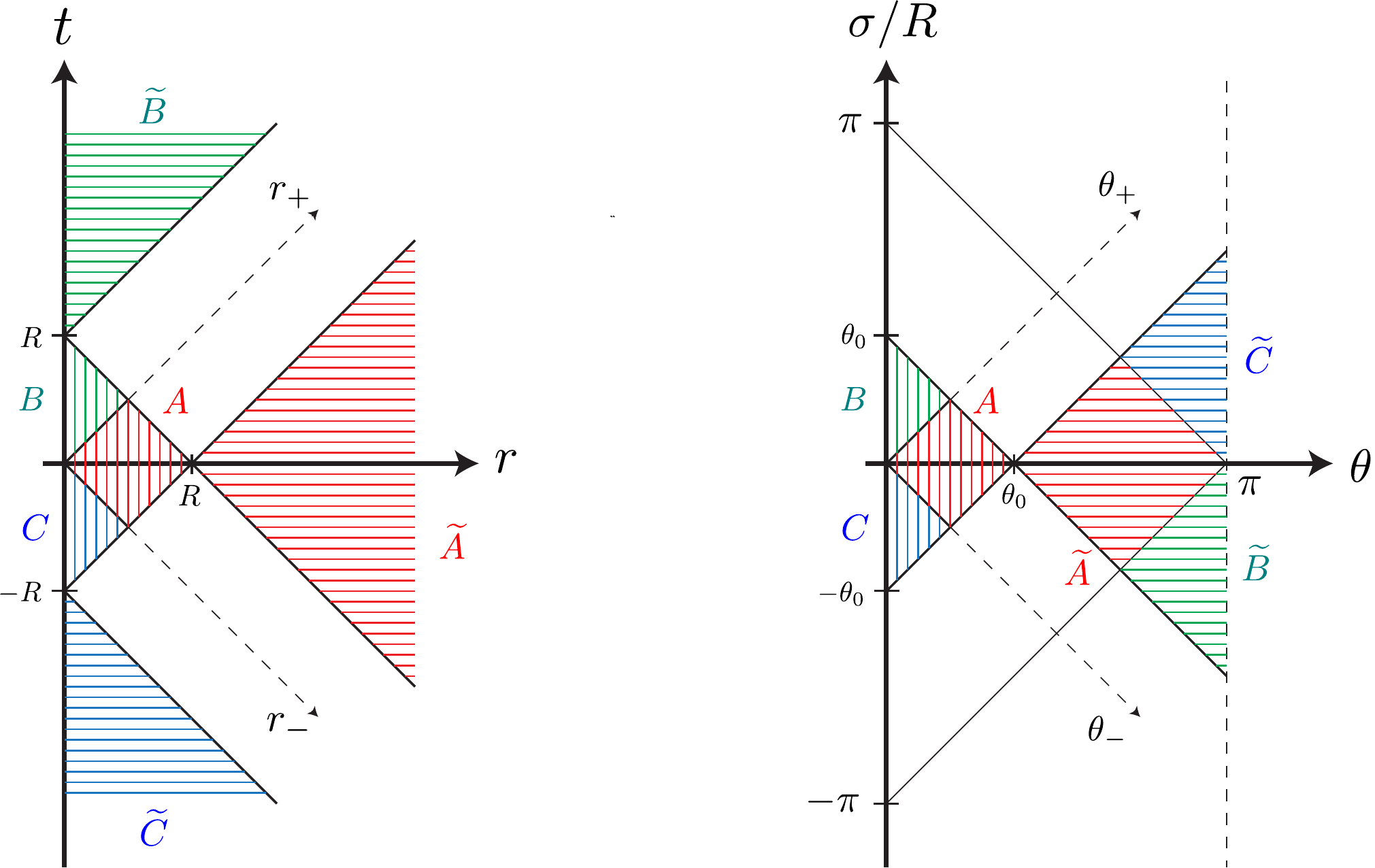}
\caption{On the left we have a diagram of the $(t,r)$ Minkowski plane with the causal domain of the ball $\mathcal{DB}=A\cup B \cup C$, where each region is marked with a different color and vertical lines. The reflected regions under the action of $\tilde{r}_\pm$ in (\ref{eq:32}) are marked with horizontal lines and the corresponding colors. On the right we have the mapping of the same regions but from the perspective of the Lorentzian cylinder with coordinates $(\sigma/R,\theta)$.}\label{fig:10}
\end{center}
\end{figure}

Let us analyze the geometric action of ${\rm CIT}$ in the causal domain of the ball, which is supposed to give the modular conjugation $\bar{J}$.  To do so it is convenient to write the reflection transformation in (\ref{eq:36}) in terms of the null radial coordinates $r_\pm=r\pm t$, which gives
\begin{equation}\label{eq:32}
\tilde{r}_\pm(r_\pm)=
  \frac{R^2}{r_\pm}\Theta(r_+r_-)-
  \frac{R^2}{r_\mp}\Theta(-r_+r_-)\ .
\end{equation}
Since this transformation is discontinuous and not well defined in the future and past null cone~${(x\cdot x)=r_+r_-=0}$, there are three regions in $\mathcal{DB}=A\cup B\cup C$ where $\tilde{r}_\pm$ acts in a distinct way (depending on the sign of $r_\pm$). In the left diagram of Fig. \ref{fig:10} we plot the three regions and their behavior under the ${\rm CIT}$ transformation in the $(t,r)$ plane.

The immediate observation is that $\widetilde{\mathcal{DB}}=\tilde{A}\cup \tilde{B}\cup\tilde{C}$ is a disconnected space-time region. This is problematic for the action of the modular conjugation since according to the Tomita-Takesaki theorem (\ref{eq:35}), $\bar{J}$ should map the algebra to its commutant. The regions $\tilde{B}$ and $\tilde{C}$ are causally connected to $\mathcal{DB}$, meaning that operators with support in $\mathcal{DB}$ and $\widetilde{\mathcal{DB}}$ do not commute with each other. Altogether this means that the mapping of the modular conjugation $J$ under this conformal transformation fails.

The origin of the problem is the same as the one discussed in Sec. \ref{sec:null_energy_bounds}: special conformal transformations are not well defined in Minkowski but on its conformal compactification, the Lorentzian cylinder. To obtain a well defined action for the modular conjugation $\bar{J}$, we must apply another mapping that takes the ${\rm CIT}$ operator to the cylinder. We can do this by using the conformal transformation in table \ref{table:2}, which we slightly modify by introducing the constant $\theta_0\in[0,\pi]$ according to
\begin{equation}\label{eq:112}
r_\pm(\theta_\pm)=
  R\frac{\tan(\theta_\pm/2)}
  {\tan(\theta_0/2)}
  \qquad \Longrightarrow \qquad
  w=\frac{\cot(\theta_0/2)}
  {2\cos(\theta_+/2)\cos(\theta_-/2)}\ ,
\end{equation}
where $w$ is the conformal factor and $\theta_\pm=\theta\pm \sigma/R$ are the null coordinates in the cylinder (\ref{eq:170}). The advantage of introducing $\theta_0$ is that $\theta_\pm=\theta_0$ corresponds to the boundary of $\mathcal{DB}$, so that the causal domain of the ball is mapped to the region in the cylinder
\begin{equation}\label{eq:109}
\bar{W}_{\theta_0}=\left\lbrace  
  (\sigma/R,\theta,\vec{v}\,)
  \in \mathbb{R}\times[0,\pi]
  \times \mathbb{R}^{d-2}:
  \quad
  |\theta_\pm|<\theta_0
  \right\rbrace\ .
\end{equation}
Although the space-time region is given by the causal domain of a cap of size $\theta_0$ around the North pole, the region in parameter space $(\sigma/R,\theta)$ is given by a wedge, see right diagram in Fig. \ref{fig:10}. We now need to obtain the mapping of $\bar{W}_{\theta_0}$ under the reflection transformation induced by ${\rm CIT}$ in (\ref{eq:36}).

One way of doing this is using the change of coordinates in (\ref{eq:10}), which take into account that a single Minkowski copy does not cover the entire cylinder. Although this is certainly possible, it is technically and conceptually more clear to take a different route based on the embedding formalism of the conformal group. In appendix \ref{zapp:Mod_conj} we use this to show that the geometric action of the modular conjugation $\bar{J}$ in the cylinder is given by the following relation
\begin{equation}\label{eq:110}
\tan(\tilde{\theta}_\pm/2)=
  \tan^2(\theta_0/2)\cot(\theta_\pm/2)\ .
\end{equation}
This transformation leaves the wedge $\theta_\pm=\theta_0$ fixed and if we apply it to $\bar{W}_{\theta_0}$ in (\ref{eq:109}) we find
\begin{equation}\label{eq:111}
\bar{W}_{\theta_0}'=
  \left\lbrace  
  (\sigma/R,\theta,\vec{v}\,)
  \in \mathbb{R}\times[0,\pi]
  \times \mathbb{R}^{d-2}:
  \quad
  |\theta_\pm|>\theta_0
  \right\rbrace\ .
\end{equation}
We plot the transformation $\bar{W}_{\theta_0}\rightarrow \bar{W}_{\theta_0}'$ in the right diagram of Fig. \ref{fig:10}. The reflection in the cylinder is exactly what we could have guessed: it reflects across a wedge in parameter space obtained by splitting the cylinder at $\theta=\theta_0$. From Fig. \ref{fig:10} we see that the issues that arise from the action of ${\rm CIT}$ in Minkowski are resolved from the perspective of the cylinder. The space-time regions~$\bar{W}_{\theta_0}=A\cup B\cup C$ and $\bar{W}_{\theta_0}'=\tilde{A}\cup \tilde{B}\cup \tilde{C}$ are the causal complements of each other, as required for the action of the modular conjugation $\bar{J}$ by the Tomita-Takesaki theory (\ref{eq:35}). 

\begin{figure}[t]
\begin{center}
\includegraphics[height=2.0 in]{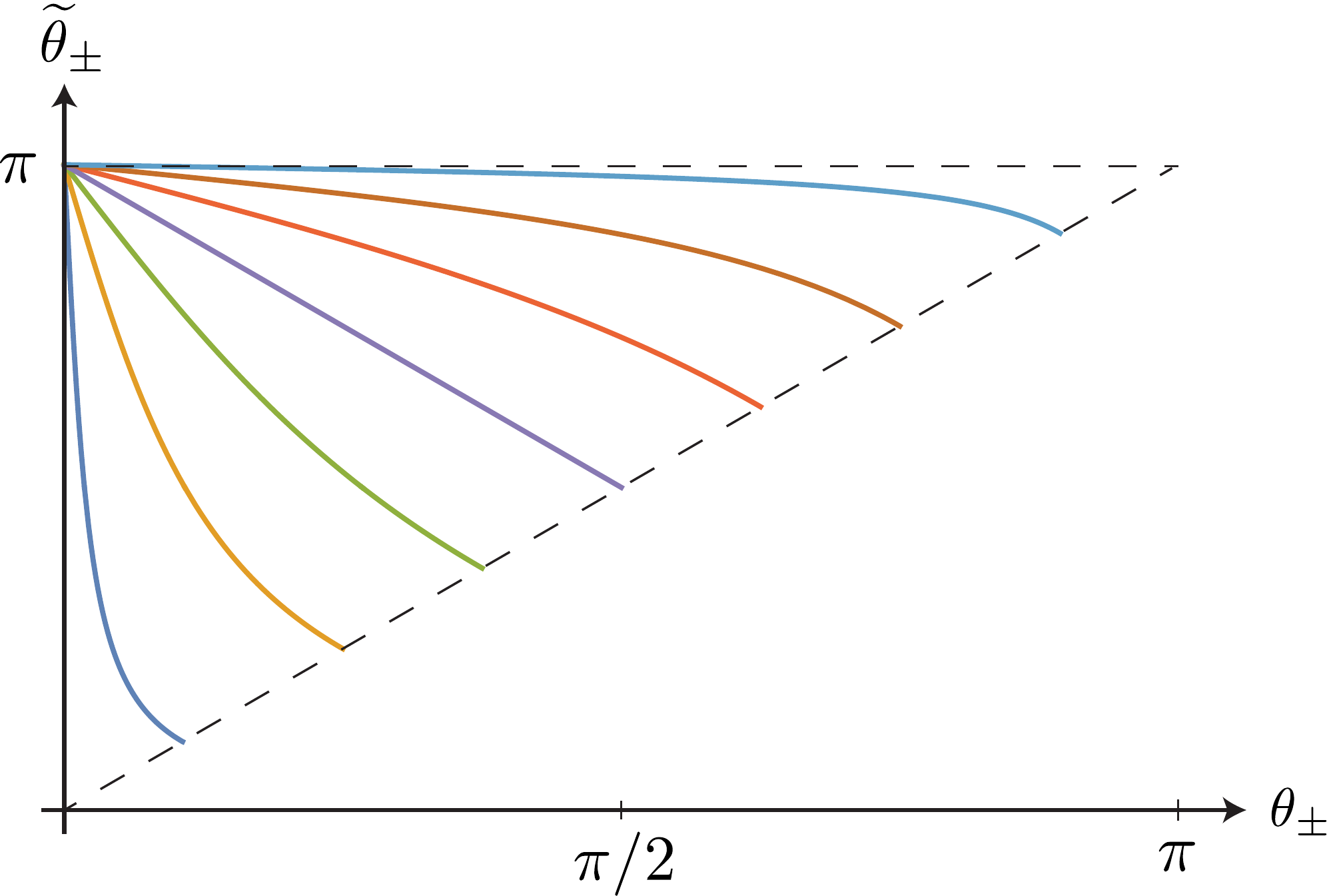}
\caption{Plot of the reflected coordinate $\tilde{\theta}_\pm$ as a function of $\theta_\pm$ for several values of $\theta_0$. Only for~$\theta_0=\pi/2$ (purple line in the center) the transformation is linear.}\label{fig:11}
\end{center}
\end{figure}

The transformation in (\ref{eq:110}) can only be explicitly solved when we split the cylinder in two wedges of equal size, \textit{i.e.} $\theta_0=\pi/2$
\begin{equation}\label{eq:113}
\tilde{\theta}_\pm(\theta_\pm)\big|_{\theta_0=\pi/2}=
  \pi-\theta_\pm\ .
\end{equation}
For $\theta_0\neq \pi/2$ the transformation is non-linear, as expected by the fact that it relates wedges of different sizes. We can still solve (\ref{eq:110}) numerically and plot it in Fig. \ref{fig:11}, where we explicitly see its non-linear behavior.

Now that we understand the mapping of the Tomita operator to the cylinder we can write the general inequality (\ref{eq:107}) and obtain wedge reflection positivity. To do so, let us first analyze the action of the modular conjugation $\bar{J}$ on primary operators, which can be obtained from the general relation~(\ref{eq:46}). The conformal factor appearing in this expression is the one relating the Minkowski coordinates $X^\mu$ to the cylinder, which is given by the product of (\ref{eq:91}) with (\ref{eq:112}), so that we find
\begin{equation}\label{eq:173}
\left|
  \frac{w(\theta_\pm)}
  {w(\tilde{\theta}_\pm)}
  \right|=
  \left|
  \frac{R^2}{r_+r_-}\times
  \frac{\cos(\tilde{\theta}_+/2)
  \cos(\tilde{\theta}_-/2)}
  {\cos(\theta_+/2)\cos(\theta_-/2)}
  \right|=
  f(\theta_+)f(\theta_-)\ ,
\end{equation}
where in the second equality we have used (\ref{eq:112}) and (\ref{eq:110}) and defined $f(\theta_\pm)$ as
\begin{equation}\label{eq:174}
f(\theta_\pm)=
  \frac{\tan(\theta_0/2)}
  {\cos(\theta_\pm/2)
  \sqrt{\tan^2(\theta_\pm/2)+\tan^4(\theta_0/2)}}
  \ge 0\ .
\end{equation}
This is non-negative since $\theta_0\in[0,\pi]$ and $\theta_\pm\in[-\pi,\pi]$ for $\bar{W}$. When the wedges are of equal size~$\theta_0=\pi/2$, this function equals to one. The Jacobian matrix associated to the reflection transformation~(\ref{eq:110}) can be written in terms of the space-time coordinates $v^\mu=(\theta_+,\theta_-,\vec{v}\,)$ using that the only non-trivial components are given by
\begin{equation}\label{eq:204}
\frac{\partial \tilde{\theta}_\pm}
  {\partial \theta_\pm}=-f(\theta_\pm)^2\ .
\end{equation}
Using all this in (\ref{eq:46}) we can explicitly write the action of the modular conjugation on a primary field of integer spin $\ell$. Moreover, the general positivity relation (\ref{eq:107}) becomes
\begin{equation}\label{eq:211}
(-1)^P\bra{\bar{0}}
  \bar{\mathcal{O}}^\dagger_{\mu_1\dots \mu_\ell}(\tilde{v}^\mu)
  \bar{\mathcal{O}}_{\nu_1\dots \nu_\ell}
  (v^\mu)
  \ket{\bar{0}}>0\ ,
  \qquad v^\mu \in \bar{W}_{\theta_0}\ ,
\end{equation}
where $P$ is the sum of $\theta_+$ indices plus $\theta_-$ indices. This proves the wedge reflection positivity of correlators in the Lorentzian cylinder. It is somewhat more interesting that Rindler positivity given that for $\theta_0\neq \pi/2$ the reflection transformation $\tilde{\theta}_\pm(\theta_\pm)$ is non-linear.

As a simple check of our calculations we can verify the validity of the identity (\ref{eq:108}) implied by~$\bar{J}\ket{\bar{0}}=\ket{\bar{0}}$. Using that the two point function of scalar primary operators of scaling dimension $\Delta$ in the cylinder is given by\footnote{We have chosen the coordinate system so that the position of the two points in the unit sphere $S^{d-2}$ is the same, \textit{i.e.} $\vec{v}_1=\vec{v}_2$. Moreover, this is the correlator for space-like separated points since for the time-like case we have an additional phase $e^{\pm i \Delta}$ depending on the ordering.}
$$\bra{\bar{0}}
  \bar{\mathcal{O}}(v_1^\mu)
  \bar{\mathcal{O}}(v_2^\mu)
  \ket{\bar{0}}=
  \left|
  4R^2
  \sin(\Delta \theta_+/2)
  \sin(\Delta \theta_-/2)
  \right|^{-\Delta},
  $$
it is straightforward to check that (\ref{eq:108}) holds for arbitrary values of $\theta_0\in[0,\pi]$.

\subsection{De Sitter}

The generalization to a CFT in de Sitter space-time is straightforward, since the conformal mapping is just given by the Weyl rescaling in (\ref{eq:154}). Since we keep the same space-time coordinates, the geometric action of the modular conjugation is still given by (\ref{eq:110}). However, the value of $\theta_0$ is restricted to $\theta_0=\pi/2$, since for other values one of the wedges in the dS diagram of Fig. \ref{fig:11} necessarily lies outside of de Sitter. The modular conjugation in dS is then characterized by $\tilde{\theta}_\pm(\theta_\pm)=\pi-\theta_\pm$, which corresponds to a reflection between the left and right de Sitter static patches.

Using that $f(\theta_\pm)\big|_{\theta_0=\pi/2}=1$ and the expressions (\ref{eq:173}) and (\ref{eq:204}) we can explicitly write the action of the modular conjugation $\bar{J}$ on any bosonic primary operator from (\ref{eq:46}).\footnote{Notice that the conformal factor $w^2(\sigma)=\cos^2(\sigma/R)$ satisfies $w^2(\sigma)/w^2(\tilde{\sigma})=1$.} Moreover, the wedge reflection positivity in de Sitter (\ref{eq:107}) is given by
\begin{equation}\label{eq:205}
(-1)^P\bra{\bar{0}}
  \bar{\mathcal{O}}^\dagger_{\mu_1\dots \mu_\ell}(\tilde{v}^\mu)
  \bar{\mathcal{O}}_{\nu_1\dots \nu_\ell}
  (v^\mu)
  \ket{\bar{0}}>0\ ,
  \qquad v^\mu \in \bar{W}_{\theta_0=\pi/2}\ ,
\end{equation}
where $P$ is the sum of $\theta_+$ indices plus $\theta_-$ indices.

\section{Discussion and future directions}
\label{sec:Discussion}

In this work we derived the ANEC for general CFTs in (A)dS and a similar novel bound for the Lorentzian cylinder. By thoroughly studing the connection of these conditions with the previous derivations of the Minkowski ANEC in Refs. \cite{Faulkner:2016mzt,Hartman:2016lgu,Kelly:2014mra} we have obtained other useful technical results. This includes null deformed modular Hamiltonians and their associated entanglement entropies in Sec. \ref{sec:null_energy_bounds}, as well as an extension of Rindler positivity to curved backgrounds in Sec. \ref{sec:ref_positivity}. Let us comment on some future research directions that would be interesting to pursue. 

\paragraph{ANEC in (A)dS beyond conformal theories:} Since our derivation of these conditions relies heavily on conformal symmetry, a natural question is whether they can be extended to general quantum field theories. For de Sitter, following Refs. \cite{Faulkner:2016mzt,Hartman:2016lgu} would require to show that the full modular Hamiltonian (\ref{eq:164}) or the wedge reflection positivity (\ref{eq:205}) are still true beyond CFTs. Since our methods used to derive both of these results rely on conformal symmetry, one would require more powerful tools to do so. For the AdS case, we have seen that both aproaches used in Refs. \cite{Faulkner:2016mzt,Hartman:2016lgu} fail even for CFTs, which suggests that a general proof of the ANEC in AdS calls for a completely new procedure.

\paragraph{ANEC in AdS from holography:} In App. \ref{zapp:ANEC_holography} we have shown how the ANEC in de Sitter can be derived for holographic CFTs described by Einstein gravity. The bound for the CFT in the cylinder~(\ref{eq:195}) has also been recently obtained through this method in Ref. \cite{Iizuka:2019ezn} for $d=3,4,5$. This suggests it might be possible to extend the holographic proof to the AdS case, although we have seen some examples where the generalization of certain results to AdS is delicate and does not work.

\paragraph{Vacuum susbstracted ANEC in the cylinder:} We have shown that a CFT in the Lorentzin cylinder satisfies the novel bound given in (\ref{eq:195}). Although we have stressed that this condition is not equivalent to the ANEC, it is still possible that the vacuum substracted ANEC is a true statement of QFTs defined in the cylinder. For the particular case of a free scalar in $\mathbb{R}\times S^{1}$ this was explicitly shown in Ref. \cite{Ford:1994bj}. In future work it would be interesting to explore other methods that could allow to extend this to more general setups.\footnote{For a field theory on a generic space-time Ref. \cite{Graham:2007va} proposed that the ANEC (without any vacuum energy subtraction) must hold along achronal null geodesics, \textit{i.e.} curves which do not contain points connected by a time-like path. Several references in the literature find evidence supporting this proposal \cite{Wall:2009wi,Kontou:2012ve,Kontou:2015yha}, while other claim to obtain counter examples \cite{Urban:2009yt,Ishibashi:2019nby}.}

\paragraph{Constraint on higher spin operators:} In the causality proof of the Minkowski ANEC in Ref.~\cite{Hartman:2016lgu} the following positivity constraint for higher spin null integrated operators was derived
\begin{equation}\label{eq:178}
\mathcal{E}^{(\ell)}(\vec{x}_\perp)\equiv
  \int_{-\infty}^{+\infty}d\lambda\,
  T^{(\ell)}_{\lambda\dots \lambda}(\lambda,\vec{x}_\perp)
  \ge 0\ ,
\end{equation}
where $T^{(\ell)}_{\mu_1\dots \mu_\ell}$ is the lowest dimension operator of even spin $\ell\ge 2$ and $(\lambda,\vec{x}_\perp)$ are the coordinates in the null plane (\ref{eq:50}). Applying the conformal transformations in Sec. \ref{sec:ANEC_mapping} we can obtain the analogous constraint in (A)dS. Since $T^{(\ell)}_{\mu_1\dots \mu_\ell}$ is a primary operator it transforms in similar way to $T_{\lambda \lambda}$ in (\ref{eq:5})
$$U
  T^{(\ell)}_{\lambda\dots \lambda}U^\dagger=
  |w_{\rm (A)dS}(\vec{x}_\perp)|^{\ell-\Delta}
  \bar{T}^{(\ell)}_{\lambda\dots \lambda}\ .$$
Integrating over $\lambda\in \mathbb{R}$, the left hand-side becomes (\ref{eq:178}) and we get
$$U
  \mathcal{E}^{(\ell)}(\vec{x}_\perp)U^\dagger=
  |w_{\rm (A)dS}(\vec{x}_\perp)|^{\ell-\Delta}
  \int_{-\infty}^{+\infty}d\lambda\,
  \bar{T}^{(\ell)}_{\lambda\dots \lambda}
  (\lambda,\vec{x}_\perp)\ ,$$
where we have used that the conformal factors $w_{\rm (A)dS}(\vec{x}_\perp)$ in (\ref{eq:197}) are independent of~$\lambda$. Since~$\lambda$ is an affine parameter in (A)dS, the higher spin Minkowski ANEC (\ref{eq:178}) implies the analogous constraint for (A)dS. The geodesics are given in (\ref{eq:188}) where in the AdS case~$\vec{x}_\perp$ is constrained to~${|\vec{x}_\perp|<R}$, so that the curves lie in the space-time. A completely analogous calculation using (\ref{eq:191}) and (\ref{eq:190}) also generalizes the bound obtained for the Lorentzian cylinder
$$U
  \mathcal{E}^{(\ell)}(\vec{x}_\perp)U^\dagger
  \propto 
  \int_{-\pi/2}^{\pi/2}
  d\bar{\lambda}
  \left(\cos(\bar{\lambda})\right)^{\Delta+\ell-2}\,
  \bar{T}^{(\ell)}_{\bar{\lambda}\dots \bar{\lambda}}
  (\bar{\lambda},\vec{x}_\perp)\ge 0\ ,$$
where the proportionality constant is positive for $\ell$ even. For the cylinder and de Sitter it should be possible to derive these higher spin constraints using the wedge reflection positivity proved in Sec.~\ref{sec:ref_positivity}. Moreover, it would be interesting to analyze the generalization of these conditions to continuous spin, as obtained for Minkowski in Ref. \cite{Kravchuk:2018htv}.

\paragraph{Witt algebra in de Sitter:} In Ref. \cite{Casini:2017roe} it was shown that it is possible to define some null integrated operators in the Minkowski null plane which satisfy the Witt algebra. More precisely, the operators\footnote{See section 4.6 of Ref. \cite{Kologlu:2019bco} for a discussion regarding some aspects of the definition of these operators.}
$$L^{(n)}(\vec{x}_\perp)=
  i\int_{-\infty}^{+\infty}d\lambda\,
  \lambda^{n+1}\,
  T_{\lambda \lambda}(\lambda,\vec{x}_\perp)\ ,
$$
where shown to satisfy the following algebra
\begin{equation}\label{eq:179}
\left[
  L^{(n)}(\vec{x}_\perp),
  L^{(m)}(\vec{y}_\perp)
  \right]=(n-m)
  \delta(\vec{x}_\perp-\vec{y}_\perp)
  L^{(n+m)}(\vec{x}_\perp)\ .
\end{equation}
We can apply the conformal transformation of Sec. \ref{sec:null_energy_bounds} from Minkowski to dS, so that using (\ref{eq:5}) the operators $L^{(n)}(\vec{x}_\perp)$ transform as
$$UL^{(n)}(\vec{x}_\perp)U^\dagger=
  \frac{R^{d-2}}{p(\vec{x}_\perp)^{d-2}}
  i\int_{-\infty}^{+\infty}d\lambda\,
  \lambda^{n+1}\bar{T}_{\lambda \lambda}(\lambda,\vec{x}_\perp)=
  \frac{R^{d-2}}{p(\vec{x}_\perp)^{d-2}}
  L_{\rm dS}^{(n)}(\vec{x}_\perp)\ ,$$
where $p(\vec{x}_\perp)=(|\vec{x}_\perp|+4R^2)/4R$ and we have defined $L_{\rm dS}^{(n)}(\vec{x}_\perp)$ in terms of $\lambda$, which is affine in de Sitter. Using this in (\ref{eq:179}), the operators $L_{\rm dS}^{(n)}(\vec{x}_\perp)$ satisfy the following algebra
$$
  \left[
  L_{\rm dS}^{(n)}(\vec{x}_\perp),
  L_{\rm dS}^{(m)}(\vec{y}_\perp)
  \right]=
  (n-m)
  \left[
  \frac{p(\vec{x}_\perp)^{d-2}}
  {R^{d-2}}\delta(\vec{x}_\perp-\vec{y}_\perp)
  \right]
  L_{\rm dS}^{(n+m)}(\vec{x}_\perp)\ .$$
The term between square brackets in the right hand side is nothing more than the Dirac delta associated to the induced metric in the null surface in de Sitter, see table \ref{table:1}. Hence, the operators~$L_{\rm dS}^{(n)}(\vec{x}_\perp)$ in this surface also satisfy the Witt algebra. It would be interesting to further explore this in the context of the calculations in Refs. \cite{Cordova:2018ygx,Huang:2019fog}.

\paragraph{Entanglement entropy beyond holography:} In appendix \ref{zapp:entanglement} we computed the entanglement entropy associated to the null deformed regions in the Lorentzian cylinder and (A)dS using AdS/CFT. Although these results are valid to all orders in the boundary CFT, it would be instructive to recover the same expressions directly in field theory. One way of doing so is by applying a similar approach as the one used in Ref. \cite{Casini:2017roe} to compute the entanglement entropy associated to the null plane and cone in Minkowski.

\paragraph{Other conformally related space-times:} In this work we have focused on the conformal transformations relating Minkowski, the Lorentzian cylinder and (A)dS. However, Ref. \cite{Candelas:1978gf} describes some additional space-times that are connected through conformal mappings which might be interesting to further explore. For instance, for a CFT in $\mathbb{R}\times \mathbb{H}^{d-1}$, with $\mathbb{H}$ a hyperbolic plane, one could use similar methods to compute both the modular Hamiltonian and associated entanglement entropy of null deformed regions.

\paragraph{Negative energy in large $d$ limit:} The energy condition obtained for the CFT in the Lorentzian cylinder (\ref{eq:195}) has a very interesting behavior in the large space-time dimension limit, where it gives a local constraint on the null projection of the stress tensor (\ref{eq:169}). This suggest that the study of negative energy in this regime might give some interesting insights. To our knowledge, the large $d$ limit of negative energy in QFT has not been systematically investigated in the literature. Since we have not been able to completely determine the limit in (\ref{eq:195}) this is an interesting result that deserves further study. 

\paragraph{Wedge reflection positivity and entropy inequalities:} In Sec. \ref{sec:ref_positivity} we derived the Rindler positivity for CFTs in the Lorentzian cylinder and de Sitter, the case of the cylinder being particularly interesting since the transformation is non-linear. Following a similar approach as in Refs. \cite{Casini:2010nn,Blanco:2019gmt} it would be interesting to explore the consequences of these properties regarding entanglement entropy inequalities.

\subsection{Comment on the QNEC}

The quantum null energy condition (QNEC) is a local constraint on the null projection of the stress tensor that has recently attracted much interest \cite{Bousso:2015mna}. For a general QFT in Minkowski the QNEC has been proven in Ref. \cite{Balakrishnan:2017bjg} and more interestingly in Ref. \cite{Ceyhan:2018zfg}, where it was shown to follow from the Minkowski ANEC. The results of this paper raise the question of whether there is a similar connection to be made between the conditions in (A)dS.

To do so let us first review the statement of the QNEC in Minkowski from the perspective of relative entropy. Consider the relative entropy between the vacuum $\sigma=\ket{0}\bra{0}$ and an arbitrary state~$\rho$ reduced to null deformations of the Rindler region. Using that the modular Hamiltonian is given by (\ref{eq:66}), the relative entropy (\ref{eq:180}) can be written as
\begin{equation}\label{eq:206}
S(\rho||\sigma)=
  2\pi 
  \int_{\mathbb{R}^{d-2}} d\vec{x}_\perp
  \int_{A(\vec{x}_\perp)}^{+\infty}d\lambda\,
  \left(
  \lambda-A(\vec{x}_\perp)
  \right)
  \big[
  \langle T_{\lambda \lambda} \rangle_\rho
  -
  \langle T_{\lambda \lambda} \rangle_{\ket{0_M}}
  \big]
  -
  \big[
  S(\rho)
  -S(\ket{0_M})
  \big]\ ,
\end{equation}
where $S(\rho)$ and $S(\ket{0_M})$ are the entanglement entropy of each state reduced to the null deformed region. Now let us consider a one parameter family of deformations labeled by $\kappa$ and given by~$A(\vec{x}_\perp;\kappa)=A(\vec{x}_\perp)+\kappa \dot{A}(\vec{x}_\perp)$ with $\dot{A}(\vec{x}_\perp)\ge 0$. The QNEC in Minkowski can be formulated as the statement that the second derivative of the relative entropy with respect to $\kappa$ is positive~$\partial^2_{\kappa}S(\rho||\sigma)\ge 0$. 

The derivative of (\ref{eq:206}) can be further simplified using that $\langle T_{\lambda \lambda} \rangle_{\ket{0_M}}$ vanishes since Minkowski is a maximally symmetric space-time (see discussion around (\ref{eq:181})). Furthermore, some symmetry considerations regarding Minkowski and the null plane given in Ref. \cite{Casini:2018kzx} show that the vacuum entanglement entropy $S(\ket{0_M})$ is independent of $A(\vec{x}_\perp)$. Altogether, the QNEC in Minkowski is given by
\begin{equation}\label{eq:182}
\frac{d^2}{d\kappa^2}
S(\rho||\sigma)\ge 0
\qquad \Longleftrightarrow \qquad
2\pi
\int_{\mathbb{R}^{d-2}} d\vec{x}_\perp
  \dot{A}(\vec{x}_\perp)^2
  \langle T_{\lambda \lambda} \rangle_\rho
  \ge 
  \frac{d^2}{d\kappa^2}
  S(\rho)\ .
\end{equation}
This was proven for general QFTs in Refs. \cite{Balakrishnan:2017bjg,Ceyhan:2018zfg}. The local version of the bound is obtained by taking $\dot{A}(\vec{x}_\perp)^2=\delta(\vec{x}_\perp-\vec{x}_\perp^{\,0})$.

Let us now discuss the case of de Sitter. The first thing we might try is to directly map the inequality on the right of (\ref{eq:182}) by applying the conformal transformation from Minkowski to dS discussed in Sec. \ref{sec:null_energy_bounds}. Using the transformation property of the stress tensor $T_{\lambda \lambda}$ in (\ref{eq:5}) and the conformal factor (\ref{eq:197}) we can map the left-hand side of the inequality and find
\begin{equation}\label{eq:208}
2\pi R^{d-2}
\int_{S^{d-2}}
  d\Omega(\vec{x}_\perp)
  \dot{A}(\vec{x}_\perp)^2
  \langle \bar{T}_{\lambda \lambda} \rangle_{\bar{\rho}}
  \ge 
  \frac{d^2}{d\kappa^2}
  S(\rho)\ ,
\end{equation}
where $d\Omega(\vec{x}_\perp)=d\vec{x}_\perp/p(\vec{x}_\perp)^{d-2}$ and $\bar{\rho}=U\rho \,U^\dagger$. The mapping of the right-hand side is more complicated since it involves the entanglement entropy. Although the entanglement entropy in quantum mechanics is invariant under a unitary transformation, this is not true in QFTs given that the entropy requires a cut-off $\epsilon$ which transforms in a non-trivial way. To our knowledge there is no standard general prescription for the transformation of the entanglement entropy.

For the particular case of holographic theories dual to Einstein gravity, Ref. \cite{Koeller:2015qmn} obtained some interesting results by using some earlier observations from Ref. \cite{Graham:1999pm}. Applying this to the mapping of Minkowski to de Sitter in Sec. \ref{sec:null_energy_bounds}, their results suggest that the transformation of the right-hand side of (\ref{eq:208}) is given by
\begin{equation}\label{eq:207}
\frac{d^2}{d\kappa^2}
  S(\rho)
  \qquad \longrightarrow \qquad
  \frac{d^2}{d\kappa^2}
  \big[
  S(\bar{\rho})-S(\ket{0_{\rm dS}})
  \big]\ ,
\end{equation}
where $S(\bar{\rho})$ is the entropy of the the mapped state $\bar{\rho}$ in the null deformed region of dS. 

A first argument supporting (\ref{eq:207}) is that it implies the saturation of the QNEC in de Sitter when evaluated in the vacuum $\ket{0_{\rm dS}}$, which we expect to be true given that it is in Minkowski. If we did not have the vacuum substraction in (\ref{eq:207}) the QNEC would not saturate given that the vacuum entanglement entropy of de Sitter (\ref{eq:171}) has a non-trivial dependence on the entangling surface.

Another argument in favor of (\ref{eq:207}) comes from relative entropy. Using the modular Hamiltonian in dS (\ref{eq:24}), we can explicitly write the the relative entropy between the states $\bar{\rho}$ and $\bar{\sigma}=\ket{0_{\rm dS}}\bra{0_{\rm dS}}$ and take its second derivative with respect to $\kappa$, so that we find
$$\frac{d^2}{d\kappa^2} S(\bar{\rho}||\bar{\sigma})\ge  0
  \qquad \Longleftrightarrow \qquad
  2\pi R^{d-2}
  \int_{S^{d-2}}d\Omega(\vec{x}_\perp)
  \dot{A}(\vec{x}_\perp)^2
  \langle
  \bar{T}_{\lambda \lambda}\rangle_{\bar{\rho}}
  \ge 
  \frac{d^2}{d\kappa^2}
  \big[
  S(\bar{\rho})-S(\ket{0_{\rm dS}})
  \big]\ .$$
To obtain this, we have written the modular Hamiltonian (\ref{eq:24}) in terms of the affine parameter~$\lambda$ using $\lambda(\eta)=p(\vec{x}_\perp)(2\eta-1)$ from table \ref{table:1}. The negativity of the second derivative of the relative entropy in dS implies precisely the same transformation property of the entropy given in (\ref{eq:207}). 

For the other space-times and surfaces studied in this paper, the treatment becomes more obscure. For AdS we have the issue that the mapping of the whole null plane does not fit inside the space-time, so that the conformal transformation of (\ref{eq:182}) becomes even more ambiguous. Moreover, the QNEC is obtained from the quantum focusing conjecture \cite{Bousso:2015mna} applied to a point $p$ and a hypersurface orthogonal surface that is locally stationary through $p$. A straightforward computation of the expansion of the null congruence of each surface considered in Sec. \ref{sec:null_energy_bounds}, show that this is only true for the case of de Sitter. This is also evident when computing the relative entropy from the modular Hamiltonians in Sec. \ref{sec:null_energy_bounds}. Since the operators in AdS (\ref{eq:18}) and the Lorentzian cylinder (\ref{eq:75}) have a much more complicated structure, their second derivative with respect to $\kappa$ is not as simple as in (\ref{eq:182}).

\bigskip
\bigskip
\leftline{\bf Acknowledgements}
\smallskip
\noindent I thank Clifford V. Johnson for comments on the manuscript and the organizers of TASI 2019 where I learned many of the tools used in this paper. This work is partially supported by DOE grant DE-SC0011687. 

\appendix
\addtocontents{toc}{\protect\setcounter{tocdepth}{1}}

\section{ANEC in de Sitter from holography}
\label{zapp:ANEC_holography}

In this section we give a proof of the ANEC for a holographic conformal field theory in de Sitter, dual to Einstein gravity. We follow the approach of Ref. \cite{Kelly:2014mra}, where the Minkowski ANEC was derived under the assumption that the gravity dual has good causal properties. More precisely, the assumption is that for two  boundary points connected by a boundary null geodesic, there is no causal curve (\textit{i.e.} time-like or null) through the bulk which travels faster than the boundary null geodesic.

\subsection{General features of bulk AdS with de Sitter boundary}

Let us start by discussing some general notions regarding AdS/CFT and asymptotically AdS$_{d+1}$ space-time. An asymptotically AdS space-time can be written in Fefferman-Graham coordinates as
\begin{equation}\label{eq:138}
ds^2=(L/z)^2\left[
  dz^2+g_{\mu \nu}(z,v)dv^\mu dv^\nu
  \right]\ ,
\end{equation}
where the AdS radius is $L$, the boundary is at $z=0$ and $z>0$ corresponds to the bulk interior. The~$d$-dimensional metric $g_{\mu \nu}(z,v)$ admits an expansion in powers of $z$ given by \cite{deHaro:2000vlm}
$$g_{\mu \nu}(z,v)=
  g_{\mu \nu}^{(0)}(v)
  +z^2g_{\mu \nu}^{(2)}(v)+
  \dots+z^d
  \ln(z^2/L^2)
  h_{\mu \nu}(v)
  +
  z^dg_{\mu \nu}^{(d)}(v)
  +o(z^{d})\ ,$$
where $h_{\mu\nu}$ is non-zero only for even $d$ and $o(z^d)$ means terms that vanish strictly faster than $z^d$. The first term in this expansion $g_{\mu \nu}^{(0)}$ gives the space-time in which the boundary CFT is defined. Since in this case we are interested in a de Sitter background, we have from (\ref{eq:154})
\begin{equation}\label{eq:146}
g_{\mu \nu}^{(0)}dv^\mu dv^\nu=
  \frac{R^2}
  {\sin^2\left[
  (\theta_+-\theta_-)/2\right]}
  \left[
  d\theta_+d\theta_-
  +\sin^2\left(\theta\right)
  d\Omega^2(\vec{v}\,)
  \right]
  \ ,
\end{equation}
where $v^\mu=(\theta_+,\theta_-,\vec{v}\,)$ with the null coordinates $\theta_\pm=\theta\pm \sigma/R$. We have written dS with the conformal factor $\sin^2(\sigma/R)$ so that $\sigma/R\in[-\pi,0]$.

The higher order terms $h_{\mu \nu}$ and $g_{\mu \nu}^{(n)}(v)$ with $n<d$ can be obtained by perturbately solving Einstein's equations. They are all written in terms of geometric quantities built from the boundary metric $g_{\mu \nu}^{({\rm dS})}$ \cite{deHaro:2000vlm}, \textit{i.e.} they are a complicated functions of the Riemann, Ricci and curvature tensor of~$g_{\mu \nu}^{({\rm dS})}$ and their covariant derivatives. For instance, the first order term is given by
\begin{equation}\label{eq:150}
g_{\mu \nu}^{(2)}=
  \frac{1}{d-2}\left[
  \frac{\mathcal{R}}{2(d-1)}g_{\mu \nu}^{({\rm dS})}-
  \mathcal{R}_{\mu \nu}
  \right]\ ,
\end{equation}
where the Ricci tensor and scalar on the right-hand side are computed from the metric $g_{\mu \nu}^{({\rm dS})}$. Given that in this particular case we are considering a de Sitter boundary, we can use the fact that it is maximally symmetric, so that the Riemann tensor is completely fixed by the metric 
$$\mathcal{R}_{\mu \nu\rho \sigma}=
  \frac{1}{R^2}\left(
  g_{\mu \rho}g_{\nu \sigma}-
  g_{\mu \sigma}g_{\nu \rho}
  \right)
  \ .$$
From this we see that (\ref{eq:150}) is proportional to the boundary metric ${g_{\mu \nu}^{(2)}=-g_{\mu \nu}^{({\rm dS})}/(2R^2)}$.\footnote{When the bulk is pure AdS the metric is Fefferman-Graham metric is given by (\ref{eq:176}) to all orders and we can explicitly check the proportionality factor $-1/(2R^2)$.} The powerful observation is that this is true for all the higher order terms $h_{\mu \nu}$ and $g^{(n)}_{\mu \nu}$ with $n<d$.\footnote{Since the Riemann is proportional to the metric, the terms in $g_{\mu \nu}^{(n)}$ involving covariant derivatives vanish.} Although the actual proportionality constants $m_n$ cannot be computed for arbitrary $d$, it will be enough to use that they are proportional
$$h_{\mu \nu}=m_dg_{\mu \nu}^{({\rm dS})}\ , \qquad \qquad
  g_{\mu \nu}^{(n)}=m_ng_{\mu \nu}^{({\rm dS})}\ ,
  \qquad n<d
  \ .$$
Using this, we can write any asymptotically AdS metric with a de Sitter boundary as
\begin{equation}\label{eq:144}
ds^2=(L/z)^2\left[
  dz^2+
  \left(m(z)
  g_{\mu \nu}^{({\rm dS})}(v)
  +z^dg_{\mu \nu}^{(d)}(v)
  \right)
  dv^\mu dv^\nu+
  o(z^{d})
  \right]\ ,
\end{equation}
where the function $m(z)$ satisfies $m(z=0)=1$ and is determined from the coefficients $m_n$ 
$$m(z)=1+m_2z^2+\dots+m_dz^d\ln(z^2/L^2)\ .$$ 
This expansion to order $o(z^{d})$ will be enough for our purposes. 

The higher order terms are determined by the particular state in the boundary CFT. The first undetermined contribution $g_{\mu\nu}^{(d)}$ is related to the expectation value of the stress tensor of the dual CFT according to the standard AdS/CFT dictionary 
\begin{equation}\label{eq:143}
\langle T_{\mu \nu} \rangle=
  \frac{dL^{d-1}}{16\pi G}
  g^{(d)}_{\mu \nu}(v)+X_{\mu \nu}[g^{(n<d)}]\ ,
\end{equation}
where $X_{\mu \nu}$ gives the anomalous term of the stress tensor in the CFT and $G$ is Newton's constant. Although in a general setup $X_{\mu \nu}$ is a functional of $g_{\mu \nu}^{(n)}$ with $n<d$, we can use the same observation as before to conclude that the anomalous terms is also proportional to the boundary metric~${X_{\mu \nu}=x_dg_{\mu \nu}^{({\rm dS})}}$. If we project the stress tensor along the null direction $\theta_-$, the anomalous terms drops out and we find
\begin{equation}\label{eq:148}
  \langle T_{--} \rangle=
  \frac{dL^{d-1}}{16\pi G}
  g_{--}^{(d)}(v)\ ,
  \qquad \qquad
  g_{--}^{(d)}=
  \frac{dv^\mu }{d\theta_-}
  \frac{dv^\nu }{d\theta_-}
  g_{\mu \nu}^{(d)}\ .
\end{equation}

\subsection{Curve ansatz and no bulk shortcut}

\begin{figure}[t]
\begin{center}
\includegraphics[height=2.4 in]{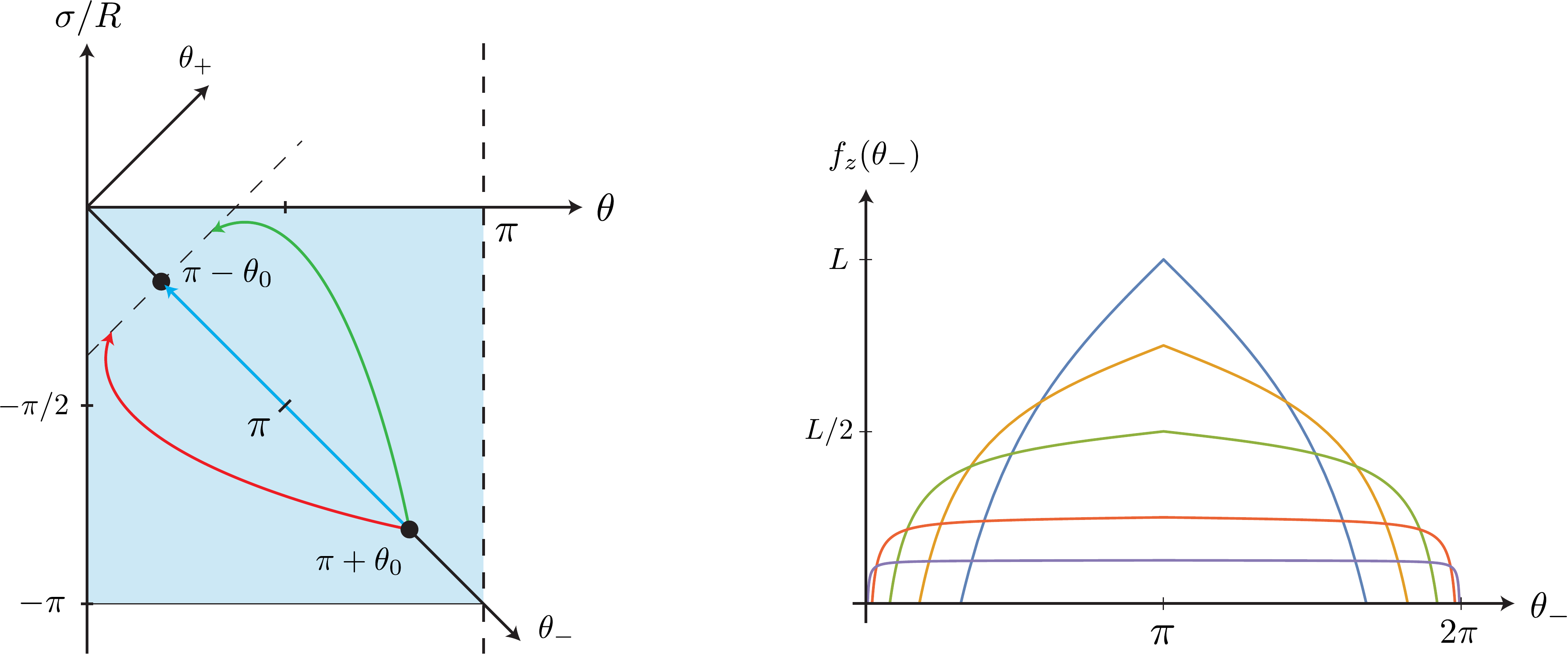}
\caption{On the left we have a diagram of the setup in the $(\sigma/R,\theta)$ plane, with the shaded blue region corresponding to de Sitter. The boundary curve is shown in blue, while in green and red we show to different bulk curves going out of the page which satisfy the boundary conditions (\ref{eq:139}). The no bulk shortcut property implies that if these bulk curves are causal, the red trajectory is forbidden. On the right we plot our ansatz for the function $f_z(\theta_-)$ which shows how the bulk curve goes into the bulk for different values of $\epsilon L$. As $\epsilon\rightarrow 0$ the depth of the curve decreases and since $\theta_0=\pi-\epsilon^{d-1}$, its range in the boundary goes to $\theta_-\in[0,2\pi]$.}
\label{fig:13}
\end{center}
\end{figure}

Let us now describe the setup that will allow us to obtain the ANEC. Consider a null geodesic in the boundary moving along the $\theta_-$ direction
\begin{equation}\label{eq:136}
v^\mu(\theta_-)=
  (0,\theta_-,\vec{v}\,)\ ,
  \qquad
  \theta_-\in[\pi-\theta_0,\pi +\theta_0]\ ,
\end{equation}
where $\vec{v}$ is fixed and the null tangent vector is given by $(0,1,\vec{0}\,)$. The parameter $\theta_0\in[0,\pi]$ determines the initial and final points of the geodesic. For $\theta_0=\pi$ the geodesic is complete, going from the South pole of de Sitter at past infinity to the North at future infinity, while for $\theta_0=0$ it is a single point. In the left diagram of Fig. \ref{fig:13} we sketch this curve in blue in the $(\sigma/R,\theta)$ plane. Although $\theta_-$ is not an affine parameter in dS, it is convenient to describe the geodesic in this way. 

We now wish to construct a bulk curve which starts at the same point as (\ref{eq:136}) at the boundary, goes into the bulk and ends in some other point at the boundary (not necessarily the same one as~(\ref{eq:136})). Consider the curve given by
$$x^A(\theta_-)=
  (z,v^\mu)=
  (
  f_z(\theta_-),f_+(\theta_-),
  \theta_-,\vec{v}\,
  )\ ,
  \qquad 
  \theta_-\in[\pi-\theta_0,\pi +\theta_0]\ ,$$
which has a tangent vector equal to
\begin{equation}\label{eq:151}
\frac{dx^A}{d\theta_-}=
  (f'_{z}(\theta_-),f'_+(\theta_-),1,\vec{0}\,)=
  (f'_{z}(\theta_-),k^\mu(\theta_-))\ .
\end{equation}
The functions $f_z(\theta_-)$ and $f_+(\theta_-)$ must satisfy the following boundary conditions
\begin{equation}\label{eq:139}
f_z(\pi\pm \theta_0)=0 \ ,
  \qquad \qquad
  f_+(\pi+\theta_0)=0\ ,
\end{equation}
which ensures that the bulk curve behaves in the way we just described. A sketch of two bulk curves in red and green are shown in the left diagram of Fig. \ref{fig:13}.

The final position of this curve in the boundary is determined from  $\theta_+^{\rm final}=f_+(\pi-\theta_0)$. The no bulk shortcut property is the statement that there is no bulk causal curve $x^A(\theta_-)$ whose end point at the boundary is at the past of the end point of (\ref{eq:136}). More concretely, it implies the following
\begin{equation}\label{eq:140}
{\rm If\,\,\,}
  g_{AB}\,
  \frac{dx^A}
  {d\theta_-}
  \frac{dx^B}
  {d\theta_-}\le 0
  \qquad \Longrightarrow \qquad
  \theta_+^{\rm final}=
  f_+(\pi-\theta_0)\ge 0\ ,
\end{equation}
where $g_{AB}$ is the full bulk metric given in (\ref{eq:144}). This forbids a causal curve as the red one shown in the left diagram of Fig. \ref{fig:13}. Violation of the no bulk shortcut property would result in causality and locality problems of the boundary theory (see Refs. \cite{Kelly:2014mra,Iizuka:2019ezn,Gao:2000ga,Witten:2019qhl} for related discussions).

The strategy is to construct a particular causal bulk curve given in (\ref{eq:151}), such that the no bulk shortcut property (\ref{eq:140}) gives the ANEC for the boundary theory. From the expansion of the bulk metric in (\ref{eq:144}), the curve is causal as long as it satisfies the following constraint
\begin{equation}\label{eq:145}
(f'_z(\theta_-))^2+
  \left[m(f_z(\theta_-))
  g_{\mu \nu}^{({\rm dS})}(\theta_-)
  +f_z(\theta_-)^dg_{\mu \nu}^{(d)}(\theta_-)
  \right]
  k^\mu(\theta_-)k^\nu(\theta_-)+
  o(z^{d})\le 0
  \  ,
\end{equation}
where $k^\mu(\theta_-)$ is given in (\ref{eq:151}). We will consider a particular bulk curve whose maximum depth in the bulk is given by $\epsilon L$ with $\epsilon$ a dimensionless quantity, and expand to leading order in $\epsilon\ll 1$. This curve must satisfy the boundary conditons in (\ref{eq:139}) to every order in $\epsilon$ as well as the causality constraint~(\ref{eq:145}) to leading order. Our ansatz is inspired by the calculations in Refs. \cite{Kelly:2014mra,Iizuka:2019ezn}.

For the function giving the $z$ coordinates $f_z(\theta_-)$ we choose
$$f_z(\theta_-)=
  \epsilon L\left(
  \frac{\tan(\theta_0/2)-|\cot(\theta_-/2)|}
  {\tan(\theta_0/2)}
  \right)\ .$$
We plot this in the right diagram of Fig. \ref{fig:13} for several values of $\epsilon$. The function is positive in the range of $\theta_-\in (\pi-\theta_0,\pi+\theta_0)$ and vanishes at the end points, so that it satisfies the boundary conditions~(\ref{eq:139}). The maximum depth of the curve into the bulk is given by $\epsilon L$. To obtain the ANEC we relate the parameter $\theta_0$ to $\epsilon$ according to
\begin{equation}\label{eq:141}
\theta_0=\pi-\epsilon^{d-1}\ .
\end{equation}
The limit of $\epsilon\ll 1$ then corresponds to a bulk curve near the boundary which covers a complete null geodesic in de Sitter (\ref{eq:136}), see Fig \ref{fig:13}. If we expand for small $\epsilon$ we find
$$f_z(\theta_-)=
  \epsilon-
  \frac{|\cot(\theta_-/2)|}{2}\epsilon^d+
  \mathcal{O}(\epsilon^{2d})\ .
  $$
From this we see that the function $f_z(\theta_-)$ is of order $\epsilon$ while its derivative goes like $\epsilon^d$. This is one of the crucial properties of the ansatz, since it ensures that the first positive term in the causality constraint (\ref{eq:145}) is subleading in the $\epsilon$ expansion.

For the remaining function $f_+(\theta_-)$ we consider the following 
$$f_+(\theta_-)=
  \frac{\epsilon^d}{R^2}
  \int_{\theta_-}^{\pi+\theta_0}
  d\theta'_-
  \sin^2(\theta_-'/2)
  g_{--}^{(d)}(0,\theta_-',\vec{v}\,)+
  Q(\theta_-)
  \left(
  \frac{\tan(\theta_0/2)+\cot(\theta_-/2)}
  {\tan(\theta_0/2)}
  \right)\epsilon^{d+\delta}
  \ ,$$
where $Q(\theta_-)$ is any regular function and $\delta$ a small positive number. This function satisfies the boundary condition in (\ref{eq:139}) since it vanishes at the initial point $\theta_-=\pi+\theta_0$. From this, we can write the tangent vector for the boundary components $k^\mu(\theta_-)$ in (\ref{eq:151}) in an expansion in $\epsilon$ as
\begin{equation}\label{eq:147}
k^\mu(\theta_-)=
  \frac{dv^\mu }{d\theta_-}+\left[
  -
  \frac{\sin^2(\theta_-/2)}{R^2}  g_{--}^{(d)}(0,\theta_-,\vec{v}\,)
  \epsilon^d+
  Q'(\theta_-)\epsilon^{d+\delta}\right]
  (1,1,\vec{0}\,)
  +
  \mathcal{O}(\epsilon^{d+1})\ .
\end{equation}

Now that we have a bulk curve which satisfies the boundary conditions in (\ref{eq:139}), we check that it is causal to leading order in $\epsilon$, \textit{i.e.} that it satisfies (\ref{eq:145}). Expanding this constraint we find
$$
  \frac{R^2 m(f_z(\theta_-))}
  {\sin^2\left[
  (f_+(\theta_-)-\theta_-)/2
  \right]}
  \left[
  -\frac{\sin^2(\theta_-/2)}{R^2}
  g_{--}^{(d)}(0,\theta_-,\vec{v}\,)
  \epsilon^d+
  Q'(\theta_-)\epsilon^{d+\delta}\right]
  +\epsilon^d g_{--}^{(d)}(0,\theta_-,\vec{v}\,)
  +
  o(\epsilon^{d})\le 0
  \  ,
$$
where we have used that the de Sitter metric is given by (\ref{eq:146}) so that the contraction of the term~$du^\mu/d\theta_-$ vanishes. Using that the function $m(f_z(\theta_-))=1+\mathcal{O}(\epsilon^2)$ and expanding the sine in the denominator we find
$$
  \left[
  -g_{--}^{(d)}(0,\theta_-,\vec{v}\,)
  \epsilon^d+
  \frac{R^2Q'(\theta_-)}
  {\sin^2\left(\theta_-/2\right)}
  \epsilon^{d+\delta}\right]
  +\epsilon^d g_{--}^{(d)}(0,\theta_-,\vec{v}\,)
  +
  o(\epsilon^{d})\le 0
  \  .
$$
The leading order in $\epsilon^d$ involving the metric $g_{--}^{(d)}$ cancels and the first non-vanishing contribution is given by $\epsilon^{d+\delta}$. Recall that $o(\epsilon^d)$ means terms that vanish strictly faster than $\epsilon^d$. This means that for any fixed bulk space-time (corresponding to a state in the boundary CFT) we can fix $\delta>0$ to be small enough so that it is the leading contribution in $\epsilon$ when compared to the unknown terms $o(\epsilon^d)$. In this way, the causality constraint reduces to the following condition on the function $Q(\theta_-)$
\begin{equation}\label{eq:149}
Q'(\theta_-)\le 0\ .
\end{equation}
By fixing this function such that it satisfies this property we are guaranteed to have a causal curve. 

Now that we have constructed the bulk causal curve, we can investigate the consequences of imposing the no bulk shortcut property in (\ref{eq:140}). Writing this explicitly we find
$$\theta_+^{\rm final}=
  \frac{\epsilon^d}{R^2}
  \int_{\pi-\theta_0}^{\pi+\theta_0}
  d\theta'_-
  \sin^2(\theta_-'/2)
  g_{--}^{(d)}(0,\theta_-',\vec{v}\,)+
  2Q(\pi-\theta_0)\epsilon^{d+\delta}
  \ge 0\ .
  $$
Since the bulk curve is causal only in the limit of $\epsilon\ll 1$ we must expand in $\epsilon$. Doing so, and using that the boundary stress tensor is related to $g_{--}^{(d)}$ according to (\ref{eq:148}) we find
$$
  \int_{0}^{2\pi}
  d\theta'_-
  \sin^2(\theta_-'/2)
  \langle T_{--}(0,\theta_-',\vec{v}\,) 
  \rangle
  \ge -
  \left(
  \frac{2R^2dL^{d-1}}{16\pi G}
  \right)
  \lim_{\epsilon\rightarrow 0}\left[
  Q(\epsilon^{d-1})\epsilon^{\delta}+
  \dots
  \right]\ .
  $$
There are three possibilities for the value of the function $Q(\theta_-)$ as $\theta_-\rightarrow 0$. 
The least interesting case is when it diverges to $+\infty$ faster that $\epsilon^\delta$ goes to zero so that the bound becomes trivial. On the contrary, if it diverges to $-\infty$ then the causality condition $Q'(\theta_-)\le 0$ in (\ref{eq:149}) is not verified and the curve is not causal. The most interesting case is when $Q(\theta_-)$ goes to a constant value, so that the right hand side vanishes and we obtain a non-trivial condition given by
$$
  \int_{0}^{2\pi}
  d\theta'_-
  \sin^2(\theta_-'/2)
  \langle T_{--}(0,\theta_-',\vec{v}\,) 
  \rangle
  \ge 0\ .
  $$
This is actually the ANEC in de Sitter, as can be seen by remembering that the parameter $\theta_-$ is not affine. If we change the integration variable to an affine parameter $\eta(\theta_-)=\cot(\theta_-/2)$ (see table \ref{table:1} noting that $\beta=\theta_-/2$), we obtain 
$$\mathcal{E}_{\rm dS}(\vec{v}\,)=
  \int_{-\infty}^{+\infty}
  d\eta\,
  T_{\eta \eta}(\eta,\vec{v}\,)\ge 0
  \ .$$

\section{Entanglement entropy of null deformed regions}
\label{zapp:entanglement}

In this appendix we compute the entanglement entropy associated to the modular Hamiltonians obtained in Sec. \ref{sec:null_energy_bounds}. The case of the null plane and null cone in Minkowski space-time have already been considered in Ref. \cite{Casini:2018kzx}. Using a similar approach we obtain explicit expressions for the entanglement entropy of null deformed regions associated to the Lorentzian cylinder and (A)dS.

\subsection{Review: Minkowski null cone}

Let us start by considering the entanglement entropy of the vacuum associated to an arbitrary surface in the null cone of Minkowski, given in (\ref{eq:120}) and following the holographic calculation in Ref. \cite{Casini:2018kzx}. Since the global state in the CFT is the Minkowski vacuum, we must consider pure AdS in Poincare coordinates
\begin{equation}\label{eq:132}
ds^2=
  \left(L/z\right)^2
  \left(dz^2
  -dt^2+dr^2+r^2d\Omega^2(\vec{v}\,)
  \right)\ ,
\end{equation}
where $L$ is the AdS radius and $d\Omega^2(\vec{v})$ is the metric on a unit sphere $S^{d-2}$ parametrized by stereographic coordinates $\vec{v}\in \mathbb{R}^{d-2}$ (\ref{eq:152}). 

According to the HRRT prescription \cite{Ryu:2006bv,Hubeny:2007xt}, the entanglement entropy  is obtained from the area of the extremal bulk surface that intersects with the boundary $z= 0$ on the entangling surface of~(\ref{eq:120}). Since the surface lies on the null cone, it is convenient to define null coordinates in the bulk $\hat{r}_\pm=\hat{r}\pm t$ obtained from
\begin{equation}\label{eq:39}
z=\hat{r}\sin(\psi)\ ,
  \qquad \qquad
  r=\hat{r}\cos(\psi)\ ,
\end{equation}
where $\hat{r}\ge 0$ and $\psi \in[0,\pi/2]$. The AdS metric in these coordinates becomes
\begin{equation}\label{eq:2}
ds^2=
  \left(
  \frac{L}{\hat{r}\sin(\psi)}
  \right)^2
  \left(
  -dt^2+d\hat{r}^2+\hat{r}^2
  \left(d\psi^2+
  \cos^2(\psi)d\Omega^2(\vec{v}\,)\right)
  \right)\ ,
\end{equation}
so that the $d$-dimensional Minkowski boundary is located at $\psi\rightarrow 0$. Since in this limit $\hat{r}\rightarrow r$, the coordinates $\hat{r}_\pm$ become the null coordinates in Minkowski $r_\pm=r\pm t$. Moreover, given that the entangling surface in the CFT (\ref{eq:120}) is located at $(r_+,r_-)=(0,2\bar{A}(\vec{v}\,))$, the boundary condition for the bulk extremal surface can be easily written as 
\begin{equation}\label{eq:125}
\lim_{\psi\rightarrow 0}
  \left(\hat{r}_+,\hat{r}_-,\vec{v}\,\right)=
  \left(0,2\bar{A}(\vec{v}\,),\vec{v}\,\right)\ ,
\end{equation}
where $\bar{A}(\vec{v}\,)>0$.

To obtain the entanglement entropy we must find the extremal codimension two surface subject to this constraint. This was computed exactly in Refs. \cite{Neuenfeld:2018dim,Casini:2018kzx} where it was shown to satisfy $\hat{r}_+=0$ not only at the boundary but at every point in the bulk. This means that the area of the extremal surface is obtained from the induced metric (\ref{eq:2}) at $\hat{r}_+=0$ 
\begin{equation}\label{eq:128}
\left.ds^2\right|_{\hat{r}_+=0}=
  L^2\left(
  \frac{d\psi^2+
  \cos^2(\psi)d\Omega^2(\vec{v}\,)}
  {\sin^2(\psi)}
  \right)\ .
\end{equation}
Using that the entropy is related to the area according to $S=2\pi {\rm Area}/\ell_p^{d-1}$, the entanglement entropy is given by
\begin{equation}\label{eq:121}
S=
\frac{4\pi a_d^*}{{\rm Vol}(S^{d-1})}
\int_{S^{d-2}} d\Omega^2(\vec{v}\,)
  \int_{0}^{\pi/2}
  d\psi
  \frac{\cos^{d-2}(\psi)}{\sin^{d-1}(\psi)}\ ,
\end{equation}
where we have conveniently defined the factor $a^*_d$ in Einstein gravity according to
$$2a_d^*={\rm Vol}(S^{d-1})(L/\ell_p)^{d-1}
  \qquad {\rm with} \qquad
  {\rm Vol}(S^{d-1})=2\pi^{d/2}/\Gamma(d/2)\ .$$
For the boundary CFT this factor is mapped to (\ref{eq:155}).

Note that the integral in (\ref{eq:121}) seems to be insensitive to the details of the extremal surface (given by $\hat{r}_-(\psi,\vec{v}\,)$) and therefore independent of the entangling surface $\bar{A}(\vec{v})$. However, this is not true since the extremal surface plays a role in regulating the integral in (\ref{eq:121}), which diverges in the limit of $\psi\rightarrow 0$.

To reproduce a divergent field theory quantity through a bulk computation we must be careful about the choice of the cut-off, since different regularizations yield distinct results. For instance, if we regulate (\ref{eq:121}) with $\psi_{\rm min}=\epsilon$ we incorrectly conclude that the entanglement entropy in the null cone is independent of the entangling surface. A field theory computation shows that this is incorrect~\cite{Casini:2018kzx}. The appropriate cut-off is dictated by holographic renormalization \cite{Skenderis:2002wp}, in which we must first write the bulk metric in Fefferman-Graham coordinates
\begin{equation}\label{eq:133}
ds^2=(L/z)^2\left[
  dz^2+g_{\mu \nu}(z,x^\mu)dx^\mu dx^\nu
  \right]\ ,
\end{equation}
where the boundary is described by $x^\mu$ and located at $z\rightarrow 0$. The metric $g_{\mu \nu}(z,x^\mu)$ admits an expansion in $z$ given by $g_{\mu \nu}(z,x^\mu)=g_{\mu \nu}^{(0)}(x^\mu)+z^2g_{\mu \nu}^{(2)}(x^\mu)+\dots$, where $g_{\mu\nu}^{(0)}(x^\mu)$ corresponds to the space-time in which the boundary CFT is defined. The appropriate cut-off $\epsilon$ is obtained from the $z$ coordinate according to $z_{\rm min}=\epsilon$.

In this case, the AdS metric as written in (\ref{eq:132}) is already in Fefferman-Graham coordiantes. We can relate $\psi$ to $z$ using (\ref{eq:39}), so that the cut-off $\epsilon$ is given by
\begin{equation}\label{eq:123}
z_{\rm min}=\epsilon=
  \sin(\psi)
  \hat{r}_-(\psi,\vec{v}\,)/2\ ,
\end{equation}
where we have evaluated at the extremal surface $\hat{r}_+=0$ and $\hat{r}_-(\psi,\vec{v}\,)$. To compute the entanglement entropy in terms of $\epsilon$, we must invert this relation to get an expansion for $\psi(\epsilon,\vec{v}\,)$ and solve the integral in (\ref{eq:121}). The function $\hat{r}_-(\psi,\vec{v}\,)$ is determined from the details of the extremal surface and has some expansion near the boundary as $\psi\rightarrow 0$
\begin{equation}\label{eq:126}
\hat{r}_-(\psi,\vec{v}\,)=
  2\bar{A}(\vec{v})+b_1(\vec{v}\,)\psi+
  b_2(\vec{v}\,)\psi^2+\dots\ ,
\end{equation}
where the first order term is fixed by the boundary condition (\ref{eq:125}) and the coefficients $b_i(\vec{v}\,)$ determine the higher order contributions. They can be obtained from the exact expressions of the extremal surface given in Refs. \cite{Neuenfeld:2018dim,Casini:2018kzx}. Using this in (\ref{eq:123}) we can invert the relation and find the expansion for $\psi(\epsilon,\vec{v}\,)$
\begin{equation}\label{eq:124}
\psi(\epsilon,\vec{v}\,)=
  \frac{\epsilon}{\bar{A}(\vec{v}\,)}
  -\frac{b_1(\vec{v}\,)}{2\bar{A}(\vec{v}\,)^3}
  \epsilon^2+\dots \ .
\end{equation}
With this expression we regulate the integral (\ref{eq:121}) and obtain the entanglement entropy. 

As usual, the entanglement entropy is dominated by a divergent area term and subleading contributions. We only compute the universal terms, \textit{i.e.} contributions that are independent of the regularization procedure. For even $d$ this is given by a logarithmic term, while for odd $d$ it is a constant term. Using (\ref{eq:124}) in (\ref{eq:121}) we find the following expansion as derived in Ref. \cite{Casini:2018kzx}
\begin{equation}\label{eq:43}
S=\frac{\mu_{d-2}}{\epsilon^{d-2}}+
  \dots+
  a_d^*\times
\begin{cases}
\displaystyle
  \frac{4(-1)^{\frac{d-2}{2}}}
  {{\rm Vol}(S^{d-2})}
  \int_{S^{d-2}}
  d\Omega(\vec{v}\,)
  \ln\left(
  2\bar{A}(\vec{v}\,)/\epsilon
  \right)\ ,
  \quad d{\rm \,\, even}
  \vspace{6pt}\\
  \qquad \qquad \qquad
  \displaystyle
  2\pi(-1)^{\frac{d-1}{2}}
  \qquad \qquad \quad \,\ ,
  \quad \, d{\rm \,\, odd}\ ,
\end{cases}
\end{equation}
where $\mu_i$ are non-universal coefficients. If we take $\bar{A}(\vec{v})=R$ we recover the well known result for the entanglement of a ball \cite{Casini:2011kv}. For odd $d$, the universal term is independent of the entangling surface~$\bar{A}(\vec{v}\,)$. In Ref. \cite{Casini:2018kzx} it was argued that this feature is not modified by quantum and higher curvature corrections in the bulk, meaning that the entanglement entropy in (\ref{eq:43}) is valid to all orders in the dual field theory.

Notice that the higher order terms in the expansion of the null surface (\ref{eq:126}) play no role in determining the universal term of the entanglement entropy. This means that the only non-trivial information we used to obtain (\ref{eq:43}) is that the whole surface satisfies $\hat{r}_+=0$. This will simplify the calculation for the curved backgrounds we consider in the following.

\subsection{Lorentzian cylinder}

We can apply a similar procedure to obtain the entanglement entropy associated to the null surface~(\ref{eq:73}) in the Lorentzian cylinder. Since the state is still given by the CFT vacuum, the bulk geometry is also pure AdS. However, we must consider a different set of coordinates which give a different conformal frame at the boundary. To do so, we define the following coordinates
\begin{equation}\label{eq:4}
\hat{r}_\pm=\hat{r}\pm t=R
  \tan(\hat{\theta}_\pm/2)
  \ ,
  \qquad \qquad
  \hat{\theta}_\pm=\hat{\theta}\pm \sigma/R\ ,
\end{equation}
so that the AdS metric (\ref{eq:2}) becomes
\begin{equation}\label{eq:127}
ds^2=
  \frac{L^2}{
  \sin^2(\psi)
  \sin^2(\hat{\theta})}
  \left[
  -(d\sigma/R)^2+d\hat{\theta}^2
  +\sin^2(\hat{\theta})
  \left(
  d\psi^2
  +\cos^2(\psi)d\Omega^2(\vec{v}\,)
  \right)
  \right]\ .
\end{equation}
As we take the boundary limit $\psi\rightarrow 0$ we recover the metric $\mathbb{R}\times S^{d-1}$, where the bulk coordinates~$\hat{\theta}_\pm$ become the null coordinates in the boundary ${\hat{\theta}_\pm\rightarrow \theta_\pm}$.

To find the entanglement entropy we look for the extremal surface with boundary conditions fixed by the entangling surface in (\ref{eq:73}), so that we have
\begin{equation}\label{eq:129}
\lim_{\psi\rightarrow 0}
  (\hat{\theta}_+,\hat{\theta}_-,\vec{v}\,)=
  (0,2\bar{A}(\vec{v}\,),\vec{v}\,)\ ,
\end{equation}
where $\bar{A}(\vec{v}\,)\in (0,\pi)$.\footnote{The function $\bar{A}(\vec{v})$ is not the same as the one for the null cone. They are related through the coordinate change~(\ref{eq:4}).}

Instead of computing the extremal surface from scratch we use the results obtained for the Minkowski null cone. The extremal surface of the Minkowski null cone is mapped under the change of coordinates (\ref{eq:4}) so that the condition $\hat{r}_+=0$ translates into $\hat{\theta}_+=0$.  The induced metric in~(\ref{eq:127}) under the constraint $\hat{\theta}_+=0$ is the same as in (\ref{eq:128}), meaning that the entanglement entropy is again determined by the integral in (\ref{eq:121}). The difference comes from the regularization procedure. To find the appropriate cut-off $z_{\rm min}=\epsilon$ we write the space-time metric (\ref{eq:127}) in Fefferman-Graham coordinates (\ref{eq:133}). The appropriate change of coordinates is given by
$$
  \cot(\psi)=
  \left(\frac{4R^2-z^2}
  {4Rz}\right)\sin(\theta)
  \ ,
  \qquad \qquad
  \cos(\hat{\theta})=
  \left(
  \frac{4R^2-z^2}{4R^2+z^2}
  \right)
  \cos(\theta)
  \ ,$$
where $z\in [0,2R]$ and $\theta\in[0,\pi]$. Inverting these relations
\begin{equation}\label{eq:12}
\frac{z}{2R}=\frac{1-
  \sqrt{1-\sin^2(\psi)\sin^2(\hat{\theta})}}
  {\sin(\psi)\sin(\hat{\theta})}
  \ ,
  \qquad \qquad
  \tan(\theta)=\cos(\psi)\tan(\hat{\theta})\ ,
\end{equation}
and applying to the AdS metric in (\ref{eq:127}) we find
$$ds^2=(L/z)^2\left[
  dz^2-\left(\frac{4R^2+z^2}{4R^2}\right)^2d\sigma^2+
  \left(\frac{4R^2-z^2}{4R^2}\right)^2
  R^2\left(
  d\theta^2+\sin^2(\theta)d\Omega^2(\vec{v}\,)
  \right)
  \right]\ ,$$
that is precisely in the Fefferman-Graham form with the cylinder metric at the boundary.

Setting $z_{\rm min}=\epsilon$ in (\ref{eq:12}) we can find the relation between $\psi$ and the cut-off by evaluating the right-hand side on the extremal surface $\hat{\theta}=\hat{\theta}_-(\psi,\vec{v}\,)/2$. This has a near boundary expansion given by
$$\hat{\theta}_-(\psi,\vec{v}\,)=
  2\bar{A}(\vec{v}\,)+b_1(\vec{v}\,)\psi+\dots\ ,$$
where the first order term is fixed by the boundary condition (\ref{eq:129}). Using this in (\ref{eq:12}) with~$z_{\rm min}=\epsilon$ and inverting we find
\begin{equation}\label{eq:131}
\psi(\epsilon,\vec{v}\,)=
  \frac{(\epsilon/R)}{\sin(\bar{A}(\vec{v}\,))}-
  \frac{b_1(\vec{v}\,)}
  {\sin^2(\bar{A}(\vec{v}\,))
  \tan(\bar{A}(\vec{v}\,))}(\epsilon/R)^2
  +\dots\ .
\end{equation}
From this we can regulate and solve the integral in (\ref{eq:121}) to obtain the universal terms of the entanglement entropy. Comparing with the expansion of $\psi(\epsilon,\vec{v}\,)$ that we got for the case of the null cone~(\ref{eq:124}) we immediately see that the entropy only gets a universal contribution from the linear term in (\ref{eq:131}). For odd $d$ it is also given by (\ref{eq:43}) while for even $d$ we obtain 
$$
S=\frac{\mu_{d-2}}{\epsilon^{d-2}}+
  \dots+
  (-1)^{\frac{d-2}{2}}
  \frac{4a_d^*}
  {{\rm Vol}(S^{d-2})}
  \int_{S^{d-2}}
  d\Omega(\vec{v}\,)
  \ln\left[
  \frac{2R}{\epsilon}
  \sin(\bar{A}(\vec{v}\,))
  \right]\ .
$$
Same as with the Minkowski null cone, we expect this result to be valid to all orders in the dual CFT. For the particular case in which $\bar{A}(\vec{v})=\theta_0$, we recover the result for a cap region of angular size $\theta_0$~\cite{Casini:2011kv}.

\subsection{De Sitter}

A similar story holds for the entanglement in de Sitter associated to the null surface in (\ref{eq:15}). Since the coordinates in the boundary are also given by $(\sigma/R,\theta,\vec{v}\,)$ we can still work with pure AdS as written in (\ref{eq:127}). To get de Sitter at the boundary we simply have to take the limit $\psi\rightarrow 0$ with the additional factor of $\sin^2(\sigma/R)$ in the conformal factor. The boundary condition of the extremal surface is obtained from (\ref{eq:15}) using this $\eta(\beta)=\cot(\beta)$ so that we find
\begin{equation}\label{eq:135}
\lim_{\psi\rightarrow 0}
  (
  \hat{\theta}_+,\hat{\theta}_-,\vec{v}\,
  )=
  \left(0,
  2\,{\rm arcot}
  \left( \bar{A}(\vec{v}\,) \right),
  \vec{v}\,\right)\ ,
\end{equation}
where in principle $\bar{A}(\vec{v})\in \mathbb{R}$.\footnote{The inverse of the cotangent function is defined so that its image is in the range $[0,\pi]$.} 

The entanglement entropy is still given by the integral in (\ref{eq:121}). To obtain the relation between the cut-off $\epsilon$ and $\psi$ we must write the metric (\ref{eq:127}) in Fefferman-Graham coordinates with $g_{\mu \nu}^{(0)}$ given by the de Sitter metric. Since the relation between the coordinates is fairly complicated it is convenient to break it up in two steps.\footnote{The easiest way to obtain these coordinate transformations is to use the embedding description of AdS and analyze the relation between the different parametrizations.} First let us consider the coordinates $\varrho\ge 0$ and $\theta\in[0,\pi]$ defined according to
\begin{equation}\label{eq:159}
\cot(\psi)=
  \frac{\varrho}{L}\sin(\theta)\ ,
  \qquad \qquad
  \cos(\hat{\theta})=
  \frac{\varrho}
  {\sqrt{\varrho^2+L^2}}\cos(\theta)\ ,
\end{equation}
which has an inverse given by
$$\varrho=L
  \frac{\sqrt{1-\sin^2(\psi)\sin^2(\hat{\theta})}}
  {\sin(\psi)\sin(\hat{\theta})}\ ,
  \qquad \qquad
  \tan(\theta)=\cos(\psi)\tan(\hat{\theta})\ .$$
The metric (\ref{eq:127}) takes the standard form of global AdS
\begin{equation}\label{eq:156}
ds^2=-
  \left(
  \frac{\varrho^2+L^2}{R^2}
  \right)d\sigma^2+
  \left(
  \frac{L^2}{\varrho^2+L^2}
  \right)d\varrho^2+
  \varrho^2\left(
  d\theta^2+\sin^2(\theta)d\Omega^2(\vec{v}\,)
  \right)\ .
\end{equation}
From this we can define the new coordinates $z\in [0,2R]$ and $\hat{\sigma}/R\in[-\pi,0]$ according to  
$$\varrho=
  -
  \left(\frac{4R^2-z^2}{4Rz}\right)
  \frac{L}{\sin(\hat{\sigma}/R)}\ ,
  \qquad \qquad
  \tan(\sigma/R)=-
  \left(\frac{4R^2-z^2}{4R^2+z^2}\right)
  \cot(\hat{\sigma}/R)
  \ .$$
Since $\hat{\sigma}$ has a finite domain we are only taking a section of the range of $\sigma$, given by $|\sigma|\le \pi R/2$. The inverse can be computed and written as
$$\frac{z}{2R}=
  \sqrt{
  (\varrho/L)^2+1}
  \cos(\sigma/R)-
  \sqrt{(\varrho/L)^2\cos^2(\sigma/R)
  -\sin^2(\sigma/R)}\ ,$$
$$
  \cos(\hat{\sigma}/R)=
  \sqrt{\frac{\varrho^2+L^2}{\varrho^2}}
  \sin(\sigma/R)
  \ .$$
The metric (\ref{eq:156}) then becomes
\begin{equation}\label{eq:176}
ds^2=
  (L/z)^2
  \left[
  dz^2+
  \left(
  \frac{4R^2-z^2}{4R^2\sin(\hat{\sigma}/R)}
  \right)^2
  \left[
  -d\hat{\sigma}^2+R^2\left(
  d\theta^2+\sin^2(\theta)d\Omega^2(\vec{v}\,)
  \right)
  \right]
  \right]\ ,
\end{equation}
which is in Fefferman-Graham coordinates with a dS boundary. Using the relations between the different coordinates we can obtain an expression for $z$ in terms of the coordinates $(\psi,\hat{\theta}_\pm)$. Imposing also the constraint $\hat{\theta}_+=0$ which is satisfied by the extremal surface we find
\begin{equation}\label{eq:157}
\frac{z}{2R}=
  \frac{\cot(\hat{\theta}_-/2)-
  \sqrt{\cot^2(\hat{\theta}_-/2)-\sin^2(\psi)}}
  {\sin(\psi)}\ ,
\end{equation}
where $\hat{\theta}_-(\psi,\vec{v}\,)$ determines the extremal surface. This has a near boundary expansion given by
$$\hat{\theta}_-(\psi,\vec{v}\,)=
  2\,{\rm arccot}\left(
  \bar{A}(\vec{v}\,)
  \right)+b_1(\vec{v}\,)\psi+\dots\ ,$$
with the first order term determined from the boundary condition (\ref{eq:135}). Evaluating this in (\ref{eq:157}) with $z_{\rm min}=\epsilon$ we get an expansion that we can invert to get $\psi(\epsilon,\vec{v}\,)$ and find
$$\psi(\epsilon,\vec{v}\,)=
  \begin{cases}
  \quad
  \displaystyle
  \bar{A}(\vec{v}\,)
  (\epsilon/R)+\mathcal{O}(\epsilon/R)^2\ ,
  \qquad \bar{A}(\vec{v}\,)>0\ , 
  \vspace{5pt}\\ \quad
  \displaystyle
  4\bar{A}(\vec{v}\,)(R/\epsilon)
  +\mathcal{O}(1)
  \quad\,\ ,
  \qquad \bar{A}(\vec{v}\,)<0\ .
  \end{cases}$$
From this we see that the case in which $\bar{A}(\vec{v}\,)<0$ is anomalous since the limit $\epsilon\rightarrow 0$ gives a divergence in $\psi$. We can understand this by noting that for $\bar{A}(\vec{v}\,)<0$ the corresponding space-time region $\mathcal{D\bar{A}}^+$ in de Sitter lies outside the space-time (see Fig. \ref{fig:6}). Since the entanglement entropy is a non-local quantity which captures global information about the region it is no surprise that the calculation breaks down in this regime. The case $\bar{A}(\vec{v}\,)=0$ is also anomalous from this perspective and corresponds to taking the space-time region $\mathcal{D\bar{A}}^+$ as the de Sitter static patch. 

If we restrict to the $\bar{A}(\vec{v}\,)>0$ case, we find that the entanglement entropy for even $d$ can be written as
$$
S=\frac{\mu_{d-2}}{\epsilon^{d-2}}+
  \dots+
  (-1)^{\frac{d-2}{2}}
  \frac{4a_d^*}
  {{\rm Vol}(S^{d-2})}
  \int_{S^{d-2}}
  d\Omega(\vec{v}\,)
  \ln\left[
  \frac{2R}{\epsilon \bar{A}(\vec{v}\,)}
  \right]\ .
$$

\subsection{Anti-de Sitter}

Finally we consider the entanglement entropy for a CFT in a fixed AdS$_d$ background given by the space-time region associated to the null surface (\ref{eq:158}). The boundary condition for the extremal surface is now given by
$$\lim_{\psi\rightarrow 0}
  (\hat{\theta}_+,\hat{\theta}_-,\vec{v}\,)=
  (0,2\,{\rm arctan}\left(\bar{A}(\vec{v}\,)\right),\vec{v}\,)\ ,$$
where $\bar{A}(\vec{v}\,)>0$. Considering the AdS$_{d+1}$ bulk metric as in (\ref{eq:127}) we get an AdS$_d$ boundary by taking the limit $\psi\rightarrow 0$ with the additional conformal factor $\cos^2(\theta)$, so that we get (\ref{eq:154}). The entropy is still given by (\ref{eq:121}) where we must regulate with an appropriate cut-off obtained from the Fefferman-Graham bulk coordinates.

To find these coordinates we first apply the transformation in (\ref{eq:159}) to (\ref{eq:127}) so that we get pure AdS$_{d+1}$ in the standard coordinates in (\ref{eq:156}). Once we have the metric in this form we can define the new coordinates $(\rho,z)$ given by
$$\varrho^2=
  \left(
  \frac{4R^2+z^2}{4 R z}
  \right)^2
  (\rho^2+L^2)
  -L^2\ ,
  \qquad \qquad
  \tan(\theta)=
  \frac{\rho}{L}
  \left(\frac{4R^2+z^2}{4R^2-z^2}\right)\ ,
  $$
where $z\in[0,2R]$ and we restrict $\theta\in[0,\pi/2]$. The inverse transformation can be written as
$$\frac{z}{2R}
  =-(\varrho/L)\cos(\theta)+
  \sqrt{(\varrho/L)^2\cos^2(\theta)+1}\ ,
  \qquad \qquad
  \rho=
  \frac{L\varrho\sin(\theta)}
  {\sqrt{\varrho^2\cos^2(\theta)+L^2}}\ .$$
This takes the AdS$_{d+1}$ bulk metric in (\ref{eq:156}) to the appropriate Fefferman-Graham coordinates
$$ds^2=
  (L/z)^2\left[
  dz^2+
  \left(\frac{4R^2+z^2}{4 L R}\right)^2
  \left[
  -\left(
  \frac{\rho^2+L^2}{R^2}\right)d\sigma^2+
  \left(\frac{L^2}{\rho^2+L^2}\right)d\rho^2+
  \rho^2d\Omega^2(\vec{v}\,)
  \right]
  \right]\ .$$
From this we can write the relation between $z$ and the coordinates $(\psi,\hat{\theta})$ as
$$
\frac{z}{2R}=
\frac{\sqrt{1+\sin^2(\hat{\theta})\tan^2(\hat{\theta})\sin^2(\psi)\cos^2(\psi)}
  -\sqrt{1-\sin^2(\hat{\theta})\sin^2(\psi)}}
  {\sin(\hat{\theta})\sin(\psi)
  \sqrt{1+\tan^2(\hat{\theta})\cos^2(\psi)}}\ .
$$
Evaluating at $z_{\rm min}=\epsilon$ and expanding for the extremal surface near the boundary 
$$\hat{\theta}(\psi,\vec{v}\,)={\rm arctan}\left(\bar{A}(\vec{v}\,)\right)+b_1(\vec{v}\,)\psi+\mathcal{O}(\psi^2)\ ,$$
we can invert the relation and find the appropriate cut-off to regulate the integral (\ref{eq:121})
$$\psi(\epsilon,\vec{v}\,)=
  \frac{(\epsilon/R)}{\bar{A}(\vec{v}\,)}  
  +\mathcal{O}(\epsilon/R)^2\ .$$
The entanglement entropy for even values of $d$ is then given by
$$
S=\frac{\mu_{d-2}}{\epsilon^{d-2}}+
  \dots+
  (-1)^{\frac{d-2}{2}}
  \frac{4a_d^*}
  {{\rm Vol}(S^{d-2})}
  \int_{S^{d-2}}
  d\Omega(\vec{v}\,)
  \ln\left[
  \frac{2R}{\epsilon}\bar{A}(\vec{v}\,)
  \right]\ .
$$

\section{Modular conjugation in cylinder from embedding formalism}
\label{zapp:Mod_conj}

In this appendix we use the embedding formalism of the conformal group to map the geometric action of ${\rm CIT}$ operator in (\ref{eq:36}) to the Lorentzian cylinder. The main idea of the embedding space formalism is to embed the space-time of the CFT into a larger space where conformal transformations act linearly. Since the conformal group is isomorphic to ${\rm SO}(d,2)$ we define the embedding coordinates~${\xi\in\mathbb{R}^{d,2}}$ 
$$\xi=(\xi^0,\xi^i,\xi^d,\xi^{d+1})\ ,$$
in the space
\begin{equation}\label{eq:94}
ds^2=
  -(d\xi^0)^2
  +\sum_{i=1}^{d-1}(d\xi^i)^2
  +\left[
  (d\xi^{d})^2-(d\xi^{d+1})^2
  \right]
  \ .
\end{equation}
Every group element $g\in {\rm SO}(d,2)$ has a representation in terms of a matrix $M_g$ which has a linear action in the embedding coordinates given by ordinary matrix multiplication $\xi'=M_g\, \xi$. The relation with the $d$-dimensional space-time of the CFT is obtained as follows. 

We first define the projective null cone as
\begin{equation}\label{eq:97}
\mathcal{PC}=
  \frac{\left\lbrace  
  \xi\in \mathbb{R}^{2,d}\,\,:
  \quad
  (\xi \cdot \xi)=0
  \right\rbrace}
  {\xi\sim c \,\xi\ ,\,\, c\in \mathbb{R}_{+}}\ ,
\end{equation}
where $(\xi\cdot \xi)$ is computed using the embedding metric (\ref{eq:94}). The denominator means that there is a gauge redundancy in the scaling of $\xi$. To obtain the $d$-dimensional Minkowski space-time we use this gauge freedom to fix $\xi^+=\xi^d+\xi^{d+1}=R$ (called the Poincare section) with $R$ an arbitrary length scale. With this gauge choice we can parametrize $\xi\in {\mathcal{PC}}$ as
\begin{equation}\label{eq:95}
\xi(x)=
  \left(x^\mu,\frac{R^2-(x\cdot x)}{2R},
  \frac{R^2+(x\cdot x)}{2R}\right)\ ,
\end{equation}
where $x^\mu=(t,\vec{x})$ and $(x\cdot x)=\eta_{\mu \nu}x^\mu x^\nu$. Using this we compute the induced metric in $\mathcal{PC}$ and obtain~$d$-dimensional Minkowski $ds^2=d\xi(x)\cdot d\xi(x)=\eta_{\mu \nu}dx^\mu dx^\nu$. By considering a different section of the projective null cone in which $\xi^+=R/w(x)$ we can obtain a different $d$-dimensional space-time that is conformally related to Minkowski.

Let us now describe how a conformal transformation is induced by the linear action of~${M_g\in {\rm SO}(d,2)}$. Since $M_g\,\xi$ might take us off the section of the projective cone that we started from (\textit{i.e.} $\xi(x)^+\neq (M_g\,\xi(x))^+$) we must also apply a rescaling, so that the overall transformation is given by
\begin{equation}\label{eq:98}
\xi(x)
  \qquad \longrightarrow \qquad
  M_g\,\xi(x)
  \qquad \longrightarrow \qquad
  \frac{\xi(x)^+}{(M_g\,\xi(x))^+}
  M_g\,\xi(x)=
  \xi(x')\ .
\end{equation}
This induces a transformation from $x\rightarrow x'$ that corresponds to a conformal transformation in the~$d$-dimensional space-time of the CFT.

We now want to apply this formalism to obtain the linear transformation in the embedding space~$\xi$ which implements the action of ${\rm CIT}$ (\ref{eq:36}) in the Poincare section. Consider the matrix $M_{\tilde{g}}$ which implements a rotation of angle $\pi$ between the embedding coordinates $(\xi^0,\xi^d)$, so that we have
\begin{equation}\label{eq:96}
M_{\tilde{g}}\,\xi=
  (-\xi^0,\xi^i,-\xi^d,\xi^{d+1})\ .
\end{equation}
Following the prescription described in (\ref{eq:98}) the transformation in the embedding coordinates is given by
$$\tilde{\xi}(x)=
  \frac{R^2}{(x\cdot x)}
  \left(
  -t,\vec{x},\frac{-R^2+(x\cdot x)}{2R},
  \frac{R^2+(x\cdot x)}{2R}
  \right)\ ,$$
where the rescaling by $R^2/(x\cdot x)$ is such that $\tilde{\xi}(x)^+=R$. Comparing this expression with $\xi(\tilde{x})$ in~(\ref{eq:95}) we find that the induced transformation in $x^\mu$ is given by
$$\tilde{x}^\mu(x)=
  \frac{R^2}{(x\cdot x)}(-t,\vec{x})\ ,$$
that is precisely the ${\rm CIT}$ reflection in (\ref{eq:36}). This shows that $M_{\tilde{g}}$ given in (\ref{eq:96}) implements the reflection transformation in the embedding space. Notice that although $M_{\tilde{g}}$ does not correspond to a conformal transformation since $M_{\tilde{g}}\not \in {\rm SO}(d,2)$, it belongs to the Euclidean conformal group~${{\rm SO}(d+1,1)}$. This is analogous to what happens with the ${\rm CRT}$ operator that is not in the Lorentz group but is part of the Euclidean group.

Using this we can easily obtain the action of ${\rm CIT}$ applied to the Lorentzian cylinder $\mathbb{R}\times S^{d-1}$. To do so, we consider a different section of the projective null cone $\mathcal{PC}$ obtained from the following parametrization
\begin{equation}\label{eq:100}
\xi(\sigma,\theta,\vec{n})=
  R
  \big(
  \sin(\sigma/R),\sin(\theta)\vec{n},
  \cos(\theta),\cos(\sigma/R)
  \big)\ ,
\end{equation}
where $\vec{n}\in \mathbb{R}^{d-1}$ such that $|\vec{n}|^2=1$. This is a vector in the projective null cone in the section given by
\begin{equation}\label{eq:99}
\xi^+=
  2R
  \cos(\theta_+/2)\cos(\theta_-/2)\ ,
\end{equation}
where $\theta_\pm=\theta\pm \sigma/R$. The $d$-dimensional induced metric in (\ref{eq:94}) is given by
\begin{equation}\label{eq:102}
ds^2=d\xi(\sigma,\theta,\vec{n})
  .d\xi(\sigma,\theta,\vec{n})=
  -d\sigma^2+R^2\left(
  d\theta^2+\sin^2(\theta)ds^2_{S^{d-2}}
  \right)\ ,
\end{equation}
that is the Lorentzian cylinder $\mathbb{R}\times S^{d-1}$. 

Considering the action of $M_{\tilde{g}}$ in (\ref{eq:96}) and following the procedure in (\ref{eq:98}), we find
$$\tilde{\xi}(\sigma,\theta,\vec{n})=
  \frac{R}{\tan(\theta_+/2)\tan(\theta_-/2)}
  \big(
  -\sin(\sigma/R),\sin(\theta)\vec{n},
  -\cos(\theta),\cos(\sigma/R)
  \big)\ ,$$
where the rescaling ensures that we remain in the section (\ref{eq:99}). Comparing with $\xi(\tilde{\sigma},\tilde{\theta},\vec{n})$\footnote{We find that it is consistent to assume that the unit vector $\vec{n}$ is not changed by the transformation.} in (\ref{eq:100}) the reflection transformation in the cylinder is given by
\begin{equation}\label{eq:101}
\tan(\tilde{\theta}_\pm)=-\tan(\theta_\pm)
  \qquad \Longrightarrow \qquad
  \tilde{\theta}_\pm(\theta_\pm)=\pi-\theta_\pm\ ,
\end{equation}
where to get rid of the trigonometric functions we imposed $\theta_++\theta_-=2\theta\in[0,2\pi]$. This is a simple linear relation that reflects points across the wedge $\theta_\pm=\pi/2$, that is the fixed point of the transformation.

Since there should not be anything special about $\theta_0=\pi/2$, we would like to generalize (\ref{eq:101}) to arbitrary values of $\theta_0\in[0,\pi]$. Inspired by some calculations in \cite{Casini:2011kv}, we can do so by considering a slight variation of the parametrization (\ref{eq:100}), given by boosting the embedding coordinates in the~$(\xi^d,\xi^{d+1})$ direction
$$
\begin{aligned}
\xi^0&=R\sin(\sigma/R)\ , \qquad \qquad \,\,
\xi^i=R\sin(\theta)n^i\ , \\
\xi^d&=R\cosh(\gamma)\cos(\theta)+R\sinh(\gamma)\cos(\sigma/R)\ ,\\
\xi^{d+1}&=R\cosh(\gamma)\cos(\sigma/R)+R\sinh(\gamma)\cos(\theta)\ ,
\end{aligned}
$$
where $\gamma\in \mathbb{R}$ is the boost parameter. Since the boost is an isometry of the embedding space, the vector $\xi$ is still null and gives the same induced metric as in (\ref{eq:102}). However the gauge condition $\xi^+$ is slightly different
\begin{equation}\label{eq:105}
\xi^+=
  2Re^{\gamma}\cos(\theta_+/2)\cos(\theta_-/2)\ .
\end{equation}
Repeating the calculation leading to (\ref{eq:101}) but for $\gamma\neq 0$, we find that the induced reflection transformation is now given by
\begin{equation}\label{eq:103}
\tan(\tilde{\theta}_\pm)=
  \frac{-\sin(\theta_\pm)}
  {\cos(\theta_\pm)\cosh(2\gamma)+\sinh(2\gamma)}\ .
\end{equation}
The value of $\gamma$ determines the size of the wedge $\theta_0$ in the cylinder where the reflection is applied. The relation between $\gamma$ and $\theta_0$ can be found by looking at the fix point of the transformation (\ref{eq:103}), so that we get $e^{\gamma}=\tan(\theta_0/2)$. Since we cannot analytically solve $\tilde{\theta}_\pm(\theta_\pm)$ for arbitrary $\theta_0\in [0,\pi]$ we compute it numerically and obtain the diagram in Fig. \ref{fig:11}. Although the relation in (\ref{eq:103}) written in terms of $\theta_0$ is quite complicated, it is straightforward to check that the same reflection transformation is obtained from the following simpler relation\footnote{This relation was obtained from the action of ${\rm CIT}$ in Minkwoski given in~(\ref{eq:32}) and applying the conformal transformation to the cylinder $r_\pm(\theta_\pm)=R\tan(\theta_\pm/2)/\tan(\theta_0/2)$. Notice that this works for the Minkowski region in which $r_+r_->0$. For the regions $r_+r_-<0$ one must carefully analyze other coordinates to the cylinder, given by~(\ref{eq:10}) with $R\rightarrow R/\tan(\theta_0/2)$. Doing so, one obtains the same result as in~(\ref{eq:104}).}
\begin{equation}\label{eq:104}
\tan(\tilde{\theta}_\pm/2)=
  \tan^2(\theta_0/2)\cot(\theta_\pm/2)\ .
\end{equation}

\bibliography{sample}
\bibliographystyle{JHEP}

\end{document}